\documentclass[times, tighten]{aastex631}

\usepackage{sansmath}

\usepackage{graphicx}
\usepackage{url}
\usepackage{epsfig}
\usepackage{upgreek}
\usepackage{xcolor}
\usepackage{textcomp}
\usepackage{booktabs}
\usepackage{ulem}
\usepackage{amsmath}
\usepackage{tikz}
\usetikzlibrary{shapes.geometric, arrows}
\usepackage{float}
\usepackage{hyperref}

\usepackage{amssymb}
\usepackage{mathrsfs}
\makeatletter
\usepackage[symbol]{footmisc}
\usepackage{multirow}

\let\oldequation\equation
\let\oldendequation\endequation
\renewenvironment{equation}
{\linenomathNonumbers\oldequation}
{\oldendequation\endlinenomath}

\def\kms{\rm km\,s^{-1}}

\def\rmd{{\rm d}}

\def\betaBH{\beta_{\bullet}}
\def\bhm{m_{\bullet}}
\def\BHM{M_{\bullet}}

\def\ergs{\rm erg~s^{-1}}

\def\gammaBH{\gamma_{\bullet}}

\def\kms{\rm km~s^{-1}}

\def\Mdot{\dot{M}_{\bullet}}
\def\mathdotM{\dot{\mathscr{M}}}

\def\mathN{\mathscr{N}}
\def\muc{\multicolumn}

\def\pBHM{\langle M_{\bullet}\rangle}

\def\Rg{R_{\rm g}}

\def\rmd{{\rm d}}

\def\Sigbh{\mathscr{M}_{m_{\bullet}}}
\def\sunm{M_{\odot}}
\def\sunmyr{\sunm\,{\rm yr}^{-1}}

\def\FCAMS{F_{\lambda,{\rm AMS}}}
\def\FCMBH{F_{\lambda,{\rm MBH}}}
\def\FCnsb{F_{\lambda,\rm NSB}}

\def\calR{{\cal{R}}}
\def\calC{{\cal{C}}}
\def\calJ{{\cal{J}}}

\def\calN{{\cal{N}}_{\rm cl}}

\def\sMBH{s\kern -1pt@\kern-1pt cMBH-disk}
\def\cMBH{cMBH}

\def\OmegaK{\Omega_{\rm K}}

\let\saved@includegraphics\includegraphics
\makeatother

\hypersetup{
colorlinks=true,
linkcolor=blue,
citecolor=blue,
urlcolor=blue
}

\usepackage{datetime2}
\usepackage{gensymb}

\usepackage[utf8]{inputenc}

\def\ihep{Key Laboratory for Particle Astrophysics, Institute of High Energy Physics,
Chinese Academy of Sciences, 19B Yuquan Road, Beijing 100049, P.\,R. China}

\def\naoc{National Astronomical Observatories of China, Chinese Academy of Sciences,
 20A Datun Road, Beijing 100020, P.\,R. China}

\def\ucas{School of Astronomy and Space Science, School of Physical Sciences, University 
of Chinese Academy of Sciences, 19A Yuquan Road, Beijing 100049, P.\,R. China}

\def\CSU{Technology and Engineering Center for Space Utilization, Chinese Academy of Sciences, Beijing 100094, P.\,R. China}

\begin{document}

\title{\large Little red dots as embryos of active galactic nuclei}
\author{Jian-Min Wang}
\affiliation{\ihep}
\affiliation{\ucas}
\affiliation{\naoc}

\author{Yi-Lin Wang}
\affiliation{\ihep}
\affiliation{\ucas}

\author{Yong-Jie Chen}
\affiliation{\CSU}

\author{Jun-Rong Liu}
\affiliation{\ihep}

\author{Yu-Yang Songsheng}
\affiliation{\ihep}

\author{Cheng Cheng}
\affiliation{\naoc}

\author{Yan-Rong Li}
\affiliation{\ihep}

\author{Pu Du}
\affiliation{\ihep}

\author{Hao Zhang}
\affiliation{\ihep}
\affiliation{\ucas}

\author{Yu Zhao}
\affiliation{\ihep}
\affiliation{\ucas}

\begin{abstract}
As an unprecedented large population in the early universe, the JWST-discovered little red dots (LRDs) have garnered much attention for formation of massive black holes and galaxies, but their nature remains a mystery. 
The LRDs appearing as ``Chimeras" like both active galactic nuclei (AGNs) and galaxies have stimulated renewed interest in the roadmap of central massive black hole (cMBH) formation in AGNs.
In this paper, we suggest that the LRDs contain $M_{\bullet}\lesssim 10^6\,\sunm$ cMBHs as demonstrated by the So{\l}tan argument and there is a large population of stellar-mass black holes (sMBHs with total mass of $\mathscr{M}_{m_{\bullet}}$) embedded inside cMBH accretion disks (cMBH-disk) as motivated by anomalous reverberations of broad H$\beta$ line in local AGNs.
This embryo structure of LRDs ($M_{\bullet}<\mathscr{M}_{m_{\bullet}}$) is formed as a consequence of gravitational collapse of primordial clouds.
In this Chimera, accretion onto sMBHs powers the rest-frame optical continuum of the LRDs but the UV continuum is jointly contributed by slim parts of the cMBH-disks and nuclear starbursts in the core of collapsing clouds governing the appearance of the observed V-shaped spectral energy distributions (SEDs).
Outflowing clumped-envelopes are unavoidably formed by radiation pressure leading to absorption features of the Balmer lines.
The present model works very well for LRDs' SEDs and avoids the issues of overly massive cMBHs.
Evolution of LRDs is briefly discussed including gravitational waves.
\end{abstract}
\subjectheadings{Active galactic nuclei (16); Galaxy formation (595); High-redshift
galaxies (734); Supermassive black holes (1663)}

\section{Introduction}
The unusual features of the LRDs \citep{Labbe2023,Maiolino2024faint,Matthee2024,Kokorev2024,Kocevski2025} challenge several fundamental aspects, and have prompted a dramatic wave of various explanations for them (see a brief summary in \S\,\ref{sec:models}).
The prominent observational characteristics of the LRDs are summarized here.
They are: 1) the optical spectra are extremely red, but a turnover of UV continuum appears in rest frame \citep{Labbe2023,Greene2024} at a rough constant frequency (in agreement with the Balmer break, i.e., \citealt{Setton2024}). This feature, known as the V-shaped continuum, is commonly accompanied by a blueshifted trough in the spectra, referred to as the absorption profile \citep[except for Abell\,2744-QSO1 with redshift trough, see][]{Ji2025Q1}. In particular, the UV continuum is not simple version of star formation in the early universe \citep{WangBJ2024}; 2) their sizes are extremely small ($\lesssim 0.1\sim$0.2\,kpc) \citep{Baggen2023}; 3) the abundances are almost $100\sim 2300$ times that of AGNs in the early universe \citep{Pizzati2025,MaYL2025}; 4) extremely low X-ray emissions \citep{Yue2024,Ananna2024}; 5) non-variable fluxes over the observed bands \citep{Zhang2024,Tee2025}; 6) the cMBHs are so massive that they deviate from the well-known Magorrian relation by a factor of $\sim100$ if the masses are estimated using the normal AGN scaling relation (Eq.\,\ref{eq:MBHmass} in Appendix \ref{sec:challenges}) \citep{Maiolino2024,LiJY2025};  7) the LRDs break down the well-known So{\l}tan argument if they are really massive (see \S\,\ref{sec:Soltan});
8) the number density of the LRDs rapidly declines after $z=3-4$, and peaks roughly at $z=4-6$, preceding the peak of the star formation history by $\Delta t\sim 10^8\,$yr \citep{Zhuang2025};
9) it has been found \citep{MaYL2025} that the LRDs have a sharp cutoff in luminosity around $L_{5100}\sim 2.5\times 10^{44}\,\ergs$, ruling out the possibility that LRDs are the counterparts of quasars.
All these properties seem diverse, but it necessitates a theoretical model for a unified explanation.
However, it is not trivial to construct such a model in light of current understanding of AGNs and galaxies.

The LRDs look like AGNs because broad emission lines dominate and they (a sample of 99 LRDs) are located in AGN regions in the BPT diagram \citep{Rinaldi2024,ZhangZJ2025}.
It has been shown that H$\alpha$ emissions are few factors brighter than normal AGNs \citep{Hviding2025}, and that they are less clustered than galaxies \citep{Carranza-Escudero2025}.
Moreover, the brightest LRD recently discovered implies that LRDs are not the counterpart of quasars \citep{MaYL2025}. 
Thus the LRDs with distinctive SEDs are not the canonical version of AGNs.
On the other hand, the LRDs are less likely to be mere galaxies because the shape of the rest-frame UV continuum is different from typical high-$z$ star forming galaxies (much flattened in LRDs, see Fig.\,3 in \citealt{Perez-Gonzalez2024,Taylor2025}).
As the astronomical version of the ``Chimeras", the LRDs may actually ``take" some parts of key ingredients of AGNs and galaxies to form their special structures in the early universe. 
These Chimeras mostly appearing in the early universe ($z\gtrsim 3\sim 4$) motivate us to think about the fates of collapsing primordial clouds (CPCs: \citealt{Hoyle1953,Hunter1962,Field1964}) along the roadmap of formation of cMBHs in quasars \citep{Rees1984}.
It has been realized that the origin of the rest-frame optical continuum is a matter of intense debate whether it is powered by AGNs, stars or a mixture of both \citep{WangBJ2024}, even something completely new.

In this paper, we will focus on the issues related to the continuum. 
The So{\l}tan argument is briefly discussed for the central massive black holes in \S\,\ref{sec:Soltan}.
Motivated by anomalous reverberations of broad-line regions in the quasar PHL\,1092 \citep[][see a brief explanation in Appendix \ref{sec:PHL1092}]{Wang2025}, we suggest a model in \S\,\ref{sec:Model} that the accretion disk of the cMBH contains a large population of stellar mass black holes (sMBHs).
The joint contributions from accretion onto both the sMBHs and the cMBH generate the V-shape spectral energy distributions (SEDs) in the LRDs.
We find that the present model of cMBH-disk works quite well for the LRDs in \S\,\ref{sec:application}.
Discussions on current models of the LRDs are provided in \S\,\ref{sec:models}, and other issues relating to the cosmic evolution of the proposed scenarios will be addressed in future work. 
We draw conclusions in the last section.

\section{A crisis of the So{\L}tan argument}\label{sec:Soltan}
First, the overabundance of the LRDs breaks down the well-known So{\l}tan argument for accretion growth of SMBHs over cosmic time \citep{Soltan1982,Yu2002} if the cMBH masses of LRD are estimated by Eq.\,(\ref{eq:MBHmass}).
Given the luminosity functions of quasars and LRDs by $\Phi_{\rm QSO}(L_{\rm Bol},z)$ and $\Phi_{\rm LRD}(L_{\rm Bol},z)$, respectively, we have the So{\l}tan argument given by 
\begin{equation}\label{eq:Soltan}
\rho_{\rm acc}(z=0)=\displaystyle{\int_{0}^{\infty}\left(\frac{\rmd t}{\rmd z}\right)\rmd z\int_0^{\infty}\Phi_{\rm QSO}\left(\frac{L_{\rm Bol}}{\eta_{\bullet} c^2}\right)\rmd L_{\rm Bol}} +\int_{0}^{\infty}\left(\frac{\rmd t}{\rmd z}\right)\rmd z\int_0^{\infty}\Phi_{\rm LRD}\left(\frac{L_{\rm Bol}}{\eta_{\bullet} c^2}\right)\rmd L_{\rm Bol},
\end{equation}   
where $t$ is the cosmic time, $\eta_{\bullet}$ is the radiation efficiency (for $\eta_{\bullet}\approx 0.1$), $L_{\rm Bol}$ is the bolometric luminosity of AGNs, and $c$ is the speed of light. 
It has been found that $\rho_{\rm acc}(z=0)=\rho_{\bullet}(z=0)$, where $\rho_{\bullet}(z=0)$ is the mass density of SMBHs in local galaxies \citep{Yu2002,Marconi2004}.
However, including the LRDs with luminosity function $\Phi_{\rm LRD}\approx \zeta_z\times \Phi_{\rm QSO}(L_{\rm Bol},z\approx 4\sim8)$ with a cutoff around $z=3\sim 4$ leads to a severe tension with the So{\l}tan argument \citep{Inayoshi2024,Jahnke2025}.
Moreover, we have the mean masses of SMBHs due to accretion
\begin{equation}
    \pBHM=\frac{\displaystyle{\int \left(\frac{\rmd t}{\rmd z}\right)\rmd z\int \dot{M}_{\rm acc} \Phi(z,L_{\rm Bol}) \rmd L_{\rm Bol}}}
    {\displaystyle{\int \rmd z\int \Phi(z,L_{\rm Bol}) \rmd L_{\rm Bol}}},\quad
\dot{M}_{\rm acc}=\frac{L_{\rm Bol}}{\eta_{\bullet} c^2},
\end{equation}
yielding 
\begin{equation}\label{eq:pBHMG}
 \pBHM_{\rm G}\int \Phi_{\rm G}(z=0,L)\rmd L=
 \left(\pBHM_{\rm QSO}+\zeta_z\pBHM_{\rm LRD}\right)\int \rmd z\int \Phi_{\rm QSO}(L_{\rm Bol},z)\rmd L_{\rm Bol}, 
\end{equation}
where $\pBHM_{\rm G}$ is the mean masses of SMBHs in local normal galaxies and $\Phi_{\rm G}(z=0,L)$ is their luminosity function.
In principle, the relation $\int \Phi_{\rm G}(z=0,L)\rmd L=
 \int \rmd z\int \Phi_{\rm QSO}(L_{\rm Bol},z)\rmd L_{\rm Bol}$ holds since all galaxies are from evolved quasars, and we have
\begin{equation}\label{eq:pSoltan}
\pBHM_{\rm G}= \pBHM_{\rm QSO}+\zeta_z\pBHM_{\rm LRD}.  
\end{equation}
Obviously, the So{\l}tan crisis arises, namely $\pBHM_{\rm G}\ll \pBHM_{\rm QSO}+\zeta_z\pBHM_{\rm LRD}\approx \zeta_z\pBHM_{\rm QSO}$ if $\pBHM_{\rm LRD}\sim \pBHM_{\rm QSO}$ from Eq.\,(\ref{eq:MBHmass}).

On the other hand, Eq.\,(\ref{eq:pSoltan}) yields
\begin{equation}\label{eq:pBHM}
\pBHM_{\rm LRD}=\zeta_z^{-1}\delta\pBHM
\lesssim 10^6\,\zeta_2^{-1}\delta\pBHM_{8}\,\sunm,   
\end{equation}
where $\delta \pBHM=\pBHM_{\rm G}-\pBHM_{\rm QSO}$ represents the fluctuations of mass estimations of supermassive black holes (SMBHs), $\pBHM_{\rm G}$ and $\pBHM_{\rm QSO}$ are the mean masses of SMBHs in local galaxies and quasars, respectively, $\zeta_2=\zeta_z/10^2$ and $\delta\pBHM_{8}=\delta \pBHM/10^8\,\sunm$. 
Given the fluctuations in both galaxies and AGNs, $\delta\pBHM\sim \pBHM_{\rm QSO}$ holds.
It has been found that $\zeta_z\approx 10^2\sim 10^3$ in \citet{Pizzati2025} and \citet{MaYL2025}.
We should note that Eq.\,(\ref{eq:pBHM}) is the minimum requirement to satisfy the So{\l}tan argument, otherwise, the So{\l}tan crisis obviously arises, namely, the accretion masses as the origin of SMBH masses will be much higher than those of local SMBHs.
Therefore, the $\lesssim 10^6\,\sunm$ cMBHs greatly reduce the significant deviations of LRDs from the local Magorrian relation \citep{Maiolino2024}.
The issue of overmassive cMBHs as a paradox can be explained.

\begin{figure*}
   \centering
   \includegraphics[trim=-30 50 10 10, clip,angle=0,width=0.9\textwidth]{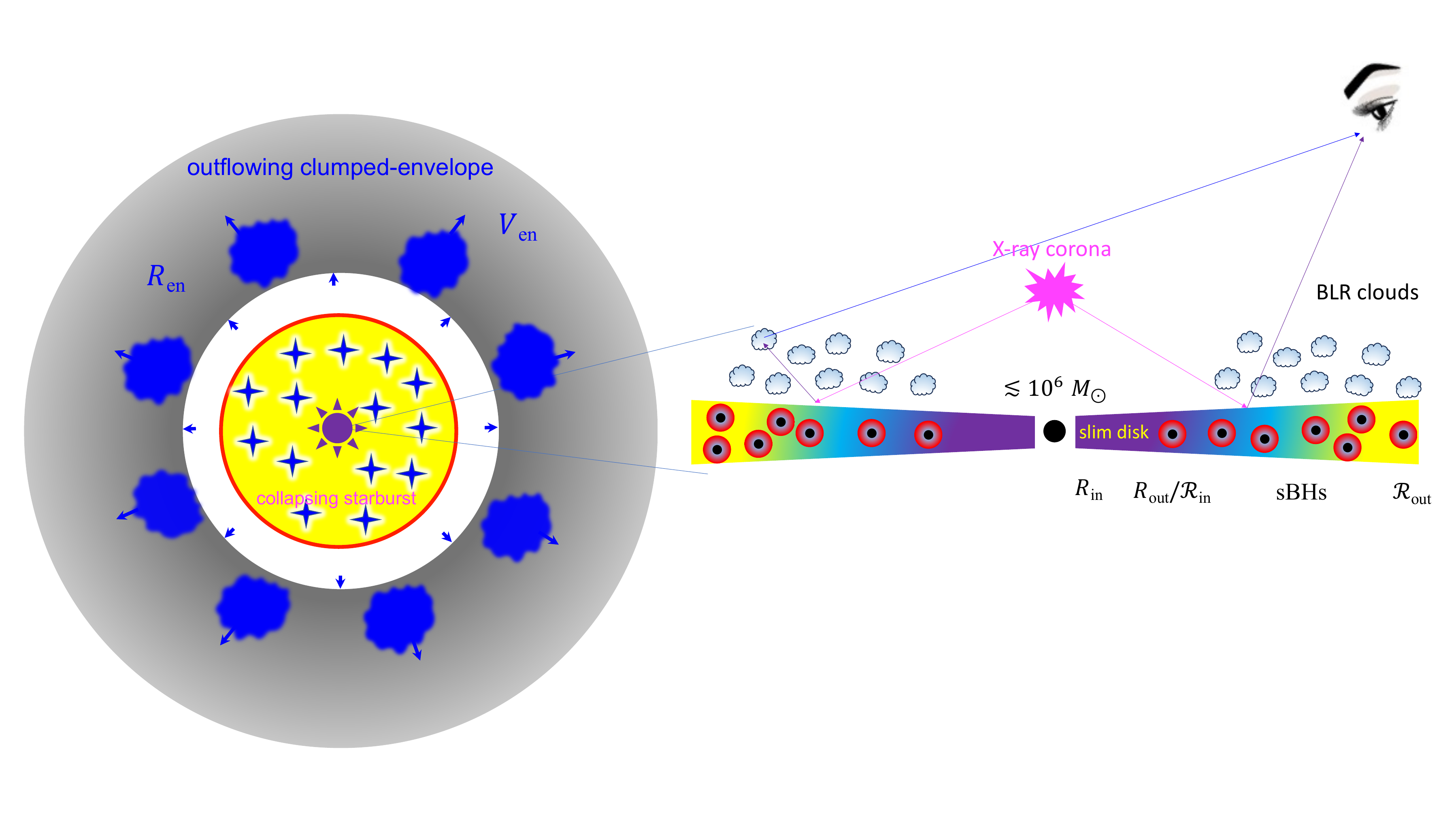}
{\caption{\footnotesize Structures of the LRDs formed during the collapse of primordial cloud proposed in this paper. 
The clumped-envelope covers the system composed of the BLR and the cMBH-disk system contributing the V-shaped SEDs of the LRDs. 
The cMBH-disk consists of two parts: 1) the slim disk of the central massive black hole (between $R_{\rm in}$ and $R_{\rm out}$); and 2) {\sMBH} represented by the part between $\calR_{\rm in}$ and $\calR_{\rm out}$ embedding a population of stellar-mass black holes.
Since the cMBHs are not significantly smaller than normal AGNs, the {\sMBH} system refers to an embryo of AGNs. 
Given the current mass density of a collapsing primordial clouds ($\rho_{\rm c}\sim 10^{-19}\,{\rm g\,cm^{-3}}$), an outflowing clumped-envelope is formed with a typical velocity of $\sim 100\,\kms$ by the radiation pressure, and has an optical depth of $\tau_{\rm abs}\approx 0.1-0.2$, explaining the absorption features when the clumps are on the line-of-sight.}
\label{fig:model}
}
\end{figure*}

This, however, raises another question of what is powering the observed rest-frame optical luminosity of LRDs greatly exceeding the Eddington luminosity of the $\lesssim 10^6\,\sunm$ cMBH accretion.
Motivated by the case of PHL\,1092 discussed in Appendix \ref{sec:PHL1092}, the cMBH-disk contains a population of sMBHs, which is denoted by the {\sMBH} system hereafter. 
In the LRDs, the total mass of sMBHs ($\mathscr{M}_{\bhm}$) dominates over that of the cMBH ($\BHM$), namely, $\mathscr{M}_{\bhm}\gtrsim\BHM$; the cMBH is still fast growing in the ``womb", and this structure is thereby named the embryos of AGNs. 
Accretion onto sMBHs powers the rest-frame optical continuum of the LRDs, while the rest-frame UV continuum is powered by super-Eddington accretion onto the cMBHs (slim disks), nuclear starbursts of the CPCs, or a combination thereof.
Fig.\,\ref{fig:model} shows the structures of the LRDs suggested in this paper.

\section{{\cMBH}-DISK and V-shaped SEDs}\label{sec:Model}
\subsection{The model}
The roadmap of cMBH formation known as the Rees' diagram \citep{Rees1984} is very motivating in the context of collapsing primordial cloud \citep{Hoyle1953,Hunter1962,Field1964} so that we have to re-consider the fates of primordial clouds whether they can evolve into LRDs along with the roadmap of the diagram \citep{Cenci2025,Kritos2025}.
Spin angular momentum (SAM) of the clouds is induced by tidal interaction with neighbors, and they acquire a dimensionless SAM of $\lambda_{\rm L}=\calJ|E_{\rm c}|^{1/2}G^{-1}M_{\rm c}^{-5/2}\approx 0.1$ if the cloud is approximated by a rigid body \citep{Peebles1969,Barnes1987}, where $\calJ$ is the cloud's SAM, $M_{\rm c}$ is the cloud mass, $G$ is the gravitational constant and $E_{\rm c}$ is the total energy.
However, the Compton drag due to cosmic background photons will efficiently remove the SAM ($\lambda_{\rm L}\lesssim 10^{-3}$) to form cMBHs in quasars \citep{Loeb1993} and a massive gaseous disk at a scale of parsecs \citep{Eisenstein1995,Choi2013}.

In core regions of the CPCs, a gaseous disk is formed since the primordial clouds have their own SAM. The disk radius is determined by the SAM, with $R_{\rm disk}\approx 2^{-1/2}\lambda_{\rm L}R_{\rm vir}=4.2\,\lambda_{\bar{3}}R_{\rm 6kpc}\,$pc \citep{Volonteri2005}, where $\lambda_{\bar{3}}=\lambda_{\rm L}/10^{-3}$ is the resultant SAM (after interaction with CMB) and $R_{\rm 6kpc}=R_{\rm vir}/6\,$kpc. Here, $R_{\rm vir}=6.6\,\left(M_{\rm H}/10^{10}h_{0.7}\sunm\right)^{1/3}\left[(1+z)/7\right]^{-1}h_{0.7}^{-1}\,$kpc \citep{Barkana2001} is the virial radius of dark matter halo, $M_{\rm H}$ is the dark matter halo mass and $h_{0.7}$ is the Hubble constant in units of $70\,{\rm km\,s^{-1}\,Mpc^{-1}}$.
Star formation takes place in such a gaseous disk \citep{Chen2024,Chen2025}, which has been studied to explain the relation between accretion rates and star formation \citep{ChenYM2009,Netzer2009} and the metallicity of AGNs \citep{Wang2010,Wang2011,Wang2012BLR,Wang2023,Fan2023}. 
Consequently, sMBHs evolved from massive stars remain inside of the accretion disks, which were originally realized as seed black holes by \citet{Eisenstein1995}. 
Despite the early recognition of these seeds more than three decades ago, their roles in the structure formation of nuclear regions and observational effects have long been overlooked.
If such an {\sMBH} system exists in AGNs, it yields anomalous reverberation of broad H$\beta$ line with respect to the varying continuum.
Actually, this structure could have been detected in the local quasar PHL\,1092 \citep{Wang2025} from the long-term SEAMBH campaign (see Appendix \ref{sec:PHL1092} for details).
Meanwhile, there is growing evidence for the presence of these black holes from LIGO gravitational wave detections and quasi-periodic ejections (QPEs) observed in X-rays (see discussions in Appendix \ref{sec:sMBH-disk}).

\renewcommand{\arraystretch}{0.9}
\begin{table*}
\centering
\footnotesize 
\caption{\bf Parameters of the \sMBH}\label{tab:parameter}
\begin{tabular}{lcll}\hline\hline
cMBH-disk&\multicolumn{2}{c}{} &Notes\\ \hline
$\BHM$    & $\ldots\ldots$ & cMBH masses & $\lesssim 10^6\,\sunm$\\ 
$\dot{M}_{\bullet}^0$&$\ldots\ldots$& accretion rates of the cMBH at $\calR_{\rm out}$ & $\sim 0.1\,\sunmyr$\\
$\alpha$ &$\ldots\ldots$& viscose parameters (keeps a constant in fitting) & 0.1\\ 
$R_{\rm in}$&$\ldots\ldots$ & radius of the inner  boundary of the slim disk & $3\,\Rg$\\
$R_{\rm out}$&$\ldots\ldots$ & radius of the outer boundary of the slim disk ($R_{\rm out}=\calR_{\rm in}$) & $\sim 10^5\,\Rg$\\
$\calR_{\rm in}$&$\ldots\ldots$ & radius of the inner boundary of the cMBH-disk & $\sim 10^5\,\Rg$\\
$\calR_{\rm out}$&$\ldots\ldots$ & radius of the outer boundary of the cMBH-disk & $\sim 10^6\,\Rg$\\ \hline
sMBHs&\multicolumn{3}{c}{} \\ \hline
$m_{\bullet}^0$& $\ldots\ldots$ & sMBH masses at $\calR_{\rm out}$, i.e.,  $m_{\bullet}=m_{\bullet}^0\left(R/\calR_{\rm out}\right)^{\betaBH}$&$\sim 1\,\sunm$ \\ 
$\Sigma_{\bullet}^0$&$\ldots\ldots$& surface number density of sMBHs at $\calR_{\rm out}$, i.e., $\Sigma_{\bullet}=\Sigma_{\bullet}^0\left(R/\calR_{\rm out}\right)^{\gammaBH}$ & $\sim 10^3\,{\rm ltd^{-2}}$ \\ 
$\betaBH$ &$\ldots\ldots$& index of sMBH mass growth migrating inward & $-0.5$\\ 
$\gammaBH$&$\ldots\ldots$& index of surface number density ($\Sigma_{\bullet}$) of sMBHs & $0.5$\\ 
$\ell_0$&$\ldots\ldots$& factor of the saturated luminosity & $1\sim 10$\\ 
$\Sigbh$&$\ldots\ldots$& total masses of sMBHs & $\sim 10^6\,\sunm$\\ \hline
Nuclear starburst&\multicolumn{3}{c}{} \\ \hline
$\alpha_0$&$\ldots\ldots$& normalization of NSB fluxes& $\sim 10^7$\\ 
$t_0$ & $\ldots\ldots$ & NSB ages& $0.1-0.5\,$Gyr \\
$Z$   & $\ldots\ldots$ & NSB metallicity (keeps a constant in fitting) & $1\,Z_{\odot}$ \\
\hline
\end{tabular}
\end{table*}

Moreover, the CPCs unavoidably undergo starbursts in the central region, forming a nuclear star cluster \citep{Neumayer2020} similar to the inside-out mode in galaxies, and the cluster grows \citep{Neumayer2020,Kritos2025}.
However, details of starbursts depend on the initial density profile of the cloud and the initial mass function of the star formation \citep{Abel2000,Abel2002} as well as the observational appearance of the CPC core \citep{Kritos2025}. 
It is far beyond the scope of this paper to self-consistently treat the nuclear starbursts (NSBs), but we include this component to fit the LRD continuum through {\tt{CIGALE}}.

Accordingly, the observed SEDs of the {\sMBH} and nuclear starbursts are determined by 
\begin{equation}\label{eq:continuum}
    F_{\lambda}^{\rm c}=\FCAMS+\FCMBH+\FCnsb,
\end{equation}
which are explained subsequently.
First, $\FCAMS$ is from the accreting sMBHs embedded inside the cMBH-disk, i.e., the {\sMBH} is the part between $(\calR_{\rm in}, \calR_{\rm out})$ and contributes the rest-frame optical continuum of the SEDs looking like a reddened AGN continuum.
Here the subscript AMS refers to accretion-modified stars \citep{Wang_outburst2021}, i.e., accreting sMBHs.
Second, $\FCMBH$, which is from the regions between ($R_{\rm in}, R_{\rm out}$), is responsible for ionizing the BLR for the broad Balmer emission lines.
Third, $\FCnsb$ is from a stellar component of nuclear starbursts (NSBs) \citep{Kritos2025}, but it might also partially originate from the \sMBH.
The observed Balmer break \citep{Setton2024} corresponds to post starbursts in the nuclear regions.
We would point out that $\FCAMS$ has never been considered before as the origin of the rest-frame optical continuum of the LRDs.
As a remarkable component, the presence of $\FCAMS$ makes the LRDs neither like a pure AGN nor a galaxy, appearing as a Chimera.

In Appendix \ref{sec:sMBH-disk}, full equations of the \sMBH\ are derived by simplifying the ecosystem composed of sMBHs inside the cMBH-disk. 
In the model, sMBHs are dynamically damped so efficiently that they co-rotate with cMBH-disks \citep{Wang_outburst2021}.
Motion effects of sMBHs on the Bondi accretion can be neglected.
Different points of the \sMBH\ system from known models of accretion disks \citep{Shakura1973} rest on: 1) extra heating of accretion onto sMBHs  ($Q_{\rm AMS}$) beyond accretion of the cMBH; 2) vertical equilibrium jointly governed by the cMBH and the sMBHs; 3) accretion rates of the cMBH decrease with radius due to accretion of sMBHs. 
For a simple illustration, the total energy budget expressed by Eq.\,(\ref{eq:Energy}) can be approximated as
\begin{equation}
\sigma_{\rm SB}T_{\rm eff}^4=Q_{\rm vis}+Q_{\rm AMS},
\end{equation}
where $Q_{\rm AMS}$ and $Q_{\rm vis}$ are given by Eqs.\,(\ref{eq:QsMBH}) and (\ref{eq:Qvis}).
Notably, the effective temperature distribution of the {\sMBH} is changed by $Q_{\rm AMS}$, resulting in a much flatter profiles than $T_{\rm eff}\propto R^{-3/4}$, leading to the rest-frame optical continuum of LRDs (i.e., $\FCAMS$).

Tab.\,\ref{tab:parameter} lists all parameters describing the present model. 
Defining $Q_{\rm AMS}^0 = \ell_0 L_{\rm Edd}^0 \Sigma_{\bullet}^0 (m_{\bullet}^0 / 1\sunm)$ (see Eq.\,\ref{eq:QsMBH}), we have only three parameters $(Q_{\rm AMS}^0,\beta_{\bullet},\gamma_{\bullet})$ except for accretion rates govern SEDs of the \sMBH\ system.
In Appendix \ref{sec:SED}, we provide structures and emergent SEDs of the \sMBH\ for different parameters. 
As shown there, V-shaped SEDs can be conveniently generated by the {\sMBH} system.
The SEDs depend mainly on the $(Q_{\rm AMS}^0,\beta_{\bullet},\gamma_{\bullet})$-parameters; in particular, $(\beta,\gamma)$ determine the turn-over frequencies of the SEDs. 

\subsection{sMBH population broadens H$\beta$ line}
As embryos of AGNs, the {\sMBH} system is of $\BHM\lesssim \Sigbh$, where sMBHs play an important role in the vertical structure of the cMBH-disk.
Details of the gravity potential of the {\sMBH} are very complicated, but we provide a simplified version of the broadening effects of the \sMBH.
The typical FWHM of H$\beta$ line near the mid-plane is given by ${\rm FWHM}\approx 2.0\times 10^3\,\left(\mathscr{M}_{m_{\bullet}}/10^8\,\sunm\right)^{1/2}\,\kms$ from Eq.\,(\ref{eq:FWHM}) in Appendix \ref{sec:profile}.
The popular estimation of the LRDs from Eq.\,(\ref{eq:MBHmass}) should be the total masses of the sMBHs and the cMBH, rather than the pure masses of the cMBH.
As the LRDs are embryos of AGNs, Eq.\,(\ref{eq:MBHmass}) doesn't apply to the estimation of cMBH masses.
A recent measurement of a lensed LRD \citep{Juodzbalis2025b} through dynamics actually still covers the sMBHs.
Moreover, the observed Balmer lines are usually fitted by two Gaussian components (e.g. \citealt{Zhuang2025,Rinaldi2024}).
The broad component is from the normal BLR, and the narrow component originates from the outflowing envelope with characteristic FWHM given by 
${\rm FWHM}=110\,\left(\Sigbh/{10^8\sunm}\right)^{1/2}\left(D_{\rm en}/{3\,{\rm pc}}\right)^{-1/2}\,\kms$.
Such a component is quite prevalent among the LRDs.

\renewcommand{\arraystretch}{1.0}
\begin{table*}
\centering
\footnotesize 
\caption{\bf Ten representative LRDs}
\begin{tabular}{lrrrcccc}\hline\hline
Name &  &{\rm RA} & {\rm Dec} & $z$ & absorption$^b$ & $v_{\rm abs}^c$ & Ref. \\  \hline
Abell\,2744-QSO1 & &3.58 & -30.40 & 7.037   &  Y & $36^{+10}_{-9}$ &1,2\\
JADES-28074 & &189.06 & 62.27 & 2.259  & N & - &3\\
RUBIES-EGS-49140 & &214.89 & 52.88 & 6.685 & Y & $\sim 50 $ &4,5  \\ \hline 
UNCOVER-4286 & &3.62 & -30.42 & 5.835  & N & - &6\\
6585-58018 & &150.14 & 2.34 & 3.928 & N & - &6\\
SDSS\,J\,1025+1402$^a$ & &156.38 & 14.04 & 0.101 & Y & $\sim 200$ &7\\
CEERS-EGS-1244 & &215.24 & 53.03 & 4.477  & N & - &8\\ \hline
RUBIES-UDS-40579 & &34.24 & -5.25 & 3.113 & N & - &6\\
RUBIES-EGS-42046 & &214.80 & 52.79 & 5.276 & Y & $\sim 160$ &6,8,9\\
RUBIES-EGS-55604 & &214.98  & 52.96 & 6.982  & Y & - &4\\ \hline
\muc{8}{l}{$^a$Local analog of LRDs.}\\
\muc{8}{l}{$^b$Absorption refers to the absorption signatures of the Balmer lines.}\\
\muc{8}{l}{$^c$Velocity of blueshifted H$\alpha$ absorption component, in units of ${\rm km\,s^{-1}}$.}\\
\muc{8}{l}{Ref. 1. \citet{Akins2025}; 2. \citet{DEugenio2025}; 3. \citet{Setton2024};}\\
\muc{8}{l}{\quad\quad 4. \citet{Hviding2025}; 5. \citet{Rusakov2025}; 6. \citet{Setton2024};}\\
\muc{8}{l}{\quad\quad 7. \citet{LinXJ2025b}; 8. \citet{Rusakov2025}; 9. \citet{Taylor2025};}
\end{tabular}
\label{tab:sample}
\end{table*}

\subsection{Absorption features of H$\alpha$ line governed by outflowing clumped-envelopes}
During the collapse of primordial clouds, an outflowing envelope is formed by radiation pressure (from AGN or nuclear starbursts; see some historical references) \citep[]{Matsuda1969,Larson1969,Unno1967}.
In Appendix \ref{sec:collapse}, we derive the characteristic radius and velocity of the envelope using a simplified model.
For the presumed density of the cloud $\rho_{\rm c}=10^{-19}\,{\rm g\,cm^{-3}}$, the derived envelope can explain the observed absorption features of the Balmer lines (which have scarcely been found in normal AGNs).
We obtain the characteristic radius of the envelope as $D_{\rm en}\approx 5\,{\rm pc}$, and the velocity of $V_{\rm en}\approx 110\,\kms$ from Eqs.\,(\ref{eq:Den}) and (\ref{eq:Ven}), respectively.
It is composed of clumps with an optical depth of the Balmer absorption as $\tau_{\rm abs}\approx 0.2$ from Eq.\,(\ref{eq:tauabs}).
The typical values of the envelope are in good agreement with absorption troughs of the LRDs.

\subsection{Other properties of LRDs}
As shown by \citet{Wang2004}, X-rays would be lowered by a factor of 10 in high Eddington ratio AGNs.
According to the magnetic field reconnection model of the hot corona, we have shown \citep{Wang2004} that the fraction of X-rays to the bolometric luminosity decreases with Eddington ratios.
Considering this fact, \citet{Leung2024,Lambrides2024,Inayoshi2025,Pacucci2024} suggest that the LRDs are undergoing super-Eddington accretion in order to explain their lack of X-rays.
The slim disk of the cMBH is radiating the ionizing photons for the broad Balmer lines.
Since super-Eddington AGNs usually have smaller amplitudes of variations \citep{Lu2019}, the LRDs should be similar to those local super-Eddington AGNs.
Including the cosmological factors, the variability amplitudes will be lowered by a factor of $(1+z)$ showing very weak variations.
Considering that the observed UV spectra are jointly contributed by nuclear starburst and {\sMBH} in some LRDs, we think that the variations are expected to be difficult to detect in both rest-frame UV and optical bands.

\section{Applications to the LRDs}\label{sec:application} 
LRDs show diverse shapes of SEDs.
We select several representative LRDs with high signal-to-noise ratios in order to illustrate applications of the present model.
The selected LRDs are classified as follows: 1) SEDs obviously need three components (slim disk, {\sMBH} and NSBs); 2) SEDs only need two components (slim disk and {\sMBH}); 3) multiple NSBs are necessary. 
Tab.\,\ref{tab:sample} lists the nine LRDs, which not only represent characteristic SED shapes but also across a wide range of redshifts. 
An analog, SDSS J1025+1402 \citep{LinXJ2025a}, is also selected.

\begin{figure*}
\centering
\includegraphics[width=0.9\textwidth]{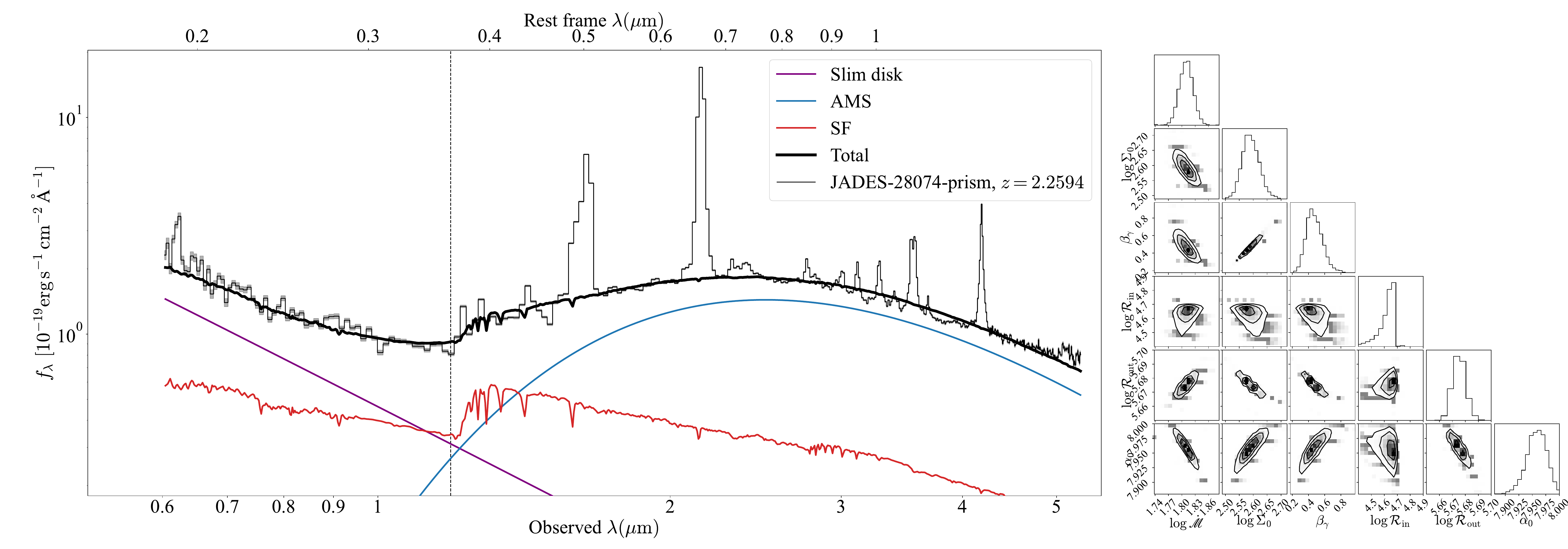} 
\includegraphics[width=0.9\textwidth]{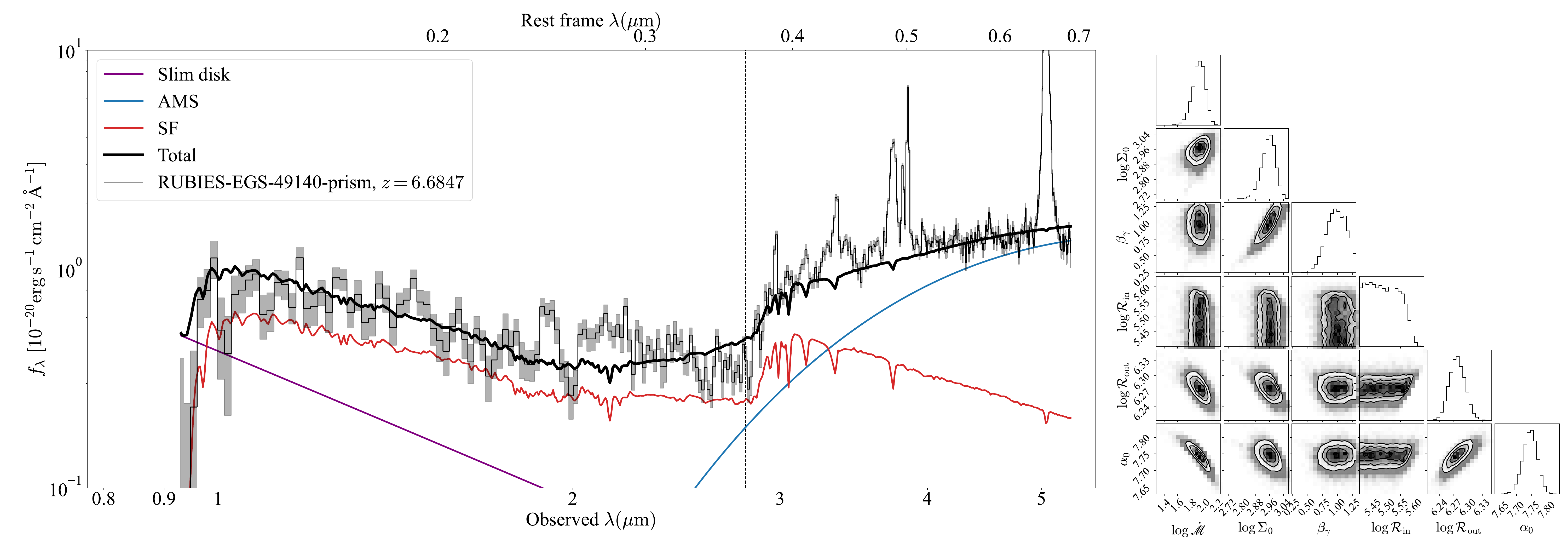}
\includegraphics[width=0.9\textwidth]{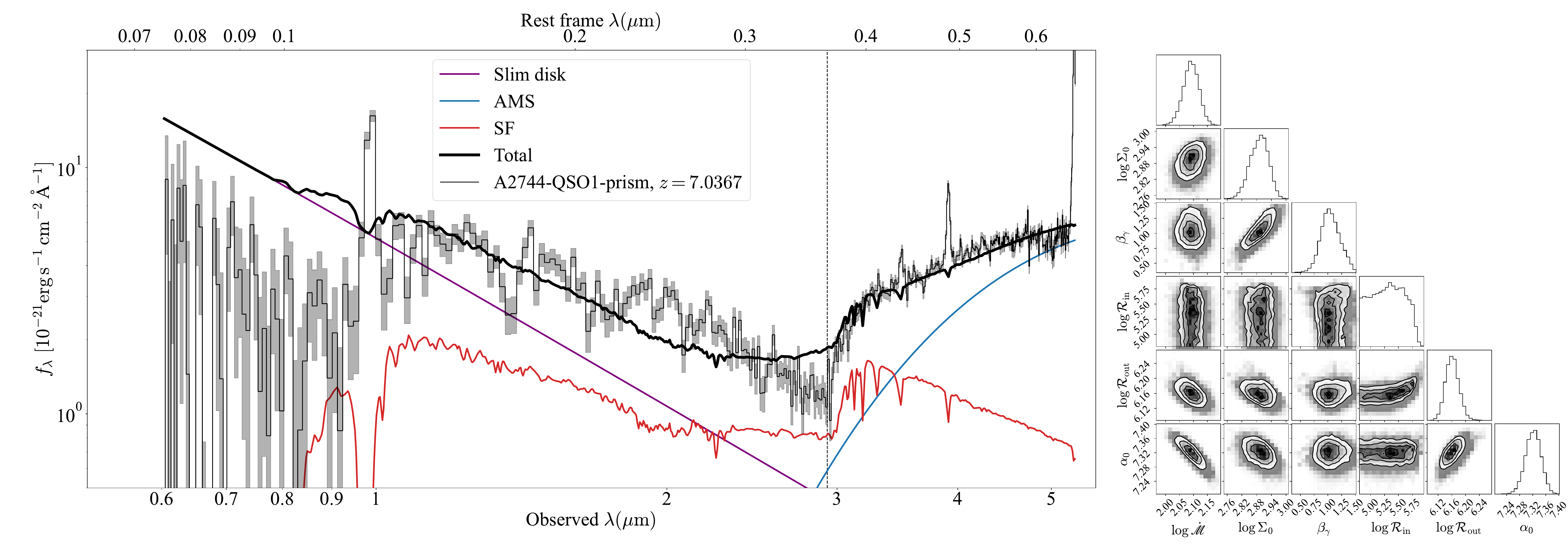}
\caption{\footnotesize Spectral energy distributions of the represent LRDs with three necessary components. NSB component composes of only single stellar population as the necessary for the part around the Balmer break.}
\label{fig:SED-fitting1}
\end{figure*}

\begin{figure*}
\centering
\includegraphics[width=0.9\textwidth]{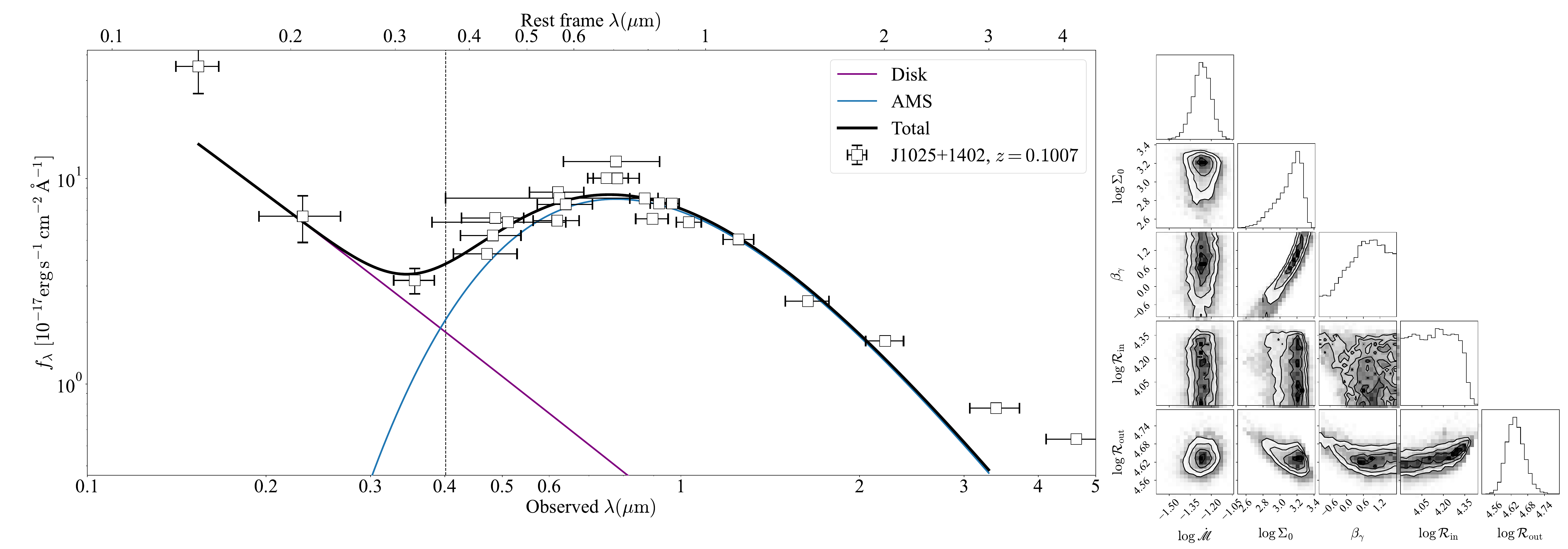}
\includegraphics[width=0.9\textwidth]{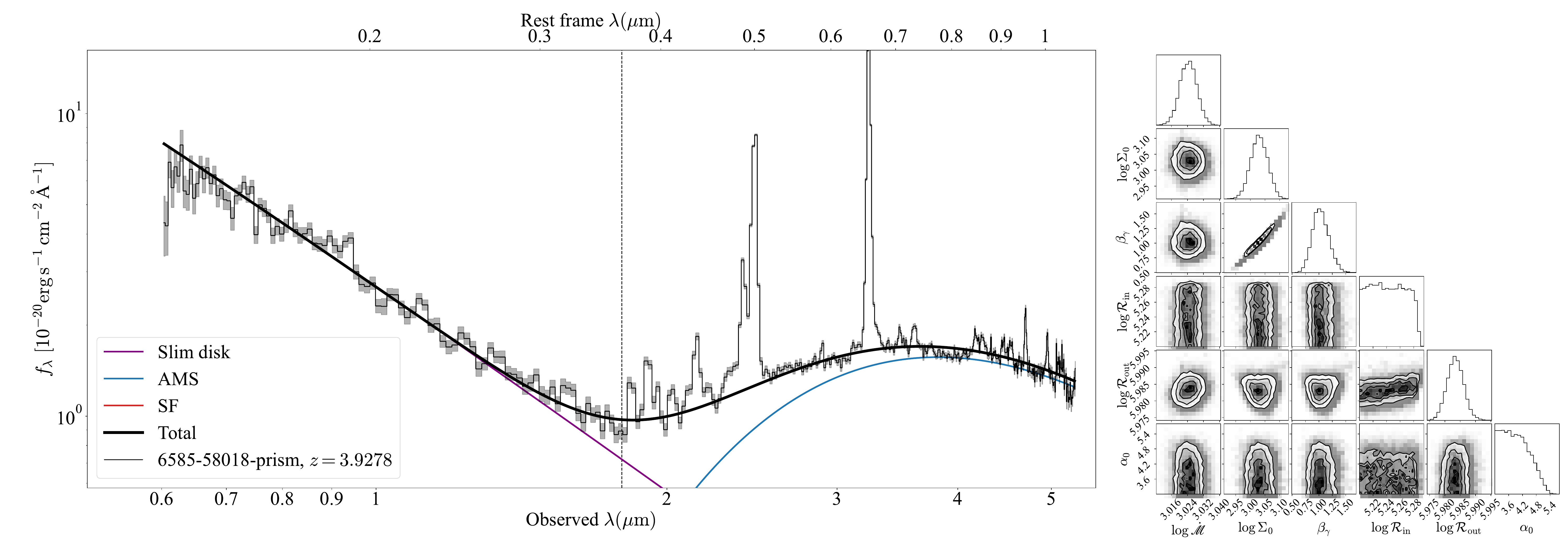}
\includegraphics[width=0.9\textwidth]{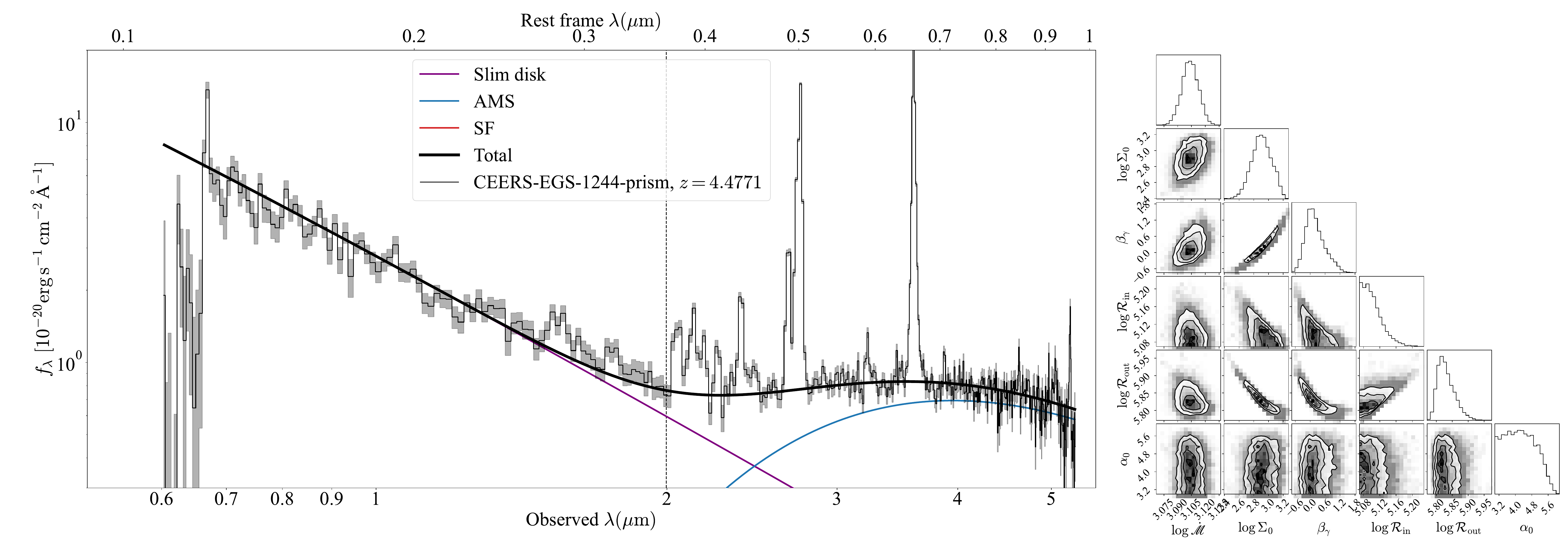}
\includegraphics[width=0.9\textwidth]{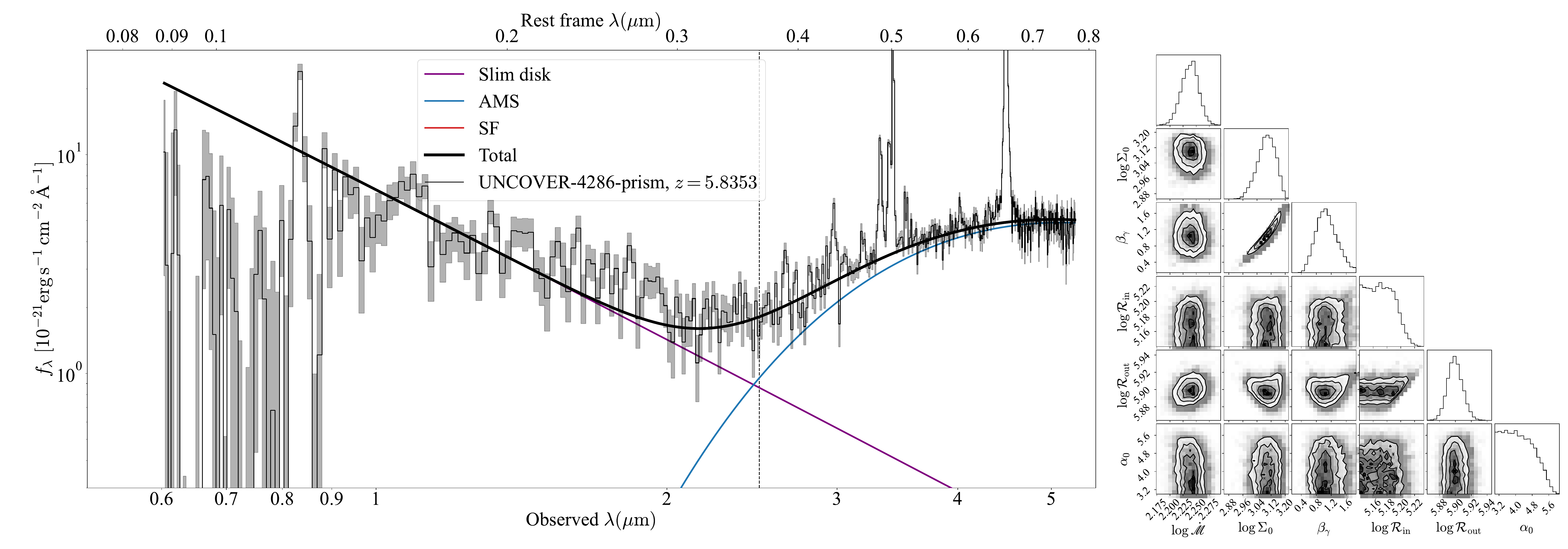}
\caption{\footnotesize Spectral energy distributions of the represent LRDs with two necessary components contributed by slim disks and {\sMBH}. NSB components are not necessary for them.
}
\label{fig:SED-fitting2}
\end{figure*}

\begin{figure*}
    \centering
    \includegraphics[width=0.9\textwidth]{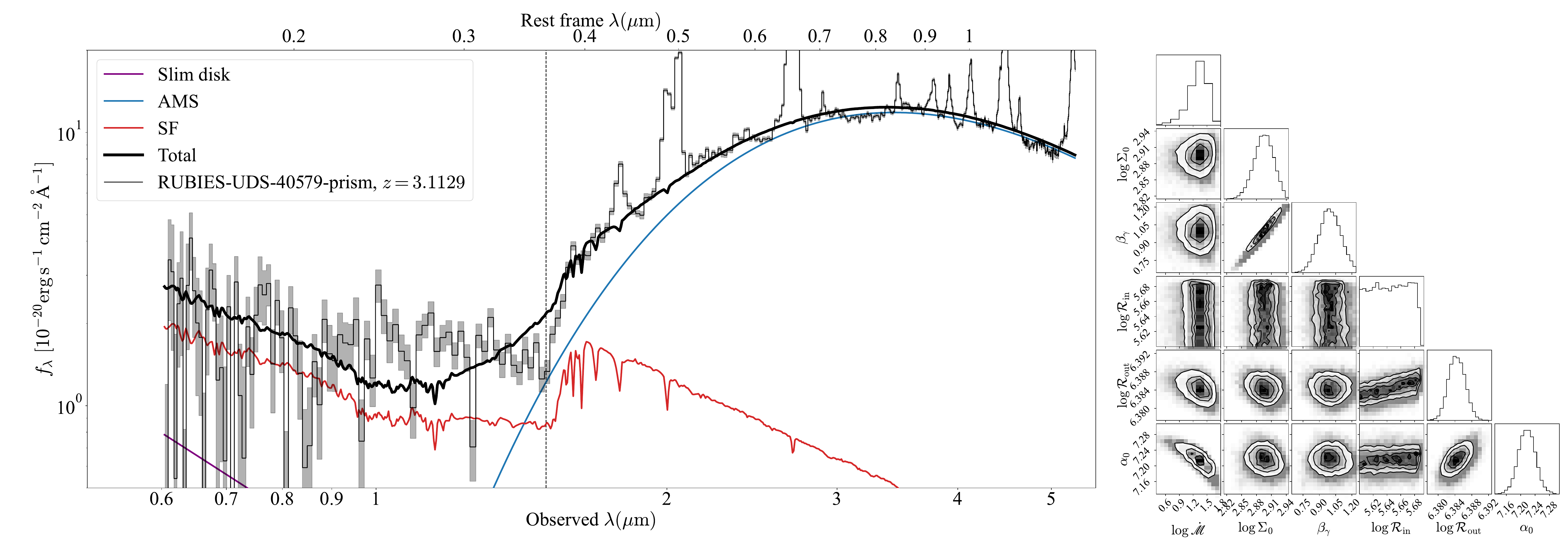}
    \includegraphics[width=0.9\textwidth]{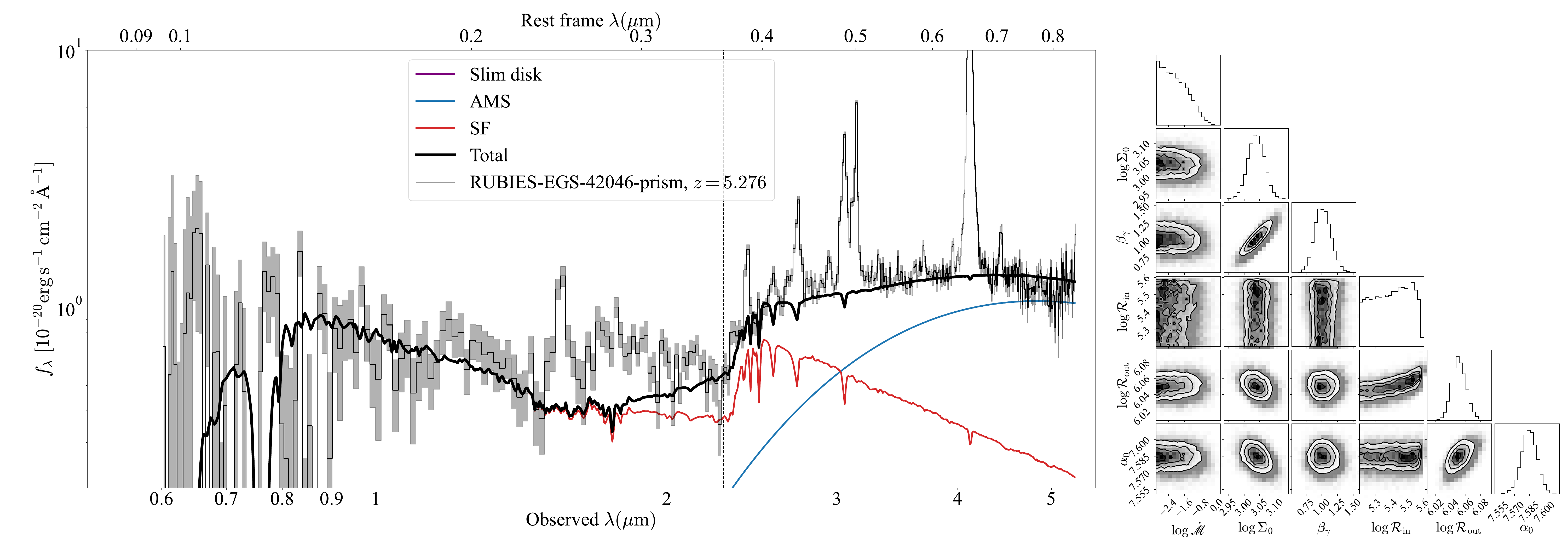}
    \includegraphics[width=0.9\textwidth]{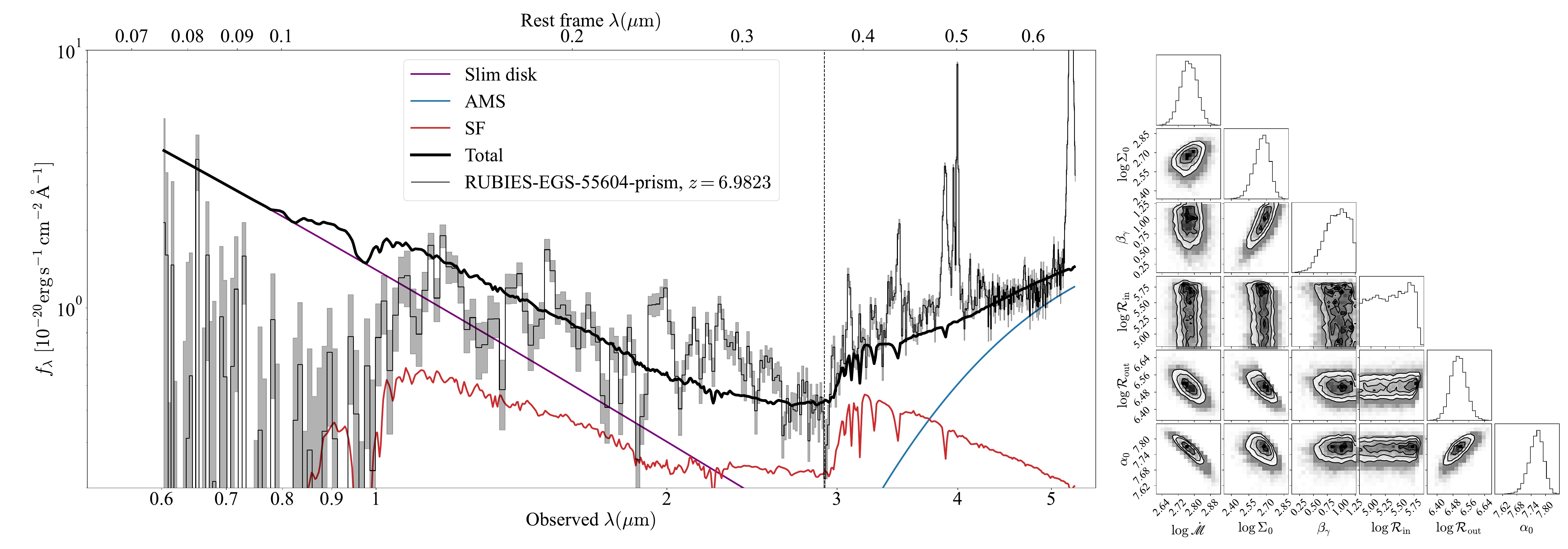}
    \caption{\footnotesize Spectral energy distributions of the third type of LRDs in which multiple NSBs are necessary between the Balmer break wavelength and $\sim 0.2\,\mu$m. Sometimes slim disks are not necessary, implying the cMBH masses are quite light (much less than $10^6\,\sunm$).
    We also note that RUBIES-EGS 42046-prisim shows P Cyg profile of the Balmer lines \citep{Hviding2025}.}
\label{fig:SED-fitting3}    
\end{figure*}

The fitting scheme is described in Appendix \ref{sec:fitting}. 
We generate $\FCAMS$ and $\FCMBH$ from the present model and $\FCnsb$ continuum through {\tt{CIGALE}} \citep{Boquien2019}.
Fitting results of three representative LRDs are shown in Fig.\,\ref{fig:SED-fitting1}, \ref{fig:SED-fitting2} and \ref{fig:SED-fitting3}, respectively.
As a local analog of the LRDs SDSS\,J1025+1402 \citep{LinXJ2025a,Ji2025} is classified as the second type shown in Fig.\,\ref{fig:SED-fitting2}.
Generally, the {\sMBH} model works very well for these LRDs, but we note that the third type needs more stellar components, implying that they underwent multiple starbursts \citep{Kritos2025}.
In a future work, we will develop a more complete model of the stellar components.

\section{Discussions}\label{sec:models}
%
There are great efforts to explain LRDs' properties,
but most of them only focus on parts of them.
On the one hand, these ideas are motivated and highly debated \citep{Juodzbalis2025}; on the other hand, we should think whether the ingredients of these models can be derived from the simplest hypothesis in order to grasp some of them in a complete model systematically explaining the properties.
We list them with brief comments.

\subsection{Dust scattering and extinction models} 
If dust particles scatter the central AGNs, and they are viewed from an edge-on direction, we will see the AGNs as the LRD spectra \citep{Labbe2025}.
On the other hand, it has been proposed that LRDs are AGNs but their continuum is highly reddened by dust \citep{Greene2024,WangBJ2024,Kokorev2024}.
Both of the models need enough dust particles for LRD properties.

On the observational side, the ALMA observations show that the dust is very dilute \citep{Casey2024}, with even non-detection of dust \citep{Labbe2025} including both hot and cold dust \citep{Setton2025}, indicating the upper limit of the dust mass is less than $10^6\,\sunm$ in 60 LRDs \citep{Casey2025}.
Moreover, a stacking analysis of LRD candidates in JADES has shown a distinct lack of hot dust \citep{Eisenstein2023} (with mean dust masses of $10^4\,\sunm$ in LRDs, see \citealt{Casey2024,Casey2025}).
On the contrary, hot dust generally exists in AGNs ($\gtrsim 10^6\,\sunm$ in PG quasars, see \citealt{Haas2003}). 
Therefore the LRDs don't have a normal dusty torus in AGNs, and the reddened AGN disk component as the red continuum could be ruled out.
Stars in the nuclear regions are so massive through accretion in a dense environment \citep{Wang2023} that they directly collapse into sMBHs \citep{Woosley2002}.
Here we show that the observed optical continuum is not necessarily reddened by dust, but it is emission from the accretion onto sMBHs in the \sMBH.

\subsection{Gas-enshrouded AGNs} 
Given the absorption features of the Balmer lines as a common feature, a dense gaseous envelope enshrouding AGNs has been suggested by several authors for the nature of the LRDs \citep{Naidu2025,Rusakov2025,Inayoshi2025} and received much attention.
However, this model suffers from the general deficit of IR-weak continuum \citep{deGraaff2025} as its reprocessing; moreover, the envelope location remains open from 40\,AU \citep{Naidu2025} to 3000\,AU \citep{LinXJ2025a}.
In addition, how to form and sustain the envelope remains open.
%
%

\subsection{Thomson broadening effects}
Thomson scattering of hot electrons is suggested to efficiently broaden the Balmer lines \citep{Rusakov2025}.
After correction of the line width, the cMBH masses are greatly reduced by factors of 10-100.
It has been suggested that the electrons could results from winds of super-Eddington accretion \citep{Liu2025} and could also account for the Balmer break.
Narrow-line Seyfert 1 galaxies are known to be super-Eddington accreting AGNs \citep{Du2019}; however, they have neither the Balmer break nor absorption features of the Balmer lines.

\subsection{Starburst} 
A nuclear pure starburst has been suggested by \citet{Baggen2023}. 
Actually, narrow-line LRDs have been claimed recently \citep{ZhangZJ2025}, indicating that dusty tori obscure AGNs (this fact conflicts with the non-detection of dust particles by ALMA observations). 
However, the observed rest-UV SEDs are different from those of star forming galaxies in the earlier universe \citep{Taylor2025}; meanwhile, they are also different from the standard accretion disk model (known as $F_{\nu}\propto \nu^{1/3}$).
The observed rest-UV SEDs could be composed of contributions from {\sMBH} and star formation .

\subsection{Self-gravitating disk} 
It has been suggested that the V-shaped SEDs can be formed by a self-gravitating disk corresponding to the Toomer parameter $Q=1$ \citep{Zhang2025,Zwick2025}.
However, the observed $\beta_{\rm opt}$ and $\beta_{\rm UV}$ are not related \citep{Zhuang2025}, in disagreement with this model.
Moreover, the outer part of accretion disks of normal AGNs and quasars has been known to be self-gravitating for many years \citep{Sirko2003}; however, $V$-shaped SEDs never appear in these objects.
This model generally mismatches the observations.

\subsection{Compactness of the LRDs}
It is a good idea to compare the densities of bulges and nuclear star clusters (NSCs) with the LRDs for their formation and evolution.
The mean densities of bulges and NSCs are $\rho_{\rm bulge}^*\sim 25\,M_{11}R_{\rm kpc}^{-3}\sunm\,{\rm pc^{-3}}$ and $\rho_{\rm NSC}^*\sim 2.5\times 10^{4}\,M_8R_{10\rm pc}^{-3}\sunm\,{\rm pc^{-3}}$, respectively, where $M_{11}=M_{\rm bulge}/10^{11}\sunm$ and $R_{\rm kpc}=R_{\rm bulge}/1\,{\rm kpc}$ are the bulge stellar masses and radii, respectively; $M_{8}=M_{\rm NSC}/10^{8}\sunm$ and $R_{\rm 10\,pc}=R_{\rm NSC}/{\rm 10\,pc}$ are the NSC stellar masses and radii \citep{Neumayer2020}.
\citet{Baggen2023} shows that the LRDs have a stellar mass of $\sim10^{9}\,\sunm$ with a mean radius of $0.1\,$kpc, showing a density of $2.5\times 10^{2}\,\sunm\,{\rm pc^{-3}}$.
This indicates $\rho_{\rm bulge}^*<\rho_{\rm LRD}^*<\rho_{\rm NSC}^*$, along with the fact that the NSCs follow the $\BHM-M_{\rm NSC}$ relation in local galaxies \citep{Ferrarese2006}, suggesting that LRDs are still undergoing growth toward quasars and NSCs.
Considering the existence of the most luminous LRDs ($L_{5100}\lesssim 2\times 10^{44}\,\ergs$), on the other hand, we may ask what the observational appearance of objects looks like beyond the maximum luminosity of the LRDs in the early universe?

We speculate about their appearance. 
Usually, the more massive primordial clouds, the higher spin angular momentum from the tidal interaction with their neighbors.  
The specific SAM is given by $j=\calJ/M_{\rm c}\propto M_{\rm c}^{2/3}$\citep[see Eq.\,(7.172) in][]{Mo2010}.
Therefore, there is a critical mass of clouds over which their SAM are not able to remove within local Hubble time by known mechanisms.
In such a case, only giant disk galaxies are formed, but the cMBHs are tiny (even without cMBHs).

\subsection{Heuristic ideas} 
Three major properties at least should be grasped in a successful model: 1) avoiding the So{\l}tan crisis; 2) a natural mechanism for an outflowing envelope is necessary for absorption features of the Balmer lines; 3) the rest-frame optical continuum as a bump without dust extinction. 
These necessary elements motivate us to propose the current model composed of three ingredients formed during the collapsing periods of the primordial clouds: 
\begin{center}
\fbox{LRDs} $=$ \fbox{outflowing envelope} $\oplus$ \fbox{{\sMBH} system} $\oplus$ \fbox{nuclear starburst}
\end{center}
Cartoon in Fig.\,\ref{fig:model} shows the structure.
While nuclear starbursts form from a rapid collapse of the clouds \citep{Vanzella2019}, the outflowing envelope is driven by the starburst or AGN activity.
With sufficient angular momentum of the nuclear regions, {\sMBH} is formed.
We would point out that the starburst could have extreme top-heavy mode so that massive stars directly collapse into sMBHs without explosion \citep{Woosley2002} yielding much fewer metal elements for dust particles in the nuclear regions.

\section{Conclusions}
We demonstrate that the cMBHs in LRDs can not exceed $10^6\sunm$ in light of the So{\l}tan argument of the current data. 
%
%
A population of sMBHs embedded inside the accretion disks of the cMBH is responsible for the rest-frame optical continuum.
This constitutes the embryo structure of AGNs and reveals the nature of LRDs. 
The sMBHs are the remnants of stellar evolution as a natural consequence of the cMBH formation through the gravitational collapse of primordial clouds. 
The cMBH-disk, or jointly with the post starburst, powers the UV continuum.
The observed V-shaped SEDs of LRDs can be well explained by the {\sMBH} scenario.
Meanwhile, the radiation pressure drives the collapsing cloud to form an envelope generating the observed absorption features of the Balmer lines.
Other properties of the LRDs, such as non-variable continuum and lack of X-ray emission, can be explained by the present model.

Internal and external factors govern the evolution of the LRDs on different timescales.
First, internal mechanisms drive mergers and inward migrations of sMBHs in the cMBH-disk \citep{Tagawa2020,Wang_merger2021,Vaccaro2025}, whose gravitational waves could be detected by the future LISA \citep{Amaro-Seoane2023}/Tianqin \citep{LuoJ2025} and the Einstein Telescope \citep{DiGiovanni2025}, even by the current LIGO, providing conclusive tests of the {\sMBH} model for LRDs.
%
Meanwhile, this leads to evolutionary decays and shifts of the V-shaped SEDs, and the Balmer absorption features since the envelope is gradually blow away, appearing either as narrow-line Seyfert 1 galaxies, or as compact galaxies. 
Successive hierarchical mergers finally yield a small number of cMBHs in the centers.
%
%
Moreover, the volume densities of LRDs are so high that their merger rates are very frequent, leading to the formation of new LRDs or galaxies \citep{Billand2025}.

\begin{acknowledgements}
{This research is supported by NSFC-12333003, -92476203, -12521005, -12573016, the National Key R\&D Program of China (2021YFA1600404,  2023YFA1607903, and 2023YFA1607904), the National Science and Technology Major Project (2024ZD0300303), the China Manned Space Project (CMS-CSST-2025-A07), and the Strategic Priority Research Program of the Chinese Academy of Sciences (XDB1160202).}
\end{acknowledgements}

\appendix

\section{Conventional estimations of the cMBH}\label{sec:challenges}
Though the LRDs are poorly understood, astronomers have to estimate the cMBH masses in light of the so-called $R-L$ relation established for AGNs \citep{Maiolino2024faint} though there are growing serious challenges to this relation \citep{Du2019,Wang2025}.
The currently popular estimations of their cMBH masses follow that for normal AGNs through H$\alpha$ or H$\beta$ lines \citep{Greene2005}
\begin{equation}\label{eq:MBHmass}   \BHM=\left(2.0,3.6\right)\times 10^6\left(\frac{L_{\rm H\alpha,H\beta}}{10^{42}\ergs}\right)^{0.56}
    \left[\frac{{\rm FWHM(H\alpha,H\beta)}}{10^3\,\kms}\right]^{2.06}\,\sunm,
\end{equation}
respectively.
We should emphasize here that Eq.\,(\ref{eq:MBHmass}) is based on the original $R-L$ relation \citep{Kaspi2000,Bentz2013,Du2019} and an assumption that there is a single cMBH surrounded by virialized motions of BLR clouds, in particular, this estimation only applies to sub-Eddington accreting AGNs \citep{Du2019}.

In the proposed model, the cMBH mass is smaller than the total masses of sMBHs embedded in the \sMBH.
Estimations of total sMBH masses become quite complicated, but we show it in Appendix \ref{sec:profile}.
According to Eq.\,(\ref{eq:MBHmass}), the LRDs have overmassive cMBHs compared with the Magorrian relation \citep{Magorrian1998}.
Recently, a direct measurement of one LRD shows evidence for an SMBH ($\sim 10^8\,\sunm$) in Abell\,2744-QSO1 \citep{Juodzbalis2025b}, but within a region of $20-150$\,pc. 
The proposed population of sMBHs may exist within the above region, and their measurements should cover the sMBHs.
The mean cMBH masses should be smaller than that given by Eq.\,(\ref{eq:MBHmass}) avoiding the crisis of the So{\l}tan argument \citep{Soltan1982,Chokshi1992,Yu2002,Jahnke2025} briefly discussed below.

\section{Inner structures of PHL\,1092} \label{sec:PHL1092}
Reverberation mapping of AGNs is a powerful tool probing their inner structures \citep{Blandford1982}. 
Since the 1980s, RM of about 300 AGNs has been done for H$\beta$ delays, but targets of long-term campaigns (longer than 6-7 years) with high cadences are quite limited (see a brief summary in \citealt{Wang2025}).
There is growing evidence for anomalous reverberations of the broad H$\beta$ line with respect to the varying continuum \citep{Wang2025}.
It has been found that the delays of the H$\beta$ line decrease with accretion rates \citep{Du2014,Wang2014,Du2019}; in particular, the quasar PHL\,1092 shows leading reverberations of H$\beta$ line \citep{Wang2025} in advance of $17-57$\,days relative to the varying 5100\,\AA\ continuum.
These extreme reverberations suggest that there are extra energy sources which spatially distribute over the accretion disks of the cMBH.
This phenomenon offers a new perspective for understanding the internal structure of AGNs.

A population of sMBHs has been suggested to be the extra energy sources \citep{Wang2025}, which efficiently heat the cMBH-disk.
SEDs emerging from such a disk depend on two key parameters (see Appendix \ref{sec:sMBH-disk}): spatial distributions of sMBHs and their mass functions.
For the LRDs, they are located in outskirts of the disks and thus gravitational energy is mainly released in optical bands.
On the other hand, migration of sMBHs towards the cMBH makes the rest-frame optical continuum shift to UV bands.
Therefore, SEDs indicate the evolutionary stage of migration of the sMBH population from embryos to normal AGNs.

\section{Structures and SEDs of the \sMBH}\label{sec:sMBH-disk}
\subsection{Evidence for sMBHs in AGNs}
Compact objects embedded inside accretion disks of SMBHs in AGNs have been considered by \citep{Cheng1999}, and later by \citep{McKernan2011,Tagawa2020,Vaccaro2025}.
LIGO may have already detected such an event (GW\,190521) \citep{Graham2020}, and more candidates \citep{Graham2023}, in particular, a merger of $137^{+22}_{-17}\,\sunm$ and $103^{+20}_{-52}\,\sunm$ was detected recently \citep{GW231123}.
Owing to its powerful spatial resolution, GRAVITY on Very Large Telescopes Interferometer (VLTI) of European Southern Observatory (ESO) depicts accurately enough the positions of flares of Sgr A* and find that they are distributed over a ring with $\sim 9\Rg$.
Such flares cannot be generated by magnetic reconnection randomly happening on the disk surface, but could be produced by an orbiting $\sim 40\,\sunm$ black hole around the cMBH \citep{Wang2023GC}.
Additionally, there is growing evidence for the presence of satellite black holes around the cMBH from X-ray observations.
In AGN 1ES\,1927+654, mHz quasi-periodic oscillations (QPOs) of X-ray variations \citep{Masterson2025}, and in GSN\,069 as a Seyfert 2 galaxy, quasi-periodic ejections (QPEs) \citep{Miniutti2019} and others \citep{Giustini2020,Arcodia2021,Chakraborty2021,Arcordia2024}, significantly enhance the validity of sMBH in AGN disk.
This also lends support to the proposition that LRDs contain a population of sMBHs in the cMBH-disk.

\subsection{Accretion onto sMBH and radiation}\label{eq:sMBH}
Considering that the \sMBH\ contains a population of sMBHs, we have to reconsider the basic equations of accretion disks \citep{Kato1998,Frank2002} of the \sMBH\ since there is additional heating of sMBHs and decreases of $\dot{M}_{\bullet}$ along with $R$ due to accretion onto sMBHs. 
\citet{Gilbaum2022,Zhou2024,Wang2025} have studied the equations of \sMBH, but we revisit all the equations of {\sMBH} by including more realistic and necessary considerations.
We first consider the heating of sMBHs.

The ideal Bondi accretion rates are extremely super-Eddington though the heating and viscosity effects efficiently reduce them \citep{Luo2025}.
For super-Eddington accretion onto sMBHs, the radiated luminosity could be saturated at a few times the Eddington luminosity with a factor of logarithmic dependence of accretion rates \citep{Abramowicz1988,Wang1999b,Mineshige2000,Yoshioka2024}.
On the other hand, we note that outflows are developed from the slim accretion disks, but the mechanical power is always less than the saturated \citep{Yoshioka2024}. 
Nevertheless, it is interesting to note that most gas is circulated by the outflows, roughly remaining the gas of the cMBH-disk there.
Keeping these uncertainties of energies released by sMBH accretion in mind, we approximately take a simple radiative luminosity from the accretion onto sMBHs as
\begin{equation}
    \ell_{\bullet}=\ell_0 L_{\rm Edd}^0\left(\frac{\bhm}{1\,\sunm}\right),
\end{equation}
where $\ell_0$ is the factor of the saturated luminosity and could be in a range of $\ell_0\approx 1\sim 10$, $L_{\rm Edd}^0=1.4\times 10^{38}\,\ergs$ is the Eddington luminosity of $1\,\sunm$ black hole.
In our previous paper \citep{Zhou2024}, we simply take the Eddington luminosity as their radiative luminosity independent of density and temperature. 
We assume that the energy from AMSs are thermalized locally.

The mass function of sMBHs is important but it depends on their accretion.
Considering that the sMBHs are growing with migration inward, we neglect the complicated details of their growth and presume a formulation of 
\begin{equation}
m_{\bullet}(R)=m_{\bullet}^0\left(\frac{R}{\calR_{\rm out}}\right)^{\betaBH},    
\end{equation}
where $m_{\bullet}^0$ is the sMBH mass at radius $\calR_{\rm out}$.
Assuming that the surface number density of sMBHs is 
\begin{equation}
\Sigma_{\bullet}=\Sigma_{\bullet}^0\left(\frac{R}{\calR_{\rm out}}\right)^{\gammaBH},
\end{equation}
with $\gammaBH>0$ (the surface number density decreases inward due to mergers), we have the heating rate of the accreting sMBHs per unit area
\begin{equation}\label{eq:QsMBH}
    Q_{\rm AMS}=\ell_0\Sigma_{\bullet}^0\left(\frac{m_{\bullet}^0}{1\,\sunm}\right)\,\left(\frac{R}{\calR_{\rm out}}\right)^{\betaBH+\gammaBH}
=Q_{\rm AMS}^0\left(\frac{R}{\calR_{\rm out}}\right)^{\betaBH+\gammaBH},
\end{equation}
where $Q_{\rm AMS}^0=\ell_0L_{\rm Edd}^0\Sigma_{\bullet}^0\left(m_{\bullet}^0/{1\,\sunm}\right)$ is a constant parameter.
%
The total numbers and masses of sMBHs are given by
\begin{equation}
    \mathN_{\bullet}=2\pi\int_{\calR_{\rm in}}^{\calR_{\rm out}}\Sigma_{\bullet}R\rmd R
    =\frac{2\pi \calR_{\rm out}^2\Sigma_{\bullet}^0}{2+\gammaBH}\left[1-\left(\frac{\calR_{\rm in}}{\calR_{\rm out}}\right)^{2+\gammaBH}\right],
\end{equation}
and
\begin{equation}\label{eq:Sigbh}
    \Sigbh=2\pi\int_{\calR_{\rm in}}^{\calR_{\rm out}}\Sigma_{\bullet}m_{\bullet}R\rmd R
 = \frac{\mathscr{M}_{m_\bullet}^0}
{1+\beta_\gamma}\left[1-\left(\frac{\calR_{\rm in}}{\calR_{\rm out}}\right)^{1+\beta_\gamma}\right],
\end{equation}
where
$\mathscr{M}_{m_\bullet}^0=2\pi \calR_{\rm out}^2 \Sigma_{\bullet}^0 \bhm^0$, $\beta_{\gamma} = 1 + \betaBH + \gammaBH$.
For typical values obtained from fittings of the 10 representative LRDs in Appendix \ref{sec:fitting}, we have $\mathN_{\bullet}\sim 10^6$, $\Sigbh\sim 10^6\,\sunm$ in light of $\calR_{\rm out}\sim 20\,$ltd, $\calR_{\rm in}\sim 1\,$ltd, and $\beta_{\gamma}\sim 0.8$.

\subsection{Mass conservation}
First of all, accretion rates of the \sMBH\ are a function of radius due to accretion on sMBHs. 
The mass conservation equation reads
\begin{equation}\label{Eq:Mass conserve}
    \Mdot(R)=-2\pi R V_{\rm R} \Sigma_{\rm gas}, 
\end{equation}
where $V_{\rm R}$ is the radial velocity of accreted gas, $\Sigma_{\rm gas}=2\rho H$ is the surface density of the disk and $\rho$ is the gas density.
Given the heating rates per unit area due to sMBHs, red the mass accretion rate onto sMBHs in a ring of width $\rmd R$ is $2\pi R Q_{\rm AMS}\rmd R /\eta_{\bullet} c^2$, which is equal to the decrease in the accretion rate of the \sMBH.
We have
\begin{equation}
    \frac{\rmd \dot{M}_{\bullet}}{\rmd R}=2\pi\eta_{\bullet}^{-1}c^{-2} RQ_{\rm AMS},
\end{equation}
yielding
\begin{equation}\label{eq:dotM}    \dot{M}_{\bullet}=\dot{M}_{\bullet}^0-\frac{2\pi Q_{\rm AMS}^0\calR_{\rm out}^2}{(1+\beta_\gamma)\eta_{\bullet} c^2}\left[1-\left(\frac{R}{\calR_{\rm out}}\right)^{1+\beta_\gamma}\right],
\end{equation}
where $\dot{M}_{\bullet}^0$ is the rates at $\calR_{\rm out}$.
$\dot{M}_{\bullet}$ decreases inward with decreasing $R$.
Considering that most of the accreted gas will be channeled into outflows \citep{Yoshioka2024}, here we assume $\eta_{\bullet}\approx 0.1$ for a simple treatment.
Eqs.\,(\ref{Eq:Mass conserve}) and (\ref{eq:dotM}) lead to $V_{\rm R}$ after determining $\Sigma_{\rm gas}$.
This effect has not been included by \citet{Gilbaum2022,Zhou2024,Xu2025}.
As a simplified treatment, this estimation accounts for the returned gas from outflow of accretion onto sMBHs by neglecting the details.

\subsection{Vertical equilibrium}
We assume that the vertical structure of the {\sMBH} system is in static equilibrium, but the vertical gravity force ($F_{\bhm}^z$) of the sMBHs should be included
\begin{equation}\label{eq:vertical}
\frac{1}{\rho}\left(\frac{\rmd P}{\rmd z}\right) =\frac{G\BHM}{R^2}\left(\frac{H}{R}\right)+F_{\bhm}^z,   
\end{equation}
The details of the $F_{\bhm}^z$ are quite complicated, but we adopt the approximation $F_{\bhm}^z\approx 2\pi f_0G\Sigma_{\bullet}$, which is from an infinite plane with homogeneous surface mass density, where $f_0\approx 0.9$ is a correction factor of the force.
We note that this approximation is valid for $H/R\lesssim 0.1$ within about $10\%$ accuracy in light of the detailed calculations \citep{Hure2011} .
Following the common approximate treatment of the vertical structure, we re-cast Eq.\,(\ref{eq:vertical})
\begin{equation}\label{eq:H}
    \frac{1}{\rho}\left(\frac{P}{H}\right) =\frac{G\BHM}{R^2}\left(\frac{H}{R}\right)+2\pi f_0G\Sigma_{\bullet}.
\end{equation}
It is clear that $F_{\bhm}^z$ stabilizes the vertical structure.

In order to obtain the energy releasing rates of the {\sMBH} and gravity-broadening effects, we have to solve the Poisson equation of $\nabla \phi=4\pi \rho_{\bullet}(r,z)$ for the {\sMBH} potential.
One analytical approximate solution with an accuracy of $10\%$ has been derived \citep{Hure2011},
\begin{equation}
    \phi(R) \approx-\frac{G\mathscr{M}_{m_\bullet}^0}{\calR_{\rm out}}\left\{\frac{1}{\beta_\gamma}+\left[0.431-\frac{1}{\beta_{\gamma}(1+\beta_\gamma)}\right]\left(\frac{R}{\calR_{\rm out}}\right)^{\beta_{\gamma}} - \frac{1}{1+\beta_{\gamma}}\left(\frac{\calR_{\rm in}}{\calR_{\rm out}}\right)^{\beta_{\gamma}} \frac{\calR_{\rm in}}{R}\right\} - \frac{G \BHM}{R},
\end{equation}
yielding angular velocity
\begin{equation}\label{eq:OmegaK}
    \Omega_{\rm K}=\left(\frac{1}{R}\frac{\rmd \phi}{\rmd R}\right)^{1/2}=\sqrt{\frac{G \mathscr{M}_{m_\bullet}^0}{\calR_{\rm out}} \frac{\mathscr{F}_R}{R^2} + \frac{G\BHM}{R^3}},
\end{equation}
where 
\begin{equation}
    \mathscr{F}_R = \left[\frac{1}{1+\beta_{\gamma}}-0.431 \beta_{\gamma}\right]\left(\frac{R}{\calR_{\rm out}}\right)^{\beta_{\gamma}} - \frac{1}{1+\beta_{\gamma}}\left(\frac{\calR_{\rm in}}{\calR_{\rm out}}\right)^{\beta_{\gamma}}\frac{\calR_{\rm in}}{R}.
\end{equation}
The roles of sMBHs in the vertical equilibrium have not been considered in \citet{Gilbaum2022} and \citet{Zhou2024}.

\subsection{Conservation of angular momentum}
The conservation equation of angular momentum reads
\begin{equation}
    R\frac{\partial(\Sigma_{\rm gas} R^2 \OmegaK)}{\partial t} + \frac{\partial}{\partial R} \left(R \Sigma_{\rm gas} V_R R^2 \OmegaK\right) = \frac{1}{2\pi} \frac{\partial {\cal{G}}}{\partial R},
\end{equation}
where ${\cal{G}}$ is the viscosity stress.
For a stationary state, by using Eq.\,(\ref{Eq:Mass conserve}), we have 
\begin{equation}\label{eq:AM}
    -\frac{\partial}{\partial R}\left(\frac{\Mdot}{2\pi} R^2 \OmegaK\right) = \frac{\partial}{\partial R}\left(R^3 \nu \Sigma_{\rm gas} \frac{\rmd\OmegaK}{\rmd R}\right).
\end{equation}
Similar to the standard model of accretion disk, we use the torque-free boundary condition at the inner edge of the {\sMBH} part, i.e., the region between $\calR_{\rm in}$ and $\calR_{\rm out}$, and we have
\begin{equation}
    \frac{1}{2\pi} \left[R^2\OmegaK(R) \Mdot(R)  - \calR_{\rm in}^2\OmegaK(\calR_{\rm in}) \Mdot(\calR_{\rm in})\right] = -R^3\nu \Sigma_{\rm gas}\left(\frac{\rmd\OmegaK}{\rmd R}\right).
\end{equation}
after integrating Eq.\,(\ref{eq:AM}).
This boundary condition holds approximately for LRDs since sMBHs are located at the disk outskirts.
Defining $\ell=R^2\OmegaK(R)$ and $\ell_{\rm in}=\calR_{\rm in}^2 \OmegaK(\calR_{\rm in})$, we have 
\begin{equation}\label{eq:angularmomentum}
    \nu \Sigma_{\rm gas} = \frac{\Mdot}{2\pi} \left[\frac{\Mdot(\calR_{\rm in}) \ell_{\rm in}}{\Mdot \ell} -1\right] \left(\frac{\rmd\ln\OmegaK}{\rmd\ln R}\right)^{-1}.
\end{equation}

\subsection{Energy balance}
The viscous heating rate per unit area can be expressed by 
\begin{equation}\label{eq:Qvis}
    Q_{\rm vis} = \int_{-H}^H \nu\rho \left(R\frac{\rmd\OmegaK}{\rmd R}\right)^2 dz = \nu\Sigma_{\rm gas} \left(R\frac{\rmd\OmegaK}{\rmd R}\right)^2=T_{r\varphi} R \frac{\rmd\OmegaK}{\rmd R},
\end{equation}
where $T_{r\varphi}=\nu\Sigma R \rmd\OmegaK/\rmd R$. Using Eq.\,(\ref{eq:QsMBH}), we have the energy equation $Q_+=Q_{\rm vis}+Q_{\rm AMS}$
\begin{equation}\label{eq:Energy}
   Q_+= Q_{\rm rad},
\end{equation}
and the surface cooling rates are given by
\begin{equation}\label{eq:Qrad}
Q_{\rm rad}=\frac{2\sigma_{\rm SB} T^4}{\tau_{\rm eff}},\quad \tau_{\rm eff}=\frac{3\tau}{8}+\frac{1}{2}+\frac{1}{4\tau}, \quad T_{\rm eff} = \tau_{\rm eff}^{-1/4} T  
\end{equation}
where $T$ is the temperature of the mid-plane of the accretion disk, $T_{\rm eff}$ is the effective temperature, and the optical depth of the {\sMBH} is $\tau=\kappa_R \rho H$.
We would point out that the first term on the right-hand side of Eq.\,(\ref{eq:Energy}) covers the energy released by cMBH-disk, including the potential of sMBHs; whereas it is not included in \citep{Gilbaum2022}.

\subsection{Viscosity and state equation}
The $r\varphi$-component of the shearing stress tensor can be expressed by
\begin{equation}
    t_{r\varphi} = \rho\nu R\frac{\rmd\OmegaK}{\rmd R} = \rho \nu \OmegaK \frac{\rmd\ln\OmegaK}{\rmd\ln R}=-\alpha P.
\end{equation}
The speed of sound is $c_{\rm s}=\sqrt{\gamma P/\rho}$, where $\gamma$ is gas index, $P$ is the total pressure, and $\rho$ is the disk density at the mid-plane.
With inclusion of the radiation pressure, we have the total pressure
\begin{equation}\label{eq:Pc}
    P=\frac{k_{\rm B}}{\mu m_{\rm p}}\rho T+\frac{\tau\sigma_{\rm SB}}{2c}T_{\rm eff}^4,
\end{equation}
where $k_{\rm B}$ is the Boltzmann constant, $m_{\rm p}$ is the proton mass, and $\mu$ is the mean molecular mass.
All these equations above formed a closed system for parameters $\rho$, $T$, $H$ and $V_R$.

\subsection{Total gas mass of the \sMBH} 
The total gas mass of the {\sMBH} is then given by
\begin{equation}
\mathscr{M}_{\rm gas}=2\pi\int_{\calR_{\rm in}}^{\calR_{\rm out}}\Sigma_{\rm gas}R\rmd R=1.4\times10^3\,\sunm,
\end{equation}
for $\mathdotM=10^2$ at $R_{\rm out}$, $\left(\calR_{\rm in},\calR_{\rm out}\right)=\left(10^5,10^6\right)\,\Rg$ ($\Rg=1.5\times 10^{10}\,M_5$\,cm and $M_5=\BHM/10^5\,\sunm$), and $\beta_{\gamma}=1$.
External supply of gas outside the {\sMBH} is necessary to maintain the disk.

As a summary, Eqs.\,(\ref{Eq:Mass conserve},\,\ref{eq:dotM},\,\ref{eq:H},\,\ref{eq:angularmomentum},\,\ref{eq:Energy},\,\ref{eq:Pc}) describe the structure and radiation of the \sMBH\ and their solutions provide results.

\subsection{Emergent spectra from the \sMBH}\label{sec:SED}
We examine the temperature of the {\sMBH} through the blackbody approximation obtained by
\begin{equation}
    T=\left(\frac{L_{\rm Bol}}{\pi \sigma_{\rm SB}R_{\rm disk}^2}\right)^{1/4}\approx 3.8\times 10^3\,L_{44}^{1/4}R_{\rm disk,20}^{-1/2}\,{\rm K},
\end{equation}
where $\sigma_{\rm SB}=5.67\times 10^{-5}$ is the Stefan-Boltzmann constant, $L_{\rm 44}=L_{\rm Bol}/10^{44}\,\ergs$ is the total bolometric luminosity of the sMBH accretion, and $R_{\rm disk,20}=R_{\rm disk}/20\,$ltd is the typical radius of the {\sMBH} (overlapping with the BLR).
In such a circumstance of $T\sim 3800\,$K, bound-bound, bound-free and free-free absorptions are important, and SEDs from this part of the disk will mainly contribute to optical bands as the origin of the V-shaped SEDs of LRDs.

The emergent spectra rely on self-consistent calculations of vertical structure of the \sMBH, such as done for the case based on standard model of the Shakura-Sunyaev disk \citep{Ross1992,Shimura1993,WangYL2025}.
However, this paper concentrates on the characteristic spectra of the {\sMBH} by taking the simple analytic approach of the vertical structure \citep{Hubeny1990}.
Combining Eqs.\,(\ref{eq:angularmomentum}), (\ref{eq:Energy}), (\ref{eq:Qrad}), we can have the effective temperature distributions for the emergent spectra without knowing the structure of the {\sMBH} unless the {\sMBH} is optically thick.
This is similar to the standard model of accretion disk because we assume the Keplerian rotation here.
The total emissions from the \sMBH\, can be obtained by
\begin{equation}\label{eq:sp_sMBH}
F_{\rm \lambda,AMS}=\frac{2\pi \cos i}{D_{\rm L}^2}\int_{\calR_{\rm in}}^{\calR_{\rm out}} B_{\lambda}(T_{\rm eff})R{\rm d}R,
\end{equation}
over the {\sMBH}, $i$ is the inclinations and $D_{\rm L}$ is the luminosity distance of a LRD.

In principle, the gas self-gravity of the \sMBH\ could be important \citep{Sirko2003,Zhang2025} but the accreting sMBHs dominate the heating of the self-gravity.  
We show that such an sMBH disk with gas is stable in Appendix \ref{sec:sMBH-disk}.
We neglect the effects of gravity in this paper.
In this paper, we take $r_{\rm in}=6\Rg$ (for a non-spinning cMBH) and $\calR_{\rm in}=r_{\rm out}$ as a free parameter bridging the two parts of the \sMBH.
On the other hand, $\calR_{\rm in}$ and $r_{\rm out}$ could be independent but they depend on $\mathdotM$ and sMBH spatial distributions and mass functions.

\subsection{Slim part of the cMBH-disks}
The cMBH has its own disk with outer radius $\calR_{\rm out}=R_{\rm in}$.
The slim disk equations are taken from \citet{Abramowicz1988} and \citet{Mineshige2000}. 
The vertical equilibrium is given by $H=c_{\rm s}/\Omega_{\rm K}$ and the mass conservation yields $\dot{M}_{\bullet}=4\pi R H \rho V_{\rm R}$, where $V_{\rm R}$ is the radial velocity of the accretion flow.
The radial motion is given by
\begin{equation}
    \frac{1}{\rho}\frac{\rmd P}{\rmd R}-\left(\Omega^2-\Omega_{\rm K}^2\right)R+V_{\rm R}\frac{\rmd V_{\rm R}}{\rmd R}=0,
\end{equation}
and angular momentum 
\begin{equation}
    \dot{M}_{\bullet}(\ell-\ell_{\rm in})=4\pi R^2H\alpha P,
\end{equation}
where $\ell$ and $\ell_{\rm in}$ are specific angular momentum at $R$ and $r_{\rm in}$ of the slim disk.
Here $\alpha$ is the $\alpha$-description of the viscosity.
The energy equation reads
\begin{equation}
    \dot{M}_{\bullet}(\ell-\ell_{\rm in})\left(-\frac{\rmd \Omega}{\rmd R}\right)+\dot{M}_{\bullet}T\left(\frac{\rmd S}{\rmd R}\right)=4\pi RF_{\rm rad},
\end{equation}
where the entropy and surface cooling rates are given by
\begin{equation}
    T\rmd S=\frac{P}{\rho}\left[\left(12-\frac{21}{2}\beta_{\rm d}\right)\frac{\rmd T}{T}-\left(4-\beta_{\rm d}\right)\frac{\rmd \rho}{\rho}\right],\quad
    F_{\rm rad}=\frac{c}{\kappa\rho}\frac{aT^4}{3H},
\end{equation}
respectively.
Here $\beta_{\rm d}=P_{\rm gas}/\left(P_{\rm gas}+P_{\rm rad}\right)$ is the fraction of gas pressure to the total, $P_{\rm rad}=aT^4/3$ is the radiation pressure in the slim disk, $\kappa$ is the opacity of the disk.
Replacing $F_{\lambda}$ with $B_{\lambda}(T)$ in Eq.\,(\ref{eq:sp_sMBH}), we get the continuum, where $B_{\lambda}$ is the Planck function.
The slim disks approach the standard accretion disks when $\mathdotM$ is small enough.

We would point out the issue of the connection between the {\sMBH} and the slim disk. 
{At large radii}, the slim disk reduces to the standard disk, but it will mismatch the {\sMBH} solution because we artificially shut off the AMS heating at $\calR_{\rm in}$.
It is therefore expected that the discontinuity in density and temperature exists between the slim disk and the {\sMBH}, but we take the accretion rates to be continuous.
This discontinuity doesn't affect SEDs significantly since little energy is released in this region.  

\section{Properties of {\cMBH}-DISK}
\begin{figure*}
\centering
\includegraphics[width=0.98\textwidth]{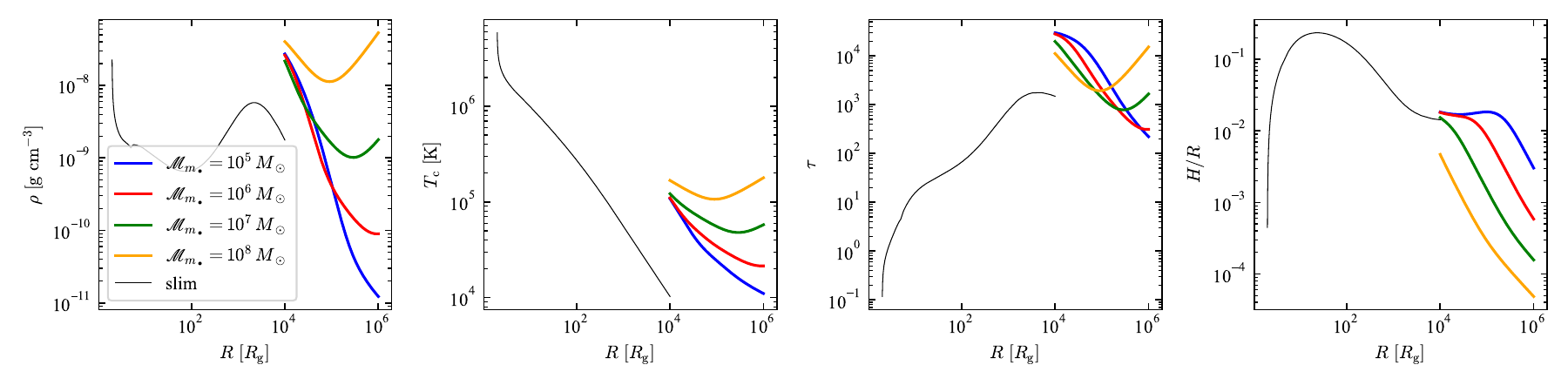}
\includegraphics[width=0.98\textwidth]{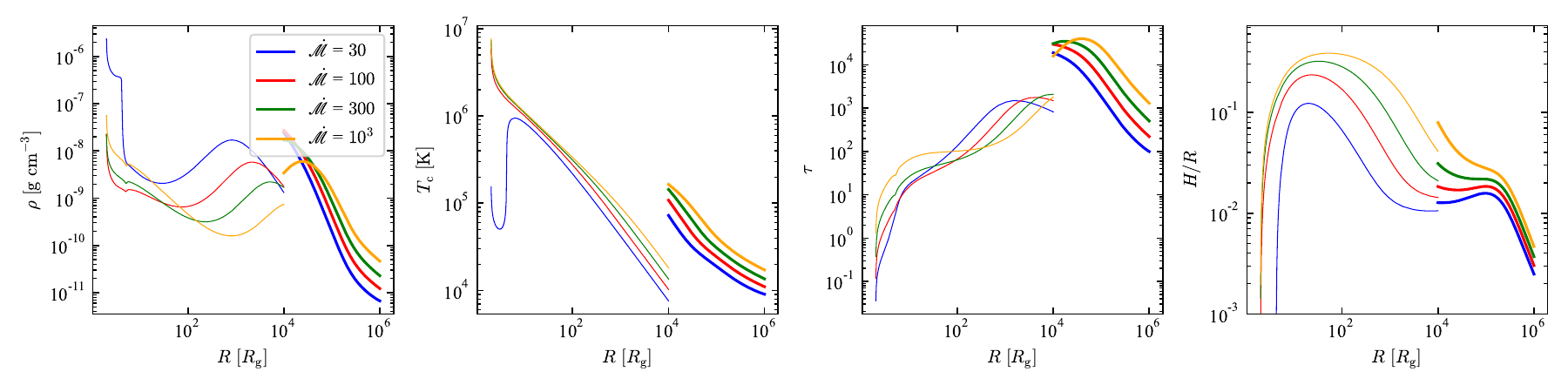}
\includegraphics[width=0.98\textwidth]{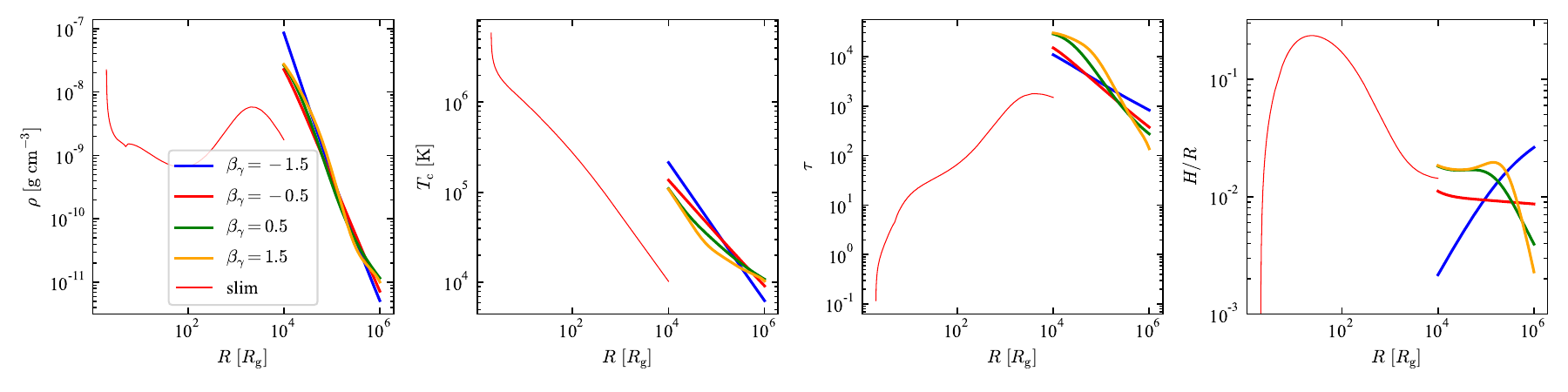}
\caption{\footnotesize Radial profiles of the disk model parameters and their dependence on $\mathscr{M}_{m_\bullet}$, $\dot{\mathscr{M}}$, and $\beta_\gamma$.
We fix the central massive black hole mass of $\BHM=10^5\,M_\odot$ for all the cases.
Here $\Rg=G\BHM/c^2$ is the gravitational radius of the cMBH.
The panels in the first row are for different $\mathscr{M}_{m_\bullet}$ and fixing $\dot{\mathscr{M}}=100$, $\beta_\gamma=1$, $\mathcal{R}_{\rm in}=10^4\,\Rg$, and $\mathcal{R}_{\rm out}=10^6\,\Rg$.
The panels in the second row are for dependence of structures on different $\dot{\mathscr{M}}$.
The panels in the third row are for dependence on $\beta_{\gamma}$.
}
\label{fig:disk-structure}
\end{figure*}

Since accretion rates decrease with radius as shown by Eq.\,(\ref{eq:dotM}), we give  $\dot{M}_{\bullet}^0$ at $\calR_{\rm out}$ as the accretion rate at the outer boundary of the \sMBH.
There is a critical value of $\dot{M}_{\bullet}^0$, with which
$\dot{M}_{\bullet}=0$ at $\calR_{\rm in}$ according to Eq.\,(\ref{eq:dotM}).
In this paper, we focus on the case in which  $\dot{M}_{\bullet}^0$ is large enough so that $\dot{M}_{\bullet}$ doesn't vanish at $\calR_{\rm in}$.
Though there are ten parameters, the structures and SEDs mainly rely on three parameters ($\mathscr{M}_{m_{\bullet}}$, $\beta_{\gamma}$ and $\dot{M}_{\bullet}^0$).
Structures of the {\sMBH} for different values of the parameters are shown in Fig.\,\ref{fig:disk-structure}.

For a fixed $\BHM=10^5\,\sunm$, the structure's dependence on $\mathscr{M}_{m_{\bullet}}$ shows interesting properties on the AMS heating. 
First, the disk densities and temperatures sensitively increase with $\mathscr{M}_{m_{\bullet}}$ whereas $H/R$ decreases with $\mathscr{M}_{m_{\bullet}}$.
AMS heating increases the temperature, and the vertical gravity of AMSs compresses the disk height, leading to increases of disk densities.
Second, disk densities, temperatures, optical depth and height monotonically increase with $\dot{\mathscr{M}}$.
All these properties are easily understood.
Third, the structure properties are quite sensitive to $\beta_{\gamma}$.
Since the mass function and spatial distributions are degenerate in the heating functions (Eq.\,\ref{eq:QsMBH}), we are not able to separate them.
The third line panels show the dependence on $\beta_{\gamma}$, in particular, the temperatures and optical depth on $\beta_{\gamma}$ as well as SEDs.
More steeper $\beta_{\gamma}$ more inner structures will be modified.

\begin{figure*}
\centering
\includegraphics[width=0.98\textwidth]{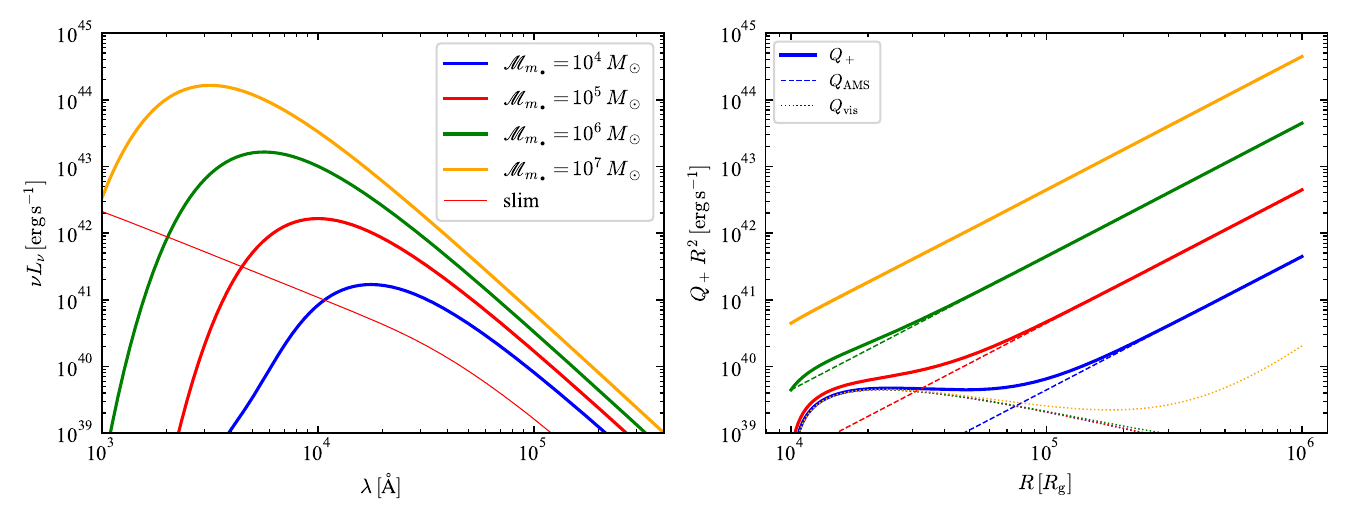}
\includegraphics[width=0.98\textwidth]{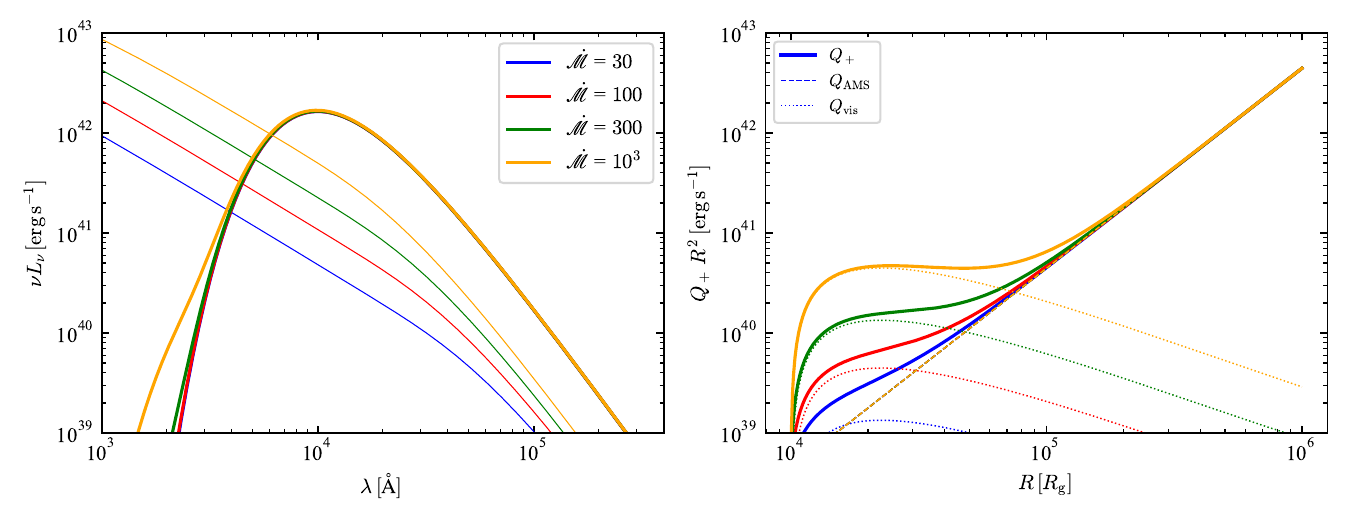}
\includegraphics[width=0.98\textwidth]{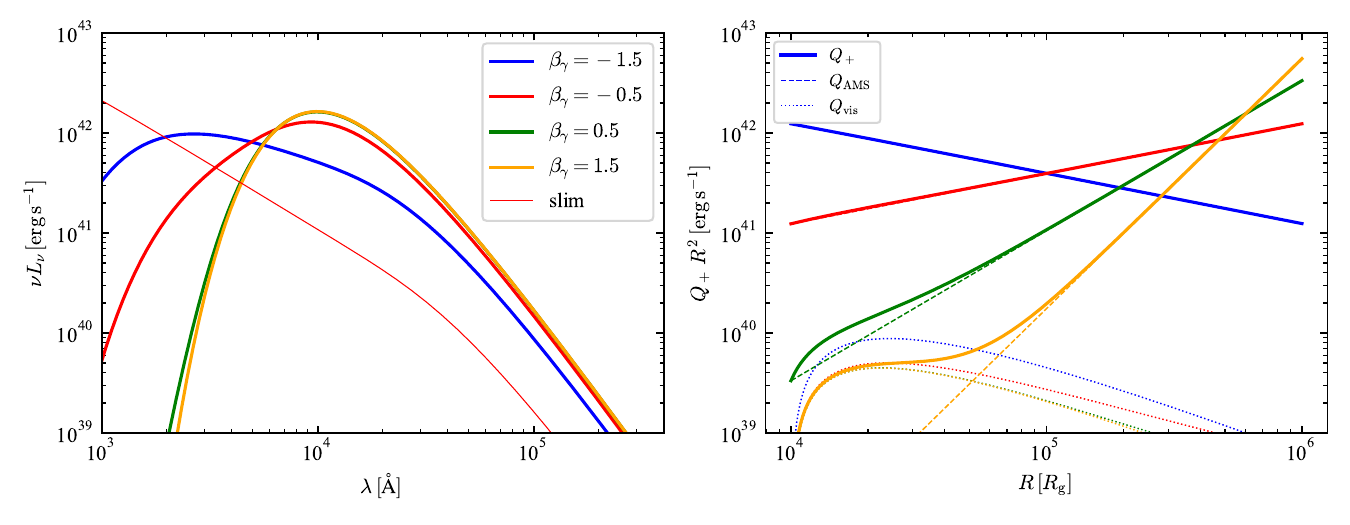}
\caption{\footnotesize Comparisons of viscosity heating and AMS, and resultant SEDs, and their dependence on $\mathscr{M}_{m_\bullet}$, $\dot{\mathscr{M}}$, and $\beta_\gamma$.
The standard parameter values are $\mathscr{M}_{m_\bullet}=10^5\,M_\odot$, $\dot{\mathscr{M}}=100$, $\beta_\gamma=1$, $\mathcal{R}_{\rm in}=10^4\,\Rg$, and $\mathcal{R}_{\rm out}=10^6\,\Rg$.
From top to bottom, the panels correspond to variations in $\mathscr{M}_{m_\bullet}$, $\dot{\mathscr{M}}$, and $\beta_\gamma$, respectively.}
\label{fig:SED1}
\end{figure*}

Fig.\,\ref{fig:SED1} shows SED dependence on the parameters.
The slim disk of the cMBH contributes to the UV continuum while the {\sMBH} provides the optical continuum.
Generally, the total SEDs show V-shapes and the turnover frequencies depend on the relative contributions of the slim disks and the {\sMBH}.
The total SEDs mainly depend on $\mathscr{M}_{m_{\bullet}}$, $\dot{\mathscr{M}}$ and $\beta_{\gamma}$.
In particular, $\beta_{\gamma}$ affects the SED shapes of {\sMBH}, as shown by the fourth row and left panel.
This is caused by the facts that energy densities of the {\sMBH} contributions increase with steeper $\beta_{\gamma}$ increasing the temperatures of the {\sMBH}.

\section{Profiles of broad H$\beta$ line}\label{sec:profile}
The observed broad H$\beta$ line is generated by the gravity of the {\sMBH} system. 
In this section, we calculate the profile of broad H$\beta$ line for a given spatial distribution of sMBHs.
The rotation velocity is given by $V_c(R)\approx \OmegaK R$, and leads to width of H$\beta$ line as
\begin{equation}\label{eq:FWHM}
    {\rm FWHM}\approx 1.0\times 10^3\left(\frac{\Sigbh^0}{10^7\sunm}\right)^{1/2}\left(\frac{R_{\rm out}}{10\,{\rm ltd}}\right)^{-1/2}\left(\frac{\mathscr{F}_{R}}{0.2}\right)^{1/2}\,\kms,
\end{equation}
from Eq.\,(\ref{eq:OmegaK}).
This is a typical width of the LRDs.
Detailed potential, with the consideration of the sMBH disk, is complicated and depends on $\betaBH$ and $\gammaBH$, but only weakly (varying by no more than 2/3: \citealt{Hure2011}), thereby shaping the profile of the broad H$\beta$ line.
Currently, the cMBH masses are estimated by the conventional formulations of Eq.\,(\ref{eq:MBHmass}) in LRDs; however, Eq.\,(\ref{eq:FWHM}) implies that the mass estimates \citep{Maiolino2024} for the cMBHs are actually close to the total masses of the sMBHs and the cMBHs (negligible compared with $\Sigbh$).

The presence of the sMBHs makes it hard to measure the cMBH masses unless the central regions can be spatially resolve.
Even the most powerful GRAVITY+/VLTI is unable to measure the CMBH, but future Kilometer-baseline interferometry in optical-infrared bands \citep{Bourdarot2024} will be capable of measuring the cMBH in local AGNs or LRD-analogs in the local universe.

\section{Collapse of primordial clouds}\label{sec:collapse}
Theoretical considerations on CPCs were originally discussed for stars and galaxies \citep{Hoyle1953,Hunter1962}, subsequently for bright quasars \citep{Field1964} and star clusters \citep{Peebles1968} in an analytical form. 
Nowadays, the collapse of primordial clouds has been extensively studied by numerical simulations for formation of star clusters \citep{Abel2000,Yoshida2003,Devecchi2009}, first stars \citep{Abel2002,Bromm2002}, and quasar's SMBHs \citep{Loeb1994}. 
However, publications about CPCs as the origin of quasars in the 1960s have gradually faded into the obscurity of historical literature.
Notwithstanding, these heuristic ideas may prove valuable for understanding the present LRDs in light of the collapsing structure, in particular, the CPC structure of a core with an envelope \citep{Unno1967,Matsuda1969,Hirasawa1969}.
The characteristic structures provide a compelling picture of the LRD's absorption features of H$\alpha$ and H$\beta$ lines.

Details of the CPCs are far beyond the scope of this paper, but we hope to grasp the radiative acceleration of the envelope for the absorption features of LRDs.
For the simplest discussion, we assume a homogeneous gas sphere with a density of $\rho_{\rm c}$.
While nuclear starburst is triggered during the collapse \citep{Field1964}, the radiation pressure of the starburst will push a gaseous envelope with a radius of $D_{\rm en}$ to expand outward. 
The kinetic energy of the envelope is given by
\begin{equation}
    \frac{1}{2}M_{\rm en}V_{\rm en}^2=\eta_*f_*M_{\rm core}c^2\tau_{\rm abs},
\end{equation}
where $M_{\rm en}$ is the envelope mass with absorption optical depth $\tau_{\rm abs}$, $V_{\rm en}$ is the velocity, $M_{\rm core}$ is the nuclear starburst mass, $\eta_*$ is the converting efficiency of the nuclear reaction and $f_*$ is the star formation efficiency.
We have the typical velocity of
\begin{equation}\label{eq:Ven}
    V_{\rm en}=112.2\,\eta_{\bar{7}}^{1/2}f_{-2}^{1/2}\tau_{0.1}^{1/2}q_{-2}^{1/2}\,\kms,
\end{equation}
where $\eta_{\bar{7}}=\eta_*/0.007$, $f_{-2}=f_*/10^{-2}$ and $q_{-2}=q/10^{-2}$ and $q=M_{\rm core}/M_{\rm en}$ is the mass ratio of the core to the envelope.
This corresponds to a distance to the center
\begin{equation}\label{eq:Den}
    D_{\rm en}\approx 4.3\,M_7V_{100}^{-2}\,{\rm pc},
\end{equation}
from $V_{\rm en}^2=G\left(\BHM+\Sigbh\right)/D_{\rm en}$, where $M_7=\left(\BHM+\Sigbh\right)/10^7\,\sunm$ and $V_{100}=V_{\rm en}/100\,\kms$.

Considering that the envelope is composed of $\calN$ clumps, we have the covering factor $\calC=\calN r_{\rm cl}^2/D_{\rm sh}^2$, where $r_{\rm cl}$ is the radius of individual clumps in the envelope.
Mass conservation for the envelope gives $\calN \rho_{\rm cl}r_{\rm cl}^3=D_{\rm en}^3\rho_{\rm c}$.
We have the optical depth of individual clouds as 
\begin{equation}\label{eq:tauabs}
    \tau_{\rm abs}=\kappa_{\rm abs}\rho_{\rm cl}r_{\rm cl}=0.2\,\calC_{0.1}^{-1}\kappa_{-2}\rho_{-19}D_{\rm sh,5},
\end{equation}
where $\kappa_{-2}=\kappa_{\rm abs}/10^{-2}$, $\rho_{-19}=\rho_{\rm c}/10^{-19}\,{\rm g\,cm^{-3}}$, $D_{\rm sh,5}=D_{\rm sh}/5\,$pc, and $\calC_{0.1}=\calC/0.1$, a value indicated by the fraction of the LRDs with absorption features.
The existence of such a thin absorbing layer is the result of the interaction between the core and the outflowing envelope.
This simple estimation of the envelope agrees with the absorption features of the Balmer lines.

Finally, we would point out that the LRD-analogs in the local universe provide interesting comparisons.
The LRDs mostly appear in the early universe \citep{MaYL2025} though there are a few local analogs \citep{LiuR2025,LinXJ2025b,ChenXY2025,Ji2025}.
J\,1025+1402 ($z=0.1$) shows extremely high equivalent width Na\,D, K\,{\sc i}, Fe\,{\sc ii}, and Ca\,{\sc ii} triplet absorption lines \citep{LinXJ2025b,Ji2025}.
Another object, J\,204837.26-002437.2 ($z=0.4$), shows a star formation rate of $400\,\sunmyr$ in extended star formation regions \citep{ChenXY2025}.
However, how are the high-$z$ LRDs similar to these local analogs remains an open question.

\section{Fitting scheme}\label{sec:fitting}
\subsection{LRD samples}
We began by collecting all LRD sources reported in previous studies \citep{Kocevski2025,Setton2024}.
After cross-matching these catalogs and removing duplicates, we obtained a unified list of LRDs. 
We then searched for the corresponding JWST/NIRSpec spectra of these sources in the DAWN JWST Archive (DJA; {\tt{https://dawn-cph.github.io/dja/}}). 
The DJA is a public database maintained by the DAWN Initiative, providing uniformly processed NIRSpec and NIRCam data from major extragalactic surveys such as CEERS, JADES, and PRIMER. 
Among the compiled candidates, those with available NIRSpec/PRISM spectra were selected for analysis. 
Tab.\,\ref{tab:sample} lists three representative kinds of LRDs with PRISM spectroscopy identified in the DJA. 
The PRISM configuration offers a broad, continuous wavelength coverage from $0.6\,{\mu}$m to $5.3\,\mu$m at low spectral resolution ($R\approx100$), enabling simultaneous measurement of the mid-infrared continuum. 
The three kinds of LRDs characterize the global spectral energy distribution of LRDs.

\subsection{Fitting}
$\FCAMS$ and $\FCMBH$ are obtained by the model described by Eq.\,(\ref{eq:continuum}) in Appendix \ref{sec:sMBH-disk}.
Continuum from nuclear starbursts can be obtained by {\tt{CIGALE}} \citep{Boquien2019}. 
The emission from a starburst depends on its age, initial mass function and total mass. 
In this paper, we use {\tt{CIGALE}} to fit the Balmer part to obtain an approximate continuum of starbursts ($f_{\lambda}$). 
Using a normalization $\alpha_0$ for the starburst continuum, we jointly fit the SEDs with the \sMBH. 
We define
\begin{equation}
    \chi^2=\sum_{i=1}^N\frac{1}{\sigma_i^2}\left(F_{\lambda}^{{\rm obs},i}-\FCAMS^i-\FCMBH^i-\FCnsb^i\right)^2,
\end{equation}
where $\FCnsb^i=\alpha_0f_{\lambda}^i$, and $\sigma_i$ are the observational errors.
We mask the emission lines in the fitting.
The fitting results are obtained when $\chi^2$ reaches its minimum.

\citet{Ji2025GN} found that GN-z11 ($z\approx 10.6$) has an excess between $0.3-0.35\,\mu$m remains open.
Here we propose that this is similar to the rest-frame optical continuum of LRDs contributed by the {\sMBH}, but $\calR_{\rm in}$ will be much smaller than in typical LRDs.
This scenario can be tested by monitoring campaigns if this component is variable.

\bibliographystyle{aasjournal}
\bibliography{reference}

@ARTICLE{Xu2025,
       author = {{Xu}, Zheng-Hao and {Chen}, Yi-Xian and {Lin}, Douglas N.~C.},
        title = "{Stellar Evolution with Radiative Feedback in AGN Disks}",
      journal = {arXiv e-prints},
         year = 2025,
        month = nov,
          eid = {arXiv:2511.03904},
        pages = {arXiv:2511.03904},
archivePrefix = {arXiv},
       eprint = {2511.03904},
 primaryClass = {astro-ph.GA},
       adsurl = {https://ui.adsabs.harvard.edu/abs/2025arXiv251103904X}
}

@ARTICLE{Vaccaro2025,
       author = {{Vaccaro}, Maria Paola and {Seif}, Yannick and {Mapelli}, Michela},
        title = "{The role of migration traps in the formation of binary black holes in AGN disks}",
      journal = {arXiv e-prints},
         year = 2025,
        month = aug,
          eid = {arXiv:2508.03637},
        pages = {arXiv:2508.03637},
          doi = {10.48550/arXiv.2508.03637},
archivePrefix = {arXiv},
       eprint = {2508.03637},
 primaryClass = {astro-ph.HE},
       adsurl = {https://ui.adsabs.harvard.edu/abs/2025arXiv250803637V}
}

@ARTICLE{Amaro-Seoane2023,
       author = {{Amaro-Seoane}, Pau and {Andrews}, Jeff and {Arca Sedda}, Manuel and {Askar}, Abbas and {Baghi}, Quentin and {Balasov}, Razvan and {Bartos}, Imre and {Bavera}, Simone S. and {Bellovary}, Jillian and {Berry}, Christopher P.~L. and {Berti}, Emanuele and {Bianchi}, Stefano and {Blecha}, Laura and {Blondin}, St{\'e}phane and {Bogdanovi{\'c}}, Tamara and {Boissier}, Samuel and {Bonetti}, Matteo and {Bonoli}, Silvia and {Bortolas}, Elisa and {Breivik}, Katelyn and {Capelo}, Pedro R. and {Caramete}, Laurentiu and {Cattorini}, Federico and {Charisi}, Maria and {Chaty}, Sylvain and {Chen}, Xian and {Chru{\'s}li{\'n}ska}, Martyna and {Chua}, Alvin J.~K. and {Church}, Ross and {Colpi}, Monica and {D'Orazio}, Daniel and {Danielski}, Camilla and {Davies}, Melvyn B. and {Dayal}, Pratika and {De Rosa}, Alessandra and {Derdzinski}, Andrea and {Destounis}, Kyriakos and {Dotti}, Massimo and {Du{\c{t}}an}, Ioana and {Dvorkin}, Irina and {Fabj}, Gaia and {Foglizzo}, Thierry and {Ford}, Saavik and {Fouvry}, Jean-Baptiste and {Franchini}, Alessia and {Fragos}, Tassos and {Fryer}, Chris and {Gaspari}, Massimo and {Gerosa}, Davide and {Graziani}, Luca and {Groot}, Paul and {Habouzit}, Melanie and {Haggard}, Daryl and {Haiman}, Zoltan and {Han}, Wen-Biao and {Istrate}, Alina and {Johansson}, Peter H. and {Khan}, Fazeel Mahmood and {Kimpson}, Tomas and {Kokkotas}, Kostas and {Kong}, Albert and {Korol}, Valeriya and {Kremer}, Kyle and {Kupfer}, Thomas and {Lamberts}, Astrid and {Larson}, Shane and {Lau}, Mike and {Liu}, Dongliang and {Lloyd-Ronning}, Nicole and {Lodato}, Giuseppe and {Lupi}, Alessandro and {Ma}, Chung-Pei and {Maccarone}, Tomas and {Mandel}, Ilya and {Mangiagli}, Alberto and {Mapelli}, Michela and {Mathis}, St{\'e}phane and {Mayer}, Lucio and {McGee}, Sean and {McKernan}, Berry and {Miller}, M. Coleman and {Mota}, David F. and {Mumpower}, Matthew and {Nasim}, Syeda S. and {Nelemans}, Gijs and {Noble}, Scott and {Pacucci}, Fabio and {Panessa}, Francesca and {Paschalidis}, Vasileios and {Pfister}, Hugo and {Porquet}, Delphine and {Quenby}, John and {Ricarte}, Angelo and {R{\"o}pke}, Friedrich K. and {Regan}, John and {Rosswog}, Stephan and {Ruiter}, Ashley and {Ruiz}, Milton and {Runnoe}, Jessie and {Schneider}, Raffaella and {Schnittman}, Jeremy and {Secunda}, Amy and {Sesana}, Alberto and {Seto}, Naoki and {Shao}, Lijing and {Shapiro}, Stuart and {Sopuerta}, Carlos and {Stone}, Nicholas C. and {Suvorov}, Arthur and {Tamanini}, Nicola and {Tamfal}, Tomas and {Tauris}, Thomas and {Temmink}, Karel and {Tomsick}, John and {Toonen}, Silvia and {Torres-Orjuela}, Alejandro and {Toscani}, Martina and {Tsokaros}, Antonios and {Unal}, Caner and {V{\'a}zquez-Aceves}, Ver{\'o}nica and {Valiante}, Rosa and {van Putten}, Maurice and {van Roestel}, Jan and {Vignali}, Christian and {Volonteri}, Marta and {Wu}, Kinwah and {Younsi}, Ziri and {Yu}, Shenghua and {Zane}, Silvia and {Zwick}, Lorenz and {Antonini}, Fabio and {Baibhav}, Vishal and {Barausse}, Enrico and {Bonilla Rivera}, Alexander and {Branchesi}, Marica and {Branduardi-Raymont}, Graziella and {Burdge}, Kevin and {Chakraborty}, Srija and {Cuadra}, Jorge and {Dage}, Kristen and {Davis}, Benjamin and {de Mink}, Selma E. and {Decarli}, Roberto and {Doneva}, Daniela and {Escoffier}, Stephanie and {Gandhi}, Poshak and {Haardt}, Francesco and {Lousto}, Carlos O. and {Nissanke}, Samaya and {Nordhaus}, Jason and {O'Shaughnessy}, Richard and {Portegies Zwart}, Simon and {Pound}, Adam and {Schussler}, Fabian and {Sergijenko}, Olga and {Spallicci}, Alessandro and {Vernieri}, Daniele and {Vigna-G{\'o}mez}, Alejandro},
        title = "{Astrophysics with the Laser Interferometer Space Antenna}",
      journal = {Living Reviews in Relativity},
         year = 2023,
        month = dec,
       volume = {26},
       number = {1},
          eid = {2},
        pages = {2},
          doi = {10.1007/s41114-022-00041-y},
archivePrefix = {arXiv},
       eprint = {2203.06016},
 primaryClass = {gr-qc},
       adsurl = {https://ui.adsabs.harvard.edu/abs/2023LRR....26....2A}
}

@ARTICLE{Liu2025,
       author = {{Liu}, Hanpu and {Jiang}, Yan-Fei and {Quataert}, Eliot and {Greene}, Jenny E. and {Ma}, Yilun},
        title = "{The Balmer Break and Optical Continuum of Little Red Dots From Super-Eddington Accretion}",
      journal = {arXiv e-prints},
         year = 2025,
        month = jul,
          eid = {arXiv:2507.07190},
        pages = {arXiv:2507.07190},
          doi = {10.48550/arXiv.2507.07190},
archivePrefix = {arXiv},
       eprint = {2507.07190},
 primaryClass = {astro-ph.GA},
       adsurl = {https://ui.adsabs.harvard.edu/abs/2025arXiv250707190L}
}

@ARTICLE{Lu2019,
       author = {{Lu}, Kai-Xing and {Huang}, Ying-Ke and {Zhang}, Zhi-Xiang and {Wang}, Kai and {Du}, Pu and {Hu}, Chen and {Xiao}, Ming and {Li}, Yan-Rong and {Bai}, Jin-Ming and {Bian}, Wei-Hao and {Yuan}, Ye-Fei and {Ho}, Luis C. and {Wang}, Jian-Min and {SEAMBH Collaboration}},
        title = "{Supermassive Black Holes with High Accretion Rates in Active Galactic Nuclei. X. Optical Variability Characteristics}",
      journal = {\apj},
         year = 2019,
        month = may,
       volume = {877},
       number = {1},
          eid = {23},
        pages = {23},
          doi = {10.3847/1538-4357/ab16e8},
archivePrefix = {arXiv},
       eprint = {1904.03393},
 primaryClass = {astro-ph.GA},
       adsurl = {https://ui.adsabs.harvard.edu/abs/2019ApJ...877...23L}
}

@ARTICLE{LiuR2025,
       author = {{Lin}, Ruqiu and {Zheng}, Zhen-Ya and {Jiang}, Chunyan and {Yuan}, Fang-Ting and {Ho}, Luis C. and {Wang}, Junxian and {Jiang}, Linhua and {Rhoads}, James E. and {Malhotra}, Sangeeta and {Barrientos}, L. Felipe and {Wold}, Isak and {Infante}, Leopoldo and {Zhu}, Shuairu and {Ji}, Xiang and {Fu}, Xiaodan},
        title = "{Discovery of Local Analogs to JWST's Little Red Dots}",
      journal = {\apjl},
         year = 2025,
        month = feb,
       volume = {980},
       number = {2},
          eid = {L34},
        pages = {L34},
          doi = {10.3847/2041-8213/adaaf1},
archivePrefix = {arXiv},
       eprint = {2412.08396},
 primaryClass = {astro-ph.GA},
       adsurl = {https://ui.adsabs.harvard.edu/abs/2025ApJ...980L..34L}
}

@ARTICLE{ChenXY2025,
       author = {{Chen}, Xiaoyang and {Ichikawa}, Kohei and {Akiyama}, Masayuki and {Inayoshi}, Kohei and {Inoue}, Akio K. and {Onoue}, Masafusa and {Toba}, Yoshiki and {Zavala}, Jorge A. and {Bakx}, Tom J. and {Kawaguchi}, Toshihiro and {Lee}, Kianhong and {Matsumoto}, Naoki and {Vijarnwannaluk}, Bovornpratch},
        title = "{A $z\simeq0.4$ Little Red Dot analog: An Extended Starburst with an Overmassive Black hole}",
      journal = {arXiv e-prints},
         year = 2025,
        month = oct,
          eid = {arXiv:2510.02801},
        pages = {arXiv:2510.02801},
          doi = {10.48550/arXiv.2510.02801},
archivePrefix = {arXiv},
       eprint = {2510.02801},
 primaryClass = {astro-ph.GA},
       adsurl = {https://ui.adsabs.harvard.edu/abs/2025arXiv251002801C}
}

@ARTICLE{Bourdarot2024,
       author = {{Bourdarot}, G. and {Eisenhauer}, F.},
        title = "{Kilometer-baseline interferometry: science drivers for the next generation instrument}",
      journal = {arXiv e-prints},
         year = 2024,
        month = oct,
          eid = {arXiv:2410.22063},
        pages = {arXiv:2410.22063},
          doi = {10.48550/arXiv.2410.22063},
archivePrefix = {arXiv},
       eprint = {2410.22063},
 primaryClass = {astro-ph.IM},
       adsurl = {https://ui.adsabs.harvard.edu/abs/2024arXiv241022063B}
}

@ARTICLE{Arcordia2024,
       author = {{Arcodia}, R. and {Liu}, Z. and {Merloni}, A. and {Malyali}, A. and {Rau}, A. and {Chakraborty}, J. and {Goodwin}, A. and {Buckley}, D. and {Brink}, J. and {Gromadzki}, M. and {Arzoumanian}, Z. and {Buchner}, J. and {Kara}, E. and {Nandra}, K. and {Ponti}, G. and {Salvato}, M. and {Anderson}, G. and {Baldini}, P. and {Grotova}, I. and {Krumpe}, M. and {Maitra}, C. and {Miller-Jones}, J.~C.~A. and {Ramos-Ceja}, M.~E.},
        title = "{The more the merrier: SRG/eROSITA discovers two further galaxies showing X-ray quasi-periodic eruptions}",
      journal = {\aap},
         year = 2024,
        month = apr,
       volume = {684},
          eid = {A64},
        pages = {A64},
          doi = {10.1051/0004-6361/202348881},
archivePrefix = {arXiv},
       eprint = {2401.17275},
 primaryClass = {astro-ph.HE},
       adsurl = {https://ui.adsabs.harvard.edu/abs/2024A&A...684A..64A}
}

@ARTICLE{Chakraborty2021,
       author = {{Chakraborty}, Joheen and {Kara}, Erin and {Masterson}, Megan and {Giustini}, Margherita and {Miniutti}, Giovanni and {Saxton}, Richard},
        title = "{Possible X-Ray Quasi-periodic Eruptions in a Tidal Disruption Event Candidate}",
      journal = {\apjl},
         year = 2021,
        month = nov,
       volume = {921},
       number = {2},
          eid = {L40},
        pages = {L40},
          doi = {10.3847/2041-8213/ac313b},
archivePrefix = {arXiv},
       eprint = {2110.10786},
 primaryClass = {astro-ph.HE},
       adsurl = {https://ui.adsabs.harvard.edu/abs/2021ApJ...921L..40C}
}

@ARTICLE{Arcodia2021,
       author = {{Arcodia}, R. and {Merloni}, A. and {Nandra}, K. and {Buchner}, J. and {Salvato}, M. and {Pasham}, D. and {Remillard}, R. and {Comparat}, J. and {Lamer}, G. and {Ponti}, G. and {Malyali}, A. and {Wolf}, J. and {Arzoumanian}, Z. and {Bogensberger}, D. and {Buckley}, D.~A.~H. and {Gendreau}, K. and {Gromadzki}, M. and {Kara}, E. and {Krumpe}, M. and {Markwardt}, C. and {Ramos-Ceja}, M.~E. and {Rau}, A. and {Schramm}, M. and {Schwope}, A.},
        title = "{X-ray quasi-periodic eruptions from two previously quiescent galaxies}",
      journal = {\nat},
         year = 2021,
        month = apr,
       volume = {592},
       number = {7856},
        pages = {704-707},
          doi = {10.1038/s41586-021-03394-6},
archivePrefix = {arXiv},
       eprint = {2104.13388},
 primaryClass = {astro-ph.HE},
       adsurl = {https://ui.adsabs.harvard.edu/abs/2021Natur.592..704A}
}

@ARTICLE{Giustini2020,
       author = {{Giustini}, Margherita and {Miniutti}, Giovanni and {Saxton}, Richard D.},
        title = "{X-ray quasi-periodic eruptions from the galactic nucleus of RX J1301.9+2747}",
      journal = {\aap},
         year = 2020,
        month = apr,
       volume = {636},
          eid = {L2},
        pages = {L2},
          doi = {10.1051/0004-6361/202037610},
archivePrefix = {arXiv},
       eprint = {2002.08967},
 primaryClass = {astro-ph.HE},
       adsurl = {https://ui.adsabs.harvard.edu/abs/2020A&A...636L...2G}
}

@ARTICLE{Miniutti2019,
       author = {{Miniutti}, G. and {Saxton}, R.~D. and {Giustini}, M. and {Alexander}, K.~D. and {Fender}, R.~P. and {Heywood}, I. and {Monageng}, I. and {Coriat}, M. and {Tzioumis}, A.~K. and {Read}, A.~M. and {Knigge}, C. and {Gandhi}, P. and {Pretorius}, M.~L. and {Ag{\'\i}s-Gonz{\'a}lez}, B.},
        title = "{Nine-hour X-ray quasi-periodic eruptions from a low-mass black hole galactic nucleus}",
      journal = {\nat},
         year = 2019,
        month = sep,
       volume = {573},
       number = {7774},
        pages = {381-384},
          doi = {10.1038/s41586-019-1556-x},
archivePrefix = {arXiv},
       eprint = {1909.04693},
 primaryClass = {astro-ph.HE},
       adsurl = {https://ui.adsabs.harvard.edu/abs/2019Natur.573..381M}
}

@ARTICLE{Masterson2025,
       author = {{Masterson}, Megan and {Kara}, Erin and {Panagiotou}, Christos and {Alston}, William N. and {Chakraborty}, Joheen and {Burdge}, Kevin and {Ricci}, Claudio and {Laha}, Sibasish and {Arcavi}, Iair and {Arcodia}, Riccardo and {Cenko}, S. Bradley and {Fabian}, Andrew C. and {Garc{\'\i}a}, Javier A. and {Giustini}, Margherita and {Ingram}, Adam and {Kosec}, Peter and {Loewenstein}, Michael and {Meyer}, Eileen T. and {Miniutti}, Giovanni and {Pinto}, Ciro and {Remillard}, Ronald A. and {Sadaula}, Dev R. and {Shuvo}, Onic I. and {Trakhtenbrot}, Benny and {Wang}, Jingyi},
        title = "{Millihertz oscillations near the innermost orbit of a supermassive black hole}",
      journal = {\nat},
         year = 2025,
        month = feb,
       volume = {638},
       number = {8050},
        pages = {370-375},
          doi = {10.1038/s41586-024-08385-x},
archivePrefix = {arXiv},
       eprint = {2501.01581},
 primaryClass = {astro-ph.HE},
       adsurl = {https://ui.adsabs.harvard.edu/abs/2025Natur.638..370M}
}

@ARTICLE{LuoJ2025,
       author = {{Luo}, Jun and {An}, Haipeng and {Bian}, Ligong and {Cai}, Rong-Gen and {Cao}, Zhoujian and {Han}, Wenbiao and {He}, Jianhua and {Hendry}, Martin A. and {Hu}, Bin and {Hu}, Yi-Ming and {Huang}, Fa Peng and {Huang}, Shun-Jia and {Kim}, Sang Pyo and {Li}, En-Kun and {Liu}, Yu-Xiao and {Milyukov}, Vadim and {Pi}, Shi and {Postnov}, Konstantin and {Sasaki}, Misao and {Shao}, Cheng-Gang and {Shao}, Lijing and {Shi}, Changfu and {Sun}, Shuo and {Wang}, Anzhong and {Wang}, Pan-Pan and {Wang}, Sai and {Wang}, Shao-Jiang and {Xianyu}, Zhong-Zhi and {Yang}, Huan and {Yang}, Tao and {Zhang}, Jian-dong and {Zhang}, Xin and {Zhao}, Wen and {Zhu}, Liang-Gui and {Mei}, Jianwei},
        title = "{Fundamental Physics and Cosmology with TianQin}",
      journal = {arXiv e-prints},
         year = 2025,
        month = feb,
          eid = {arXiv:2502.20138},
        pages = {arXiv:2502.20138},
          doi = {10.48550/arXiv.2502.20138},
archivePrefix = {arXiv},
       eprint = {2502.20138},
 primaryClass = {gr-qc},
       adsurl = {https://ui.adsabs.harvard.edu/abs/2025arXiv250220138L}
}

@ARTICLE{DiGiovanni2025,
       author = {{Di Giovanni}, Matteo},
        title = "{Einstein Telescope and Cosmic Explorer}",
      journal = {arXiv e-prints},
         year = 2025,
        month = may,
          eid = {arXiv:2505.11033},
        pages = {arXiv:2505.11033},
          doi = {10.48550/arXiv.2505.11033},
archivePrefix = {arXiv},
       eprint = {2505.11033},
 primaryClass = {gr-qc},
       adsurl = {https://ui.adsabs.harvard.edu/abs/2025arXiv250511033D}
}

@ARTICLE{Choi2013,
       author = {{Choi}, Jun-Hwan and {Shlosman}, Isaac and {Begelman}, Mitchell C.},
        title = "{Supermassive Black Hole Formation at High Redshifts via Direct Collapse: Physical Processes in the Early Stage}",
      journal = {\apj},
         year = 2013,
        month = sep,
       volume = {774},
       number = {2},
          eid = {149},
        pages = {149},
          doi = {10.1088/0004-637X/774/2/149},
archivePrefix = {arXiv},
       eprint = {1304.1369},
 primaryClass = {astro-ph.CO},
       adsurl = {https://ui.adsabs.harvard.edu/abs/2013ApJ...774..149C}
}

@ARTICLE{LinXJ2025b,
       author = {{Lin}, Xiaojing and {Fan}, Xiaohui and {Cai}, Zheng and {Bian}, Fuyan and {Liu}, Hanpu and {Sun}, Fengwu and {Ma}, Yilun and {Greene}, Jenny E. and {Strauss}, Michael A. and {Green}, Richard and {Lyu}, Jianwei and {Champagne}, Jaclyn B. and {Goulding}, Andy D. and {Inayoshi}, Kohei and {Jin}, Xiangyu and {Leung}, Gene C.~K. and {Li}, Mingyu and {Liu}, Yichen and {Mao}, Junjie and {Pudoka}, Maria Anne and {Tee}, Wei Leong and {Wang}, Ben and {Wang}, Feige and {Wu}, Yunjing and {Yang}, Jinyi and {Zhang}, Haowen and {Zhu}, Yongda},
        title = "{The Discovery of Little Red Dots in the Local Universe: Signatures of Cool Gas Envelopes}",
      journal = {arXiv e-prints},
         year = 2025,
        month = jul,
          eid = {arXiv:2507.10659},
        pages = {arXiv:2507.10659},
          doi = {10.48550/arXiv.2507.10659},
archivePrefix = {arXiv},
       eprint = {2507.10659},
 primaryClass = {astro-ph.GA},
       adsurl = {https://ui.adsabs.harvard.edu/abs/2025arXiv250710659L}
}

@ARTICLE{Wang2023GC,
       author = {{Wang}, Jian-Min and {Liu}, Jun-Rong and {Li}, Yan-Rong and {Songsheng}, Yu-Yang and {Yuan}, Ye-Fei and {Ho}, Luis C.},
        title = "{Accretion-modified Stars in Accretion Disks of Active Galactic Nuclei: The Low-luminosity Cases and an Application to Sgr A*}",
      journal = {\apjl},
         year = 2023,
        month = dec,
       volume = {958},
       number = {2},
          eid = {L40},
        pages = {L40},
          doi = {10.3847/2041-8213/ad0bd9},
archivePrefix = {arXiv},
       eprint = {2311.06781},
 primaryClass = {astro-ph.HE},
       adsurl = {https://ui.adsabs.harvard.edu/abs/2023ApJ...958L..40W}
}

@ARTICLE{Ji2025,
       author = {{Ji}, Xihan and {D'Eugenio}, Francesco and {Juod{\v{z}}balis}, Ignas and {Walton}, Dom and {Fabian}, Andrew C. and {Maiolino}, Roberto and {Ramos Almeida}, Cristina and {Acosta Pulido}, Jose A. and {Belokurov}, Vasily A. and {Isobe}, Yuki and {Jones}, Gareth and {Maraston}, Claudia and {Scholtz}, Jan and {Simmonds}, Charlotte and {Tacchella}, Sandro and {Terlevich}, Elena and {Terlevich}, Roberto},
        title = "{Lord of LRDs: Insights into a ``Little Red Dot'' with a low-ionization spectrum at z = 0.1}",
      journal = {arXiv e-prints},
         year = 2025,
        month = jul,
          eid = {arXiv:2507.23774},
        pages = {arXiv:2507.23774},
          doi = {10.48550/arXiv.2507.23774},
archivePrefix = {arXiv},
       eprint = {2507.23774},
 primaryClass = {astro-ph.GA},
       adsurl = {https://ui.adsabs.harvard.edu/abs/2025arXiv250723774J}
}

@ARTICLE{Peebles1969,
       author = {{Peebles}, P.~J.~E.},
        title = "{Origin of the Angular Momentum of Galaxies}",
      journal = {\apj},
         year = 1969,
        month = feb,
       volume = {155},
        pages = {393},
          doi = {10.1086/149876},
       adsurl = {https://ui.adsabs.harvard.edu/abs/1969ApJ...155..393P}
}

@ARTICLE{Wang2011,
       author = {{Wang}, Jian-Min and {Ge}, Jun-Qiang and {Hu}, Chen and {Baldwin}, Jack A. and {Li}, Yan-Rong and {Ferland}, Gary J. and {Xiang}, Fei and {Yan}, Chang-Shuo and {Zhang}, Shu},
        title = "{Star Formation in Self-gravitating Disks in Active Galactic Nuclei. I. Metallicity Gradients in Broad-line Regions}",
      journal = {\apj},
         year = 2011,
        month = sep,
       volume = {739},
       number = {1},
          eid = {3},
        pages = {3},
          doi = {10.1088/0004-637X/739/1/3},
archivePrefix = {arXiv},
       eprint = {1107.3620},
 primaryClass = {astro-ph.GA},
       adsurl = {https://ui.adsabs.harvard.edu/abs/2011ApJ...739....3W}
}

@ARTICLE{Wang2010,
       author = {{Wang}, Jian-Min and {Yan}, Chang-Shuo and {Gao}, Han-Qin and {Hu}, Chen and {Li}, Yan-Rong and {Zhang}, Shu},
        title = "{Accretion Disks in Active Galactic Nuclei: Gas Supply Driven by Star Formation}",
      journal = {\apjl},
         year = 2010,
        month = aug,
       volume = {719},
       number = {2},
        pages = {L148-L152},
          doi = {10.1088/2041-8205/719/2/L148},
archivePrefix = {arXiv},
       eprint = {1007.4060},
 primaryClass = {astro-ph.CO},
       adsurl = {https://ui.adsabs.harvard.edu/abs/2010ApJ...719L.148W}
}

@ARTICLE{Volonteri2005,
       author = {{Volonteri}, Marta and {Rees}, Martin J.},
        title = "{Rapid Growth of High-Redshift Black Holes}",
      journal = {\apj},
         year = 2005,
        month = nov,
       volume = {633},
       number = {2},
        pages = {624-629},
          doi = {10.1086/466521},
archivePrefix = {arXiv},
       eprint = {astro-ph/0506040},
 primaryClass = {astro-ph},
       adsurl = {https://ui.adsabs.harvard.edu/abs/2005ApJ...633..624V}
}

@ARTICLE{Billand2025,
       author = {{Billand}, Jean-Baptiste and {Elbaz}, David and {Gentile}, Fabrizio and {Tarrasse}, Maxime and {Franco}, Maximilien and {Magnelli}, Benjamin and {Daddi}, Emanuele and {Lyu}, Yipeng and {Dekel}, Avishai and {Pacucci}, Fabio and {Sangalli}, Valentina and {Dickinson}, Mark and {Giavalisco}, Mauro and {Holwerda}, Benne W. and {Kocevski}, Dale D. and {Koekemoer}, Anton M. and {Kokorev}, Vasily and {Lucas}, Ray A. and {P{\'e}rez-Gonz{\'a}lez}, Pablo G.},
        title = "{Investigating the Growth of Little Red Dot Descendants at $z<4$ with the JWST}",
      journal = {arXiv e-prints},
         year = 2025,
        month = jul,
          eid = {arXiv:2507.04011},
        pages = {arXiv:2507.04011},
          doi = {10.48550/arXiv.2507.04011},
archivePrefix = {arXiv},
       eprint = {2507.04011},
 primaryClass = {astro-ph.GA},
       adsurl = {https://ui.adsabs.harvard.edu/abs/2025arXiv250704011B}
}

@BOOK{Mo2010,
       author = {{Mo}, Houjun and {van den Bosch}, Frank C. and {White}, Simon},
        title = "{Galaxy Formation and Evolution}",
         year = 2010,
          doi = {10.1017/CBO9780511807244},
       adsurl = {https://ui.adsabs.harvard.edu/abs/2010gfe..book.....M}
}

@ARTICLE{Maiolino2024faint,
       author = {{Maiolino}, Roberto and {Scholtz}, Jan and {Curtis-Lake}, Emma and {Carniani}, Stefano and {Baker}, William and {de Graaff}, Anna and {Tacchella}, Sandro and {{\"U}bler}, Hannah and {D'Eugenio}, Francesco and {Witstok}, Joris and {Curti}, Mirko and {Arribas}, Santiago and {Bunker}, Andrew J. and {Charlot}, St{\'e}phane and {Chevallard}, Jacopo and {Eisenstein}, Daniel J. and {Egami}, Eiichi and {Ji}, Zhiyuan and {Jones}, Gareth C. and {Lyu}, Jianwei and {Rawle}, Tim and {Robertson}, Brant and {Rujopakarn}, Wiphu and {Perna}, Michele and {Sun}, Fengwu and {Venturi}, Giacomo and {Williams}, Christina C. and {Willott}, Chris},
        title = "{JADES: The diverse population of infant black holes at $4 < z < 11$: Merging, tiny, poor, but mighty}",
      journal = {\aap},
         year = 2024,
        month = nov,
       volume = {691},
          eid = {A145},
        pages = {A145},
          doi = {10.1051/0004-6361/202347640},
archivePrefix = {arXiv},
       eprint = {2308.01230},
 primaryClass = {astro-ph.GA},
       adsurl = {https://ui.adsabs.harvard.edu/abs/2024A&A...691A.145M}
}

@ARTICLE{Ji2025Q1,
       author = {{Ji}, Xihan and {Maiolino}, Roberto and {{\"U}bler}, Hannah and {Scholtz}, Jan and {D'Eugenio}, Francesco and {Sun}, Fengwu and {Perna}, Michele and {Turner}, Hannah and {Arribas}, Santiago and {Bennett}, Jake S. and {Bunker}, Andrew and {Carniani}, Stefano and {Charlot}, St{\'e}phane and {Cresci}, Giovanni and {Curti}, Mirko and {Egami}, Eiichi and {Fabian}, Andy and {Inayoshi}, Kohei and {Isobe}, Yuki and {Jones}, Gareth and {Juod{\v{z}}balis}, Ignas and {Kumari}, Nimisha and {Lyu}, Jianwei and {Mazzolari}, Giovanni and {Parlanti}, Eleonora and {Robertson}, Brant and {Rodr{\'\i}guez Del Pino}, Bruno and {Schneider}, Raffaella and {Sijacki}, Debora and {Tacchella}, Sandro and {Trinca}, Alessandro and {Valiante}, Rosa and {Venturi}, Giacomo and {Volonteri}, Marta and {Willott}, Chris and {Witten}, Callum and {Witstok}, Joris},
        title = "{BlackTHUNDER -- A non-stellar Balmer break in a black hole-dominated little red dot at $z=7.04$}",
      journal = {arXiv e-prints},
         year = 2025,
        month = jan,
          eid = {arXiv:2501.13082},
        pages = {arXiv:2501.13082},
          doi = {10.48550/arXiv.2501.13082},
archivePrefix = {arXiv},
       eprint = {2501.13082},
 primaryClass = {astro-ph.GA},
       adsurl = {https://ui.adsabs.harvard.edu/abs/2025arXiv250113082J}
}

@ARTICLE{Ji2025GN,
       author = {{Ji}, Xihan and {Maiolino}, Roberto and {Ferland}, Gary and {D'Eugenio}, Francesco and {Bhatawdekar}, Rachana and {Charlot}, St{\'e}phane and {Chevallard}, Jacopo and {Curti}, Mirko and {Curtis-Lake}, Emma and {Hainline}, Kevin and {Ji}, Zhiyuan and {Robertson}, Brant and {Rodr{\'\i}guez Del Pino}, Bruno and {Scholtz}, Jan and {Tacchella}, Sandro and {Williams}, Christina C. and {Witstok}, Joris},
        title = "{JADES {\textendash} the small blue bump in GN-z11: insights into the nuclear region of a galaxy at z = 10.6}",
      journal = {\mnras},
         year = 2025,
        month = aug,
       volume = {541},
       number = {3},
        pages = {2134-2161},
          doi = {10.1093/mnras/staf1083},
archivePrefix = {arXiv},
       eprint = {2405.05772},
 primaryClass = {astro-ph.GA},
       adsurl = {https://ui.adsabs.harvard.edu/abs/2025MNRAS.541.2134J}
}

@ARTICLE{McKernan2011,
       author = {{McKernan}, B. and {Ford}, K.~E.~S. and {Lyra}, W. and {Perets}, H.~B. and {Winter}, L.~M. and {Yaqoob}, T.},
        title = "{On rapid migration and accretion within discs around supermassive black holes}",
      journal = {\mnras},
         year = 2011,
        month = oct,
       volume = {417},
       number = {1},
        pages = {L103-L107},
          doi = {10.1111/j.1745-3933.2011.01132.x},
archivePrefix = {arXiv},
       eprint = {1108.1787},
 primaryClass = {astro-ph.CO},
       adsurl = {https://ui.adsabs.harvard.edu/abs/2011MNRAS.417L.103M}
}

@ARTICLE{Cheng1999,
       author = {{Cheng}, K.~S. and {Wang}, Jian-Min},
        title = "{The Formation and Merger of Compact Objects in the Central Engine of Active Galactic Nuclei and Quasars: Gamma-Ray Burst and Gravitational Radiation}",
      journal = {\apj},
         year = 1999,
        month = aug,
       volume = {521},
       number = {2},
        pages = {502-508},
          doi = {10.1086/307572},
archivePrefix = {arXiv},
       eprint = {astro-ph/9908228},
 primaryClass = {astro-ph},
       adsurl = {https://ui.adsabs.harvard.edu/abs/1999ApJ...521..502C}
}

@ARTICLE{Woosley2002,
       author = {{Woosley}, S.~E. and {Heger}, A. and {Weaver}, T.~A.},
        title = "{The evolution and explosion of massive stars}",
      journal = {Reviews of Modern Physics},
         year = 2002,
        month = nov,
       volume = {74},
       number = {4},
        pages = {1015-1071},
          doi = {10.1103/RevModPhys.74.1015},
       adsurl = {https://ui.adsabs.harvard.edu/abs/2002RvMP...74.1015W}
}

@ARTICLE{deGraaff2025,
       author = {{de Graaff}, Anna and {Brammer}, Gabriel and {Weibel}, Andrea and {Lewis}, Zach and {Maseda}, Michael V. and {Oesch}, Pascal A. and {Bezanson}, Rachel and {Boogaard}, Leindert A. and {Cleri}, Nikko J. and {Cooper}, Olivia R. and {Gottumukkala}, Rashmi and {Greene}, Jenny E. and {Hirschmann}, Michaela and {Hviding}, Raphael E. and {Katz}, Harley and {Labb{\'e}}, Ivo and {Leja}, Joel and {Matthee}, Jorryt and {McConachie}, Ian and {Miller}, Tim B. and {Naidu}, Rohan P. and {Price}, Sedona H. and {Rix}, Hans-Walter and {Setton}, David J. and {Suess}, Katherine A. and {Wang}, Bingjie and {Whitaker}, Katherine E. and {Williams}, Christina C.},
        title = "{RUBIES: A complete census of the bright and red distant Universe with JWST/NIRSpec}",
      journal = {\aap},
         year = 2025,
        month = may,
       volume = {697},
          eid = {A189},
        pages = {A189},
          doi = {10.1051/0004-6361/202452186},
archivePrefix = {arXiv},
       eprint = {2409.05948},
 primaryClass = {astro-ph.GA},
       adsurl = {https://ui.adsabs.harvard.edu/abs/2025A&A...697A.189D}
}

@ARTICLE{Hviding2025,
       author = {{Hviding}, Raphael E. and {de Graaff}, Anna and {Miller}, Tim B. and {Setton}, David J. and {Greene}, Jenny E. and {Labb{\'e}}, Ivo and {Brammer}, Gabriel and {Bezanson}, Rachel and {Boogaard}, Leindert A. and {Cleri}, Nikko J. and {Leja}, Joel and {Maseda}, Michael V. and {McConachie}, Ian and {Matthee}, Jorryt and {Naidu}, Rohan P. and {Oesch}, Pascal A. and {Wang}, Bingjie and {Whitaker}, Katherine E. and {Williams}, Christina C.},
        title = "{RUBIES: A spectroscopic census of little red dots: All point sources with v-shaped continua have broad lines}",
      journal = {\aap},
         year = 2025,
        month = oct,
       volume = {702},
          eid = {A57},
        pages = {A57},
          doi = {10.1051/0004-6361/202555816},
       adsurl = {https://ui.adsabs.harvard.edu/abs/2025A&A...702A..57H}
}

@ARTICLE{GW231123,
       author = {{The LIGO Scientific Collaboration} and {the Virgo Collaboration} and {the KAGRA Collaboration}},
        title = "{GW231123: a Binary Black Hole Merger with Total Mass 190-265 $M_{\odot}$}",
      journal = {arXiv e-prints},
         year = 2025,
        month = jul,
          eid = {arXiv:2507.08219},
        pages = {arXiv:2507.08219},
archivePrefix = {arXiv},
       eprint = {2507.08219},
 primaryClass = {astro-ph.HE},
       adsurl = {https://ui.adsabs.harvard.edu/abs/2025arXiv250708219T}
}

@ARTICLE{Inayoshi2024,
       author = {{Inayoshi}, Kohei and {Ichikawa}, Kohei},
        title = "{Birth of Rapidly Spinning, Overmassive Black Holes in the Early Universe}",
      journal = {\apjl},
         year = 2024,
        month = oct,
       volume = {973},
       number = {2},
          eid = {L49},
        pages = {L49},
          doi = {10.3847/2041-8213/ad74e2},
archivePrefix = {arXiv},
       eprint = {2402.14706},
 primaryClass = {astro-ph.GA},
       adsurl = {https://ui.adsabs.harvard.edu/abs/2024ApJ...973L..49I}
}

@ARTICLE{Jahnke2025,
       author = {{Jahnke}, Knud},
        title = "{The Soltan argument at redshift 6: UV-luminous quasars contribute less than 10\% to early black hole mass growth}",
      journal = {The Open Journal of Astrophysics},
         year = 2025,
        month = jan,
       volume = {8},
          eid = {9},
        pages = {9},
          doi = {10.33232/001c.129063},
archivePrefix = {arXiv},
       eprint = {2411.03184},
 primaryClass = {astro-ph.GA},
       adsurl = {https://ui.adsabs.harvard.edu/abs/2025OJAp....8E...9J}
}

@ARTICLE{Tagawa2020,
       author = {{Tagawa}, Hiromichi and {Haiman}, Zolt{\'a}n and {Kocsis}, Bence},
        title = "{Formation and Evolution of Compact-object Binaries in AGN Disks}",
      journal = {\apj},
         year = 2020,
        month = jul,
       volume = {898},
       number = {1},
          eid = {25},
        pages = {25},
          doi = {10.3847/1538-4357/ab9b8c},
archivePrefix = {arXiv},
       eprint = {1912.08218},
 primaryClass = {astro-ph.GA},
       adsurl = {https://ui.adsabs.harvard.edu/abs/2020ApJ...898...25T}
}

@ARTICLE{Magorrian1998,
       author = {{Magorrian}, John and {Tremaine}, Scott and {Richstone}, Douglas and {Bender}, Ralf and {Bower}, Gary and {Dressler}, Alan and {Faber}, S.~M. and {Gebhardt}, Karl and {Green}, Richard and {Grillmair}, Carl and {Kormendy}, John and {Lauer}, Tod},
        title = "{The Demography of Massive Dark Objects in Galaxy Centers}",
      journal = {\aj},
         year = 1998,
        month = jun,
       volume = {115},
       number = {6},
        pages = {2285-2305},
          doi = {10.1086/300353},
archivePrefix = {arXiv},
       eprint = {astro-ph/9708072},
 primaryClass = {astro-ph},
       adsurl = {https://ui.adsabs.harvard.edu/abs/1998AJ....115.2285M}
}

@BOOK{Frank2002,
       author = {{Frank}, Juhan and {King}, Andrew and {Raine}, Derek J.},
        title = "{Accretion Power in Astrophysics: Third Edition}",
         year = 2002,
       adsurl = {https://ui.adsabs.harvard.edu/abs/2002apa..book.....F}
}

@ARTICLE{Hirasawa1969,
       author = {{Hirasawa}, T.},
        title = "{Formation of Protogalaxies and Molecular Processes in Hydrogen Gas}",
      journal = {Progress of Theoretical Physics},
         year = 1969,
        month = sep,
       volume = {42},
       number = {3},
        pages = {523-543},
          doi = {10.1143/PTP.42.523},
       adsurl = {https://ui.adsabs.harvard.edu/abs/1969PThPh..42..523H}
}

@ARTICLE{Unno1967,
       author = {{Unno}, Wasaburo and {Kato}, Shoji and {Osaki}, Yoji},
        title = "{Contraction and Pulsation of Very Massive Stars}",
      journal = {\zap},
         year = 1967,
        month = jan,
       volume = {65},
        pages = {327},
       adsurl = {https://ui.adsabs.harvard.edu/abs/1967ZA.....65..327U}
}

@ARTICLE{Matsuda1969,
       author = {{Matsuda}, T. and {Sat{\={o}}}, H.},
        title = "{Hydrodynamical Behaviour of Gas Spheres with Masses of {}10$^{4}$ M$_{{\ensuremath{\odot}}}$ to {}10$^{20}$ M$_{{\ensuremath{\odot}}}$}",
      journal = {Progress of Theoretical Physics},
         year = 1969,
        month = apr,
       volume = {41},
       number = {4},
        pages = {1021-1040},
          doi = {10.1143/PTP.41.1021},
       adsurl = {https://ui.adsabs.harvard.edu/abs/1969PThPh..41.1021M}
}

@ARTICLE{Loeb1994,
       author = {{Loeb}, Abraham and {Rasio}, Frederic A.},
        title = "{Collapse of Primordial Gas Clouds and the Formation of Quasar Black Holes}",
      journal = {\apj},
         year = 1994,
        month = sep,
       volume = {432},
        pages = {52},
          doi = {10.1086/174548},
archivePrefix = {arXiv},
       eprint = {astro-ph/9401026},
 primaryClass = {astro-ph},
       adsurl = {https://ui.adsabs.harvard.edu/abs/1994ApJ...432...52L}
}

@ARTICLE{Bromm2002,
       author = {{Bromm}, Volker and {Coppi}, Paolo S. and {Larson}, Richard B.},
        title = "{The Formation of the First Stars. I. The Primordial Star-forming Cloud}",
      journal = {\apj},
         year = 2002,
        month = jan,
       volume = {564},
       number = {1},
        pages = {23-51},
          doi = {10.1086/323947},
archivePrefix = {arXiv},
       eprint = {astro-ph/0102503},
 primaryClass = {astro-ph},
       adsurl = {https://ui.adsabs.harvard.edu/abs/2002ApJ...564...23B}
}

@ARTICLE{Larson1969,
       author = {{Larson}, Richard B.},
        title = "{Numerical calculations of the dynamics of collapsing proto-star}",
      journal = {\mnras},
         year = 1969,
        month = jan,
       volume = {145},
        pages = {271},
          doi = {10.1093/mnras/145.3.271},
       adsurl = {https://ui.adsabs.harvard.edu/abs/1969MNRAS.145..271L}
}

@ARTICLE{Devecchi2009,
       author = {{Devecchi}, B. and {Volonteri}, M.},
        title = "{Formation of the First Nuclear Clusters and Massive Black Holes at High Redshift}",
      journal = {\apj},
         year = 2009,
        month = mar,
       volume = {694},
       number = {1},
        pages = {302-313},
          doi = {10.1088/0004-637X/694/1/302},
archivePrefix = {arXiv},
       eprint = {0810.1057},
 primaryClass = {astro-ph},
       adsurl = {https://ui.adsabs.harvard.edu/abs/2009ApJ...694..302D}
}

@ARTICLE{Yoshida2003,
       author = {{Yoshida}, Naoki and {Abel}, Tom and {Hernquist}, Lars and {Sugiyama}, Naoshi},
        title = "{Simulations of Early Structure Formation: Primordial Gas Clouds}",
      journal = {\apj},
         year = 2003,
        month = aug,
       volume = {592},
       number = {2},
        pages = {645-663},
          doi = {10.1086/375810},
archivePrefix = {arXiv},
       eprint = {astro-ph/0301645},
 primaryClass = {astro-ph},
       adsurl = {https://ui.adsabs.harvard.edu/abs/2003ApJ...592..645Y}
}

@ARTICLE{Abel2002,
       author = {{Abel}, Tom and {Bryan}, Greg L. and {Norman}, Michael L.},
        title = "{The Formation of the First Star in the Universe}",
      journal = {Science},
         year = 2002,
        month = jan,
       volume = {295},
       number = {5552},
        pages = {93-98},
          doi = {10.1126/science.1063991},
archivePrefix = {arXiv},
       eprint = {astro-ph/0112088},
 primaryClass = {astro-ph},
       adsurl = {https://ui.adsabs.harvard.edu/abs/2002Sci...295...93A}
}

@ARTICLE{Abel2000,
       author = {{Abel}, Tom and {Bryan}, Greg L. and {Norman}, Michael L.},
        title = "{The Formation and Fragmentation of Primordial Molecular Clouds}",
      journal = {\apj},
         year = 2000,
        month = sep,
       volume = {540},
       number = {1},
        pages = {39-44},
          doi = {10.1086/309295},
archivePrefix = {arXiv},
       eprint = {astro-ph/0002135},
 primaryClass = {astro-ph},
       adsurl = {https://ui.adsabs.harvard.edu/abs/2000ApJ...540...39A}
}

@ARTICLE{Peebles1968,
       author = {{Peebles}, P.~J.~E. and {Dicke}, R.~H.},
        title = "{Origin of the Globular Star Clusters}",
      journal = {\apj},
         year = 1968,
        month = dec,
       volume = {154},
        pages = {891},
          doi = {10.1086/149811},
       adsurl = {https://ui.adsabs.harvard.edu/abs/1968ApJ...154..891P}
}

@ARTICLE{Field1964,
       author = {{Field}, George B.},
        title = "{Quasi-Stellar Radio Sources as Spherical Galaxies in the Process of Formation.}",
      journal = {\apj},
         year = 1964,
        month = nov,
       volume = {140},
        pages = {1434},
          doi = {10.1086/148048},
       adsurl = {https://ui.adsabs.harvard.edu/abs/1964ApJ...140.1434F}
}

@ARTICLE{Hunter1962,
       author = {{Hunter}, C.},
        title = "{The Instability of the Collapse of a Self-Gravitating Gas Cloud.}",
      journal = {\apj},
         year = 1962,
        month = sep,
       volume = {136},
        pages = {594},
          doi = {10.1086/147410},
       adsurl = {https://ui.adsabs.harvard.edu/abs/1962ApJ...136..594H}
}

@ARTICLE{Hoyle1953,
       author = {{Hoyle}, F.},
        title = "{On the Fragmentation of Gas Clouds Into Galaxies and Stars.}",
      journal = {\apj},
         year = 1953,
        month = nov,
       volume = {118},
        pages = {513},
          doi = {10.1086/145780},
       adsurl = {https://ui.adsabs.harvard.edu/abs/1953ApJ...118..513H}
}

@ARTICLE{Loeb1993,
       author = {{Loeb}, Abraham},
        title = "{Cosmological Formation of Quasar Black Holes}",
      journal = {\apj},
         year = 1993,
        month = feb,
       volume = {403},
        pages = {542},
          doi = {10.1086/172224},
       adsurl = {https://ui.adsabs.harvard.edu/abs/1993ApJ...403..542L}
}

@ARTICLE{Perez-Gonzalez2024,
       author = {{P{\'e}rez-Gonz{\'a}lez}, Pablo G. and {Barro}, Guillermo and {Rieke}, George H. and {Lyu}, Jianwei and {Rieke}, Marcia and {Alberts}, Stacey and {Williams}, Christina C. and {Hainline}, Kevin and {Sun}, Fengwu and {Pusk{\'a}s}, D{\'a}vid and {Annunziatella}, Marianna and {Baker}, William M. and {Bunker}, Andrew J. and {Egami}, Eiichi and {Ji}, Zhiyuan and {Johnson}, Benjamin D. and {Robertson}, Brant and {Rodr{\'\i}guez Del Pino}, Bruno and {Rujopakarn}, Wiphu and {Shivaei}, Irene and {Tacchella}, Sandro and {Willmer}, Christopher N.~A. and {Willott}, Chris},
        title = "{What Is the Nature of Little Red Dots and what Is Not, MIRI SMILES Edition}",
      journal = {\apj},
         year = 2024,
        month = jun,
       volume = {968},
       number = {1},
          eid = {4},
        pages = {4},
          doi = {10.3847/1538-4357/ad38bb},
archivePrefix = {arXiv},
       eprint = {2401.08782},
 primaryClass = {astro-ph.GA},
       adsurl = {https://ui.adsabs.harvard.edu/abs/2024ApJ...968....4P}
}

@ARTICLE{Ferrarese2006,
       author = {{Ferrarese}, Laura and {C{\^o}t{\'e}}, Patrick and {Dalla Bont{\`a}}, Elena and {Peng}, Eric W. and {Merritt}, David and {Jord{\'a}n}, Andr{\'e}s and {Blakeslee}, John P. and {Ha{\c{s}}egan}, Monica and {Mei}, Simona and {Piatek}, Slawomir and {Tonry}, John L. and {West}, Michael J.},
        title = "{A Fundamental Relation between Compact Stellar Nuclei, Supermassive Black Holes, and Their Host Galaxies}",
      journal = {\apjl},
         year = 2006,
        month = jun,
       volume = {644},
       number = {1},
        pages = {L21-L24},
          doi = {10.1086/505388},
archivePrefix = {arXiv},
       eprint = {astro-ph/0603840},
 primaryClass = {astro-ph},
       adsurl = {https://ui.adsabs.harvard.edu/abs/2006ApJ...644L..21F}
}

@ARTICLE{Rinaldi2024,
       author = {{Rinaldi}, P. and {Bonaventura}, N. and {Rieke}, G.~H. and {Alberts}, S. and {Caputi}, K.~I. and {Baker}, W.~M. and {Baum}, S. and {Bhatawdekar}, R. and {Bunker}, A.~J. and {Carniani}, S. and {Curtis-Lake}, E. and {D'Eugenio}, F. and {Egami}, E. and {Ji}, Z. and {Hainline}, K. and {Helton}, J.~M. and {Lin}, X. and {Lyu}, J. and {Johnson}, B.~D. and {Ma}, Z. and {Maiolino}, R. and {P{\'e}rez-Gonz{\'a}lez}, P.~G. and {Rieke}, M. and {Robertson}, B.~E. and {Shivaei}, I. and {Stone}, M. and {Sun}, Y. and {Tacchella}, S. and {{\"U}bler}, H. and {Williams}, C.~C. and {Willmer}, C.~N.~A. and {Willott}, C. and {Zhang}, J. and {Zhu}, Y.},
        title = "{Not Just a Dot: the complex UV morphology and underlying properties of Little Red Dots}",
      journal = {arXiv e-prints},
         year = 2024,
        month = nov,
          eid = {arXiv:2411.14383},
        pages = {arXiv:2411.14383},
          doi = {10.48550/arXiv.2411.14383},
archivePrefix = {arXiv},
       eprint = {2411.14383},
 primaryClass = {astro-ph.GA},
       adsurl = {https://ui.adsabs.harvard.edu/abs/2024arXiv241114383R}
}

@ARTICLE{ZhangZJ2025,
       author = {{Zhang}, Zijian and {Jiang}, Linhua and {Liu}, Weiyang and {Ho}, Luis C. and {Inayoshi}, Kohei},
        title = "{JWST Insights into Narrow-line Little Red Dots}",
      journal = {arXiv e-prints},
         year = 2025,
        month = jun,
          eid = {arXiv:2506.04350},
        pages = {arXiv:2506.04350},
          doi = {10.48550/arXiv.2506.04350},
archivePrefix = {arXiv},
       eprint = {2506.04350},
 primaryClass = {astro-ph.GA},
       adsurl = {https://ui.adsabs.harvard.edu/abs/2025arXiv250604350Z}
}

@ARTICLE{Netzer2009,
       author = {{Netzer}, Hagai},
        title = "{Accretion and star formation rates in low-redshift type II active galactic nuclei}",
      journal = {\mnras},
         year = 2009,
        month = nov,
       volume = {399},
       number = {4},
        pages = {1907-1920},
          doi = {10.1111/j.1365-2966.2009.15434.x},
archivePrefix = {arXiv},
       eprint = {0907.3575},
 primaryClass = {astro-ph.GA},
       adsurl = {https://ui.adsabs.harvard.edu/abs/2009MNRAS.399.1907N}
}

@ARTICLE{MaYL2025,
       author = {{Ma}, Yilun and {Greene}, Jenny E. and {Volonteri}, Marta and {Goulding}, Andy D. and {Setton}, David J. and {Annunziatella}, Marianna and {Egami}, Eiichi and {Fan}, Xiaohui and {Kokorev}, Vasily and {Labbe}, Ivo and {Lin}, Xiaojing and {Marchesini}, Danilo and {Matthee}, Jorryt and {Nanayakkara}, Themiya and {Robbins}, Luke and {Sajina}, Anna and {Sawicki}, Marcin},
        title = "{No Luminous Little Red Dots: A Sharp Cutoff in Their Luminosity Function}",
      journal = {arXiv e-prints},
         year = 2025,
        month = sep,
          eid = {arXiv:2509.02662},
        pages = {arXiv:2509.02662},
          doi = {10.48550/arXiv.2509.02662},
archivePrefix = {arXiv},
       eprint = {2509.02662},
 primaryClass = {astro-ph.GA},
       adsurl = {https://ui.adsabs.harvard.edu/abs/2025arXiv250902662M}
}

@ARTICLE{Juodzbalis2025b,
       author = {{Juod{\v{z}}balis}, Ignas and {Marconcini}, Cosimo and {D'Eugenio}, Francesco and {Maiolino}, Roberto and {Marconi}, Alessandro and {{\"U}bler}, Hannah and {Scholtz}, Jan and {Ji}, Xihan and {Arribas}, Santiago and {Bennett}, Jake S. and {Bromm}, Volker and {Bunker}, Andrew J. and {Carniani}, Stefano and {Charlot}, St{\'e}phane and {Cresci}, Giovanni and {Dayal}, Pratika and {Egami}, Eiichi and {Fabian}, Andrew and {Inayoshi}, Kohei and {Isobe}, Yuki and {Ivey}, Lucy and {Jones}, Gareth C. and {Koudmani}, Sophie and {Laporte}, Nicolas and {Liu}, Boyuan and {Lyu}, Jianwei and {Mazzolari}, Giovanni and {Monty}, Stephanie and {Parlanti}, Eleonora and {P{\'e}rez-Gonz{\'a}lez}, Pablo G. and {Perna}, Michele and {Robertson}, Brant and {Schneider}, Raffaella and {Sijacki}, Debora and {Tacchella}, Sandro and {Trinca}, Alessandro and {Valiante}, Rosa and {Volonteri}, Marta and {Witstok}, Joris and {Zhang}, Saiyang},
        title = "{A direct black hole mass measurement in a Little Red Dot at the Epoch of Reionization}",
      journal = {arXiv e-prints},
         year = 2025,
        month = aug,
          eid = {arXiv:2508.21748},
        pages = {arXiv:2508.21748},
          doi = {10.48550/arXiv.2508.21748},
archivePrefix = {arXiv},
       eprint = {2508.21748},
 primaryClass = {astro-ph.GA},
       adsurl = {https://ui.adsabs.harvard.edu/abs/2025arXiv250821748J}
}

@ARTICLE{Carranza-Escudero2025,
       author = {{Carranza-Escudero}, Mar{\'\i}a and {Conselice}, Christopher J. and {Adams}, Nathan and {Harvey}, Thomas and {Austin}, Duncan and {Behroozi}, Peter and {Ferreira}, Leonardo and {Ormerod}, Katherine and {Duan}, Qiao and {Trussler}, James and {Li}, Qiong and {Westcott}, Lewi and {Windhorst}, Rogier A. and {Coe}, Dan and {Cohen}, Seth H. and {Cheng}, Cheng and {Driver}, Simon P. and {Frye}, Brenda and {Furtak}, Lukas J. and {Grogin}, Norman A. and {Hathi}, Nimish P. and {Jansen}, Rolf A. and {Koekemoer}, Anton M. and {Marshall}, Madeline A. and {O'Brien}, Rosalia and {Pirzkal}, Norbert and {Polletta}, Maria and {Robotham}, Aaron and {Rutkowski}, Michael J. and {Summers}, Jake and {Wilkins}, Stephen M. and {Willmer}, Christopher N.~A. and {Yan}, Haojing and {Zitrin}, Adi},
        title = "{Lonely Little Red Dots: Challenges to the Active Galactic Nucleus Nature of Little Red Dots through Their Clustering and Spectral Energy Distributions}",
      journal = {\apjl},
         year = 2025,
        month = aug,
       volume = {989},
       number = {2},
          eid = {L50},
        pages = {L50},
          doi = {10.3847/2041-8213/adf73d},
archivePrefix = {arXiv},
       eprint = {2506.04004},
 primaryClass = {astro-ph.GA},
       adsurl = {https://ui.adsabs.harvard.edu/abs/2025ApJ...989L..50C}
}

@ARTICLE{Boquien2019,
       author = {{Boquien}, M. and {Burgarella}, D. and {Roehlly}, Y. and {Buat}, V. and {Ciesla}, L. and {Corre}, D. and {Inoue}, A.~K. and {Salas}, H.},
        title = "{CIGALE: a python Code Investigating GALaxy Emission}",
      journal = {\aap},
         year = 2019,
        month = feb,
       volume = {622},
          eid = {A103},
        pages = {A103},
          doi = {10.1051/0004-6361/201834156},
archivePrefix = {arXiv},
       eprint = {1811.03094},
 primaryClass = {astro-ph.GA},
       adsurl = {https://ui.adsabs.harvard.edu/abs/2019A&A...622A.103B}
}

@ARTICLE{Baggen2023,
       author = {{Baggen}, Josephine F.~W. and {van Dokkum}, Pieter and {Labb{\'e}}, Ivo and {Brammer}, Gabriel and {Miller}, Tim B. and {Bezanson}, Rachel and {Leja}, Joel and {Wang}, Bingjie and {Whitaker}, Katherine E. and {Suess}, Katherine A. and {Nelson}, Erica J.},
        title = "{Sizes and Mass Profiles of Candidate Massive Galaxies Discovered by JWST at $7 < z < 9$: Evidence for Very Early Formation of the Central 100 pc of Present-day Ellipticals}",
      journal = {\apjl},
         year = 2023,
        month = sep,
       volume = {955},
       number = {1},
          eid = {L12},
        pages = {L12},
          doi = {10.3847/2041-8213/acf5ef},
archivePrefix = {arXiv},
       eprint = {2305.17162},
 primaryClass = {astro-ph.GA},
       adsurl = {https://ui.adsabs.harvard.edu/abs/2023ApJ...955L..12B}
}

@ARTICLE{Barkana2001,
       author = {{Barkana}, R. and {Loeb}, A.},
        title = "{In the beginning: the first sources of light and the reionization of the universe}",
      journal = {\physrep},
         year = 2001,
        month = jul,
       volume = {349},
       number = {2},
        pages = {125-238},
          doi = {10.1016/S0370-1573(01)00019-9},
archivePrefix = {arXiv},
       eprint = {astro-ph/0010468},
 primaryClass = {astro-ph},
       adsurl = {https://ui.adsabs.harvard.edu/abs/2001PhR...349..125B}
}

@ARTICLE{Neumayer2020,
       author = {{Neumayer}, Nadine and {Seth}, Anil and {B{\"o}ker}, Torsten},
        title = "{Nuclear star clusters}",
      journal = {\aapr},
         year = 2020,
        month = jul,
       volume = {28},
       number = {1},
          eid = {4},
        pages = {4},
          doi = {10.1007/s00159-020-00125-0},
archivePrefix = {arXiv},
       eprint = {2001.03626},
 primaryClass = {astro-ph.GA},
       adsurl = {https://ui.adsabs.harvard.edu/abs/2020A&ARv..28....4N}
}

@PROCEEDINGS{Kato1998,
        title = "{Black-hole accretion disks}",
    booktitle = {Black-hole accretion disks. Edited by Shoji Kato},
         year = 1998,
       editor = {{Kato}, Shoji and {Fukue}, Jun and {Mineshige}, Shin},
        month = jan,
       adsurl = {https://ui.adsabs.harvard.edu/abs/1998bhad.conf.....K}
}

@ARTICLE{LinXJ2025a,
       author = {{Lin}, Xiaojing and {Fan}, Xiaohui and {Wang}, Feige and {Sun}, Fengwu and {Champagne}, Jaclyn B. and {Egami}, Eiichi and {Kakiichi}, Koki and {Lyu}, Jianwei and {Tee}, Wei Leong and {Yang}, Jinyi and {Bian}, Fuyan and {Bosman}, Sarah E.~I. and {Cai}, Zheng and {Casey}, Caitlin M. and {Decarli}, Roberto and {Faisst}, Andreas L. and {Fujimoto}, Seiji and {Harish}, Santosh and {Ilbert}, Olivier and {Inoue}, Akio K. and {Jin}, Xiangyu and {Kartaltepe}, Jeyhan S. and {Kocevski}, Dale D. and {Li}, Mingyu and {Liu}, Weizhe and {Liu}, Yichen and {Schindler}, Jan-Torge and {Shuntov}, Marko and {Tanaka}, Takumi S. and {Vestergaard}, Marianne and {Wu}, Yunjing and {Zhang}, Haowen and {Zhang}, Zijian},
        title = "{Bridging Quasars and Little Red Dots: Insights into Broad-Line AGNs at $z=5-8$ from the First JWST COSMOS-3D Dataset}",
      journal = {arXiv e-prints},
         year = 2025,
        month = apr,
          eid = {arXiv:2504.08039},
        pages = {arXiv:2504.08039},
          doi = {10.48550/arXiv.2504.08039},
archivePrefix = {arXiv},
       eprint = {2504.08039},
 primaryClass = {astro-ph.GA},
       adsurl = {https://ui.adsabs.harvard.edu/abs/2025arXiv250408039L}
}

@ARTICLE{Setton2025,
       author = {{Setton}, David J. and {Greene}, Jenny E. and {Spilker}, Justin S. and {Williams}, Christina C. and {Labbe}, Ivo and {Ma}, Yilun and {Wang}, Bingjie and {Whitaker}, Katherine E. and {Leja}, Joel and {de Graaff}, Anna and {Alberts}, Stacey and {Bezanson}, Rachel and {Boogaard}, Leindert A. and {Brammer}, Gabriel and {Cutler}, Sam E. and {Cleri}, Nikko J. and {Cooper}, Olivia R. and {Dayal}, Pratika and {Fujimoto}, Seiji and {Furtak}, Lukas J. and {Goulding}, Andy D. and {Hirschmann}, Michaela and {Kokorev}, Vasily and {Maseda}, Michael V. and {McConachie}, Ian and {Matthee}, Jorryt and {Miller}, Tim B. and {Naidu}, Rohan P. and {Oesch}, Pascal A. and {Pan}, Richard and {Price}, Sedona H. and {Suess}, Katherine A. and {Weaver}, John R. and {Xiao}, Mengyuan and {Zhang}, Yunchong and {Zitrin}, Adi},
        title = "{A confirmed deficit of hot and cold dust emission in the most luminous Little Red Dots}",
      journal = {arXiv e-prints},
         year = 2025,
        month = mar,
          eid = {arXiv:2503.02059},
        pages = {arXiv:2503.02059},
          doi = {10.48550/arXiv.2503.02059},
archivePrefix = {arXiv},
       eprint = {2503.02059},
 primaryClass = {astro-ph.GA},
       adsurl = {https://ui.adsabs.harvard.edu/abs/2025arXiv250302059S}
}

@ARTICLE{Greene2005,
       author = {{Greene}, Jenny E. and {Ho}, Luis C.},
        title = "{Estimating Black Hole Masses in Active Galaxies Using the H{\ensuremath{\alpha}} Emission Line}",
      journal = {\apj},
         year = 2005,
        month = sep,
       volume = {630},
       number = {1},
        pages = {122-129},
          doi = {10.1086/431897},
archivePrefix = {arXiv},
       eprint = {astro-ph/0508335},
 primaryClass = {astro-ph},
       adsurl = {https://ui.adsabs.harvard.edu/abs/2005ApJ...630..122G}
}

@ARTICLE{Taylor2025,
       author = {{Taylor}, Anthony J. and {Kokorev}, Vasily and {Kocevski}, Dale D. and {Akins}, Hollis B. and {Cullen}, Fergus and {Dickinson}, Mark and {Finkelstein}, Steven L. and {Arrabal Haro}, Pablo and {Bromm}, Volker and {Giavalisco}, Mauro and {Inayoshi}, Kohei and {Juneau}, Stephanie and {Leung}, Gene C.~K. and {Perez-Gonzalez}, Pablo G. and {Somerville}, Rachel S. and {Trump}, Jonathan R. and {Amorin}, Ricardo O. and {Barro}, Guillermo and {Burgarella}, Denis and {Brooks}, Madisyn and {Carnall}, Adam and {Casey}, Caitlin M. and {Cheng}, Yingjie and {Chisholm}, John and {Chworowsky}, Katherine and {Davis}, Kelcey and {Donnan}, Callum T. and {Dunlop}, James S. and {Ellis}, Richard S. and {Fernandez}, Vital and {Fujimoto}, Seiji and {Grogin}, Norman A. and {Gupta}, Ansh R. and {Hathi}, Nimish P. and {Jung}, Intae and {Hirschmann}, Michaela and {Kartaltepe}, Jeyhan S. and {Koekemoer}, Anton M. and {Larson}, Rebecca L. and {Leung}, Ho-Hin and {Llerena}, Mario and {Lucas}, Ray A. and {McLeod}, Derek J. and {McLure}, Ross and {Napolitano}, Lorenzo and {Papovich}, Casey and {Stanton}, Thomas M. and {Tripodi}, Roberta and {Wang}, Xin and {Wilkins}, Stephen M. and {Yung}, L.~Y. Aaron and {Zavala}, Jorge A.},
        title = "{CAPERS-LRD-z9: A Gas Enshrouded Little Red Dot Hosting a Broad-line AGN at z=9.288}",
      journal = {arXiv e-prints},
         year = 2025,
        month = may,
          eid = {arXiv:2505.04609},
        pages = {arXiv:2505.04609},
          doi = {10.48550/arXiv.2505.04609},
archivePrefix = {arXiv},
       eprint = {2505.04609},
 primaryClass = {astro-ph.GA},
       adsurl = {https://ui.adsabs.harvard.edu/abs/2025arXiv250504609T}
}

@ARTICLE{Haas2003,
       author = {{Haas}, M. and {Klaas}, U. and {M{\"u}ller}, S.~A.~H. and {Bertoldi}, F. and {Camenzind}, M. and {Chini}, R. and {Krause}, O. and {Lemke}, D. and {Meisenheimer}, K. and {Richards}, P.~J. and {Wilkes}, B.~J.},
        title = "{The ISO view of Palomar-Green quasars}",
      journal = {\aap},
         year = 2003,
        month = apr,
       volume = {402},
        pages = {87-111},
          doi = {10.1051/0004-6361:20030110},
       adsurl = {https://ui.adsabs.harvard.edu/abs/2003A&A...402...87H}
}

@ARTICLE{Eisenstein2023,
       author = {{Eisenstein}, Daniel J. and {Willott}, Chris and {Alberts}, Stacey and {Arribas}, Santiago and {Bonaventura}, Nina and {Bunker}, Andrew J. and {Cameron}, Alex J. and {Carniani}, Stefano and {Charlot}, Stephane and {Curtis-Lake}, Emma and {D'Eugenio}, Francesco and {Endsley}, Ryan and {Ferruit}, Pierre and {Giardino}, Giovanna and {Hainline}, Kevin and {Hausen}, Ryan and {Jakobsen}, Peter and {Johnson}, Benjamin D. and {Maiolino}, Roberto and {Rieke}, Marcia and {Rieke}, George and {Rix}, Hans-Walter and {Robertson}, Brant and {Stark}, Daniel P. and {Tacchella}, Sandro and {Williams}, Christina C. and {Willmer}, Christopher N.~A. and {Baker}, William M. and {Baum}, Stefi and {Bhatawdekar}, Rachana and {Boyett}, Kristan and {Chen}, Zuyi and {Chevallard}, Jacopo and {Circosta}, Chiara and {Curti}, Mirko and {Danhaive}, A. Lola and {DeCoursey}, Christa and {de Graaff}, Anna and {Dressler}, Alan and {Egami}, Eiichi and {Helton}, Jakob M. and {Hviding}, Raphael E. and {Ji}, Zhiyuan and {Jones}, Gareth C. and {Kumari}, Nimisha and {L{\"u}tzgendorf}, Nora and {Laseter}, Isaac and {Looser}, Tobias J. and {Lyu}, Jianwei and {Maseda}, Michael V. and {Nelson}, Erica and {Parlanti}, Eleonora and {Perna}, Michele and {Pusk{\'a}s}, D{\'a}vid and {Rawle}, Tim and {Rodr{\'\i}guez Del Pino}, Bruno and {Sandles}, Lester and {Saxena}, Aayush and {Scholtz}, Jan and {Sharpe}, Katherine and {Shivaei}, Irene and {Silcock}, Maddie S. and {Simmonds}, Charlotte and {Skarbinski}, Maya and {Smit}, Renske and {Stone}, Meredith and {Suess}, Katherine A. and {Sun}, Fengwu and {Tang}, Mengtao and {Topping}, Michael W. and {{\"U}bler}, Hannah and {Villanueva}, Natalia C. and {Wallace}, Imaan E.~B. and {Whitler}, Lily and {Witstok}, Joris and {Woodrum}, Charity},
        title = "{Overview of the JWST Advanced Deep Extragalactic Survey (JADES)}",
      journal = {arXiv e-prints},
         year = 2023,
        month = jun,
          eid = {arXiv:2306.02465},
        pages = {arXiv:2306.02465},
          doi = {10.48550/arXiv.2306.02465},
archivePrefix = {arXiv},
       eprint = {2306.02465},
 primaryClass = {astro-ph.GA},
       adsurl = {https://ui.adsabs.harvard.edu/abs/2023arXiv230602465E}
}

@ARTICLE{Hure2011,
       author = {{Hur{\'e}}, J. -M. and {Hersant}, F.},
        title = "{The Newtonian potential of thin disks}",
      journal = {\aap},
         year = 2011,
        month = jul,
       volume = {531},
          eid = {A36},
        pages = {A36},
          doi = {10.1051/0004-6361/201015854},
archivePrefix = {arXiv},
       eprint = {1104.5079},
 primaryClass = {astro-ph.IM},
       adsurl = {https://ui.adsabs.harvard.edu/abs/2011A&A...531A..36H}
}

@ARTICLE{Lambrides2024,
       author = {{Lambrides}, Erini and {Garofali}, Kristen and {Larson}, Rebecca and {Ptak}, Andrew and {Chiaberge}, Marco and {Long}, Arianna S. and {Hutchison}, Taylor A. and {Norman}, Colin and {McKinney}, Jed and {Akins}, Hollis B. and {Berg}, Danielle A. and {Chisholm}, John and {Civano}, Francesca and {Cloonan}, Aidan P. and {Endsley}, Ryan and {Faisst}, Andreas L. and {Gilli}, Roberto and {Gillman}, Steven and {Hirschmann}, Michaela and {Kartaltepe}, Jeyhan S. and {Kocevski}, Dale D. and {Kokorev}, Vasily and {Pacucci}, Fabio and {Richardson}, Chris T. and {Stiavelli}, Massimo and {Whalen}, Kelly E.},
        title = "{The Case for Super-Eddington Accretion: Connecting Weak X-ray and UV Line Emission in JWST Broad-Line AGN During the First Gyr of Cosmic Time}",
      journal = {arXiv e-prints},
         year = 2024,
        month = sep,
          eid = {arXiv:2409.13047},
        pages = {arXiv:2409.13047},
          doi = {10.48550/arXiv.2409.13047},
archivePrefix = {arXiv},
       eprint = {2409.13047},
 primaryClass = {astro-ph.HE},
       adsurl = {https://ui.adsabs.harvard.edu/abs/2024arXiv240913047L}
}

@ARTICLE{Leung2024,
       author = {{Leung}, Gene C.~K. and {Finkelstein}, Steven L. and {P{\'e}rez-Gonz{\'a}lez}, Pablo G. and {Morales}, Alexa M. and {Taylor}, Anthony J. and {Barro}, Guillermo and {Kocevski}, Dale D. and {Akins}, Hollis B. and {Carnall}, Adam C. and {Ch{\'a}vez Ortiz}, {\'O}scar A. and {Cleri}, Nikko J. and {Cullen}, Fergus and {Donnan}, Callum T. and {Dunlop}, James S. and {Ellis}, Richard S. and {Grogin}, Norman A. and {Hirschmann}, Michaela and {Koekemoer}, Anton M. and {Kokorev}, Vasily and {Lucas}, Ray A. and {McLeod}, Derek J. and {Papovich}, Casey and {Yung}, L.~Y. Aaron},
        title = "{Exploring the Nature of Little Red Dots: Constraints on AGN and Stellar Contributions from PRIMER MIRI Imaging}",
      journal = {arXiv e-prints},
         year = 2024,
        month = nov,
          eid = {arXiv:2411.12005},
        pages = {arXiv:2411.12005},
          doi = {10.48550/arXiv.2411.12005},
archivePrefix = {arXiv},
       eprint = {2411.12005},
 primaryClass = {astro-ph.GA},
       adsurl = {https://ui.adsabs.harvard.edu/abs/2024arXiv241112005L}
}

@ARTICLE{Inayoshi2025,
       author = {{Inayoshi}, Kohei and {Maiolino}, Roberto},
        title = "{Extremely Dense Gas around Little Red Dots and High-redshift Active Galactic Nuclei: A Nonstellar Origin of the Balmer Break and Absorption Features}",
      journal = {\apjl},
         year = 2025,
        month = feb,
       volume = {980},
       number = {2},
          eid = {L27},
        pages = {L27},
          doi = {10.3847/2041-8213/adaebd},
archivePrefix = {arXiv},
       eprint = {2409.07805},
 primaryClass = {astro-ph.GA},
       adsurl = {https://ui.adsabs.harvard.edu/abs/2025ApJ...980L..27I}
}

@ARTICLE{Pacucci2024,
       author = {{Pacucci}, Fabio and {Narayan}, Ramesh},
        title = "{Mildly Super-Eddington Accretion onto Slowly Spinning Black Holes Explains the X-Ray Weakness of the Little Red Dots}",
      journal = {\apj},
         year = 2024,
        month = nov,
       volume = {976},
       number = {1},
          eid = {96},
        pages = {96},
          doi = {10.3847/1538-4357/ad84f7},
archivePrefix = {arXiv},
       eprint = {2407.15915},
 primaryClass = {astro-ph.HE},
       adsurl = {https://ui.adsabs.harvard.edu/abs/2024ApJ...976...96P}
}

@ARTICLE{Wang2004,
       author = {{Wang}, Jian-Min and {Watarai}, Ken-Ya and {Mineshige}, Shin},
        title = "{The Hot Disk Corona and Magnetic Turbulence in Radio-quiet Active Galactic Nuclei: Observational Constraints}",
      journal = {\apjl},
         year = 2004,
        month = jun,
       volume = {607},
       number = {2},
        pages = {L107-L110},
          doi = {10.1086/421906},
archivePrefix = {arXiv},
       eprint = {astro-ph/0407160},
 primaryClass = {astro-ph},
       adsurl = {https://ui.adsabs.harvard.edu/abs/2004ApJ...607L.107W}
}

@ARTICLE{Shakura1973,
       author = {{Shakura}, N.~I. and {Sunyaev}, R.~A.},
        title = "{Black holes in binary systems. Observational appearance.}",
      journal = {\aap},
         year = 1973,
        month = jan,
       volume = {24},
        pages = {337-355},
       adsurl = {https://ui.adsabs.harvard.edu/abs/1973A&A....24..337S}
}

@ARTICLE{WangBJ2024,
       author = {{Wang}, Bingjie and {Leja}, Joel and {Labb{\'e}}, Ivo and {Bezanson}, Rachel and {Whitaker}, Katherine E. and {Brammer}, Gabriel and {Furtak}, Lukas J. and {Weaver}, John R. and {Price}, Sedona H. and {Zitrin}, Adi and {Atek}, Hakim and {Coe}, Dan and {Cutler}, Sam E. and {Dayal}, Pratika and {van Dokkum}, Pieter and {Feldmann}, Robert and {Marchesini}, Danilo and {Franx}, Marijn and {F{\"o}rster Schreiber}, Natascha and {Fujimoto}, Seiji and {Geha}, Marla and {Glazebrook}, Karl and {de Graaff}, Anna and {Greene}, Jenny E. and {Juneau}, St{\'e}phanie and {Kassin}, Susan and {Kriek}, Mariska and {Khullar}, Gourav and {Maseda}, Michael and {Mowla}, Lamiya A. and {Muzzin}, Adam and {Nanayakkara}, Themiya and {Nelson}, Erica J. and {Oesch}, Pascal A. and {Pacifici}, Camilla and {Pan}, Richard and {Papovich}, Casey and {Setton}, David J. and {Shapley}, Alice E. and {Smit}, Renske and {Stefanon}, Mauro and {Suess}, Katherine A. and {Taylor}, Edward N. and {Williams}, Christina C.},
        title = "{The UNCOVER Survey: A First-look HST+JWST Catalog of Galaxy Redshifts and Stellar Population Properties Spanning 0.2 {\ensuremath{\lesssim}} z {\ensuremath{\lesssim}} 15}",
      journal = {\apjs},
         year = 2024,
        month = jan,
       volume = {270},
       number = {1},
          eid = {12},
        pages = {12},
          doi = {10.3847/1538-4365/ad0846},
archivePrefix = {arXiv},
       eprint = {2310.01276},
 primaryClass = {astro-ph.GA},
       adsurl = {https://ui.adsabs.harvard.edu/abs/2024ApJS..270...12W}
}

@ARTICLE{Casey2025,
       author = {{Casey}, Caitlin M. and {Akins}, Hollis B. and {Finkelstein}, Steven L. and {Franco}, Maximilien and {Fujimoto}, Seiji and {Liu}, Daizhong and {Long}, Arianna S. and {Magdis}, Georgios and {Manning}, Sinclaire M. and {McKinney}, Jed and {Shuntov}, Marko and {Tanaka}, Takumi S.},
        title = "{An upper limit of 10$^6$ M$_\odot$ in dust from ALMA observations in 60 Little Red Dots}",
      journal = {arXiv e-prints},
         year = 2025,
        month = may,
          eid = {arXiv:2505.18873},
        pages = {arXiv:2505.18873},
          doi = {10.48550/arXiv.2505.18873},
archivePrefix = {arXiv},
       eprint = {2505.18873},
 primaryClass = {astro-ph.GA},
       adsurl = {https://ui.adsabs.harvard.edu/abs/2025arXiv250518873C}
}

@ARTICLE{Kaspi2000,
       author = {{Kaspi}, Shai and {Smith}, Paul S. and {Netzer}, Hagai and {Maoz}, Dan and {Jannuzi}, Buell T. and {Giveon}, Uriel},
        title = "{Reverberation Measurements for 17 Quasars and the Size-Mass-Luminosity Relations in Active Galactic Nuclei}",
      journal = {\apj},
         year = 2000,
        month = apr,
       volume = {533},
       number = {2},
        pages = {631-649},
          doi = {10.1086/308704},
archivePrefix = {arXiv},
       eprint = {astro-ph/9911476},
 primaryClass = {astro-ph},
       adsurl = {https://ui.adsabs.harvard.edu/abs/2000ApJ...533..631K}
}

@ARTICLE{Bentz2013,
       author = {{Bentz}, Misty C. and {Denney}, Kelly D. and {Grier}, Catherine J. and {Barth}, Aaron J. and {Peterson}, Bradley M. and {Vestergaard}, Marianne and {Bennert}, Vardha N. and {Canalizo}, Gabriela and {De Rosa}, Gisella and {Filippenko}, Alexei V. and {Gates}, Elinor L. and {Greene}, Jenny E. and {Li}, Weidong and {Malkan}, Matthew A. and {Pogge}, Richard W. and {Stern}, Daniel and {Treu}, Tommaso and {Woo}, Jong-Hak},
        title = "{The Low-luminosity End of the Radius-Luminosity Relationship for Active Galactic Nuclei}",
      journal = {\apj},
         year = 2013,
        month = apr,
       volume = {767},
       number = {2},
          eid = {149},
        pages = {149},
          doi = {10.1088/0004-637X/767/2/149},
archivePrefix = {arXiv},
       eprint = {1303.1742},
 primaryClass = {astro-ph.CO},
       adsurl = {https://ui.adsabs.harvard.edu/abs/2013ApJ...767..149B}
}

@ARTICLE{Juodzbalis2025,
       author = {{Juod{\v{z}}balis}, Ignas and {Maiolino}, Roberto and {Baker}, William M. and {Lake}, Emma Curtis and {Scholtz}, Jan and {D'Eugenio}, Francesco and {Trefoloni}, Bartolomeo and {Isobe}, Yuki and {Tacchella}, Sandro and {Bunker}, Andrew J. and {Carniani}, Stefano and {Charlot}, St{\'e}phane and {Jones}, Gareth C. and {Parlanti}, Eleonora and {Perna}, Michele and {Rinaldi}, Pierluigi and {Robertson}, Brant and {{\"U}bler}, Hannah and {Venturi}, Giacomo and {Willott}, Chris},
        title = "{JADES: comprehensive census of broad-line AGN from Reionization to Cosmic Noon revealed by JWST}",
      journal = {arXiv e-prints},
         year = 2025,
        month = apr,
          eid = {arXiv:2504.03551},
        pages = {arXiv:2504.03551},
          doi = {10.48550/arXiv.2504.03551},
archivePrefix = {arXiv},
       eprint = {2504.03551},
 primaryClass = {astro-ph.GA},
       adsurl = {https://ui.adsabs.harvard.edu/abs/2025arXiv250403551J}
}

@ARTICLE{Maiolino2024,
       author = {{Maiolino}, Roberto and {Scholtz}, Jan and {Witstok}, Joris and {Carniani}, Stefano and {D'Eugenio}, Francesco and {de Graaff}, Anna and {{\"U}bler}, Hannah and {Tacchella}, Sandro and {Curtis-Lake}, Emma and {Arribas}, Santiago and {Bunker}, Andrew and {Charlot}, St{\'e}phane and {Chevallard}, Jacopo and {Curti}, Mirko and {Looser}, Tobias J. and {Maseda}, Michael V. and {Rawle}, Timothy D. and {Rodr{\'\i}guez del Pino}, Bruno and {Willott}, Chris J. and {Egami}, Eiichi and {Eisenstein}, Daniel J. and {Hainline}, Kevin N. and {Robertson}, Brant and {Williams}, Christina C. and {Willmer}, Christopher N.~A. and {Baker}, William M. and {Boyett}, Kristan and {DeCoursey}, Christa and {Fabian}, Andrew C. and {Helton}, Jakob M. and {Ji}, Zhiyuan and {Jones}, Gareth C. and {Kumari}, Nimisha and {Laporte}, Nicolas and {Nelson}, Erica J. and {Perna}, Michele and {Sandles}, Lester and {Shivaei}, Irene and {Sun}, Fengwu},
        title = "{A small and vigorous black hole in the early Universe}",
      journal = {\nat},
         year = 2024,
        month = mar,
       volume = {627},
       number = {8002},
        pages = {59-63},
          doi = {10.1038/s41586-024-07052-5},
archivePrefix = {arXiv},
       eprint = {2305.12492},
 primaryClass = {astro-ph.GA},
       adsurl = {https://ui.adsabs.harvard.edu/abs/2024Natur.627...59M}
}

@ARTICLE{Hubeny1990,
       author = {{Hubeny}, I.},
        title = "{Vertical Structure of Accretion Disks: A Simplified Analytical Model}",
      journal = {\apj},
         year = 1990,
        month = mar,
       volume = {351},
        pages = {632},
          doi = {10.1086/168501},
       adsurl = {https://ui.adsabs.harvard.edu/abs/1990ApJ...351..632H}
}

@ARTICLE{WangYL2025,
       author = {{Wang}, Yi-Lin and {Liu}, Jun-Rong and {Wang}, Jian-Min},
        title = "{Continuum reverberation mapping of accretion disks depending on the vertical structures in active galactic nuclei}",
      journal = {\aap},
         year = 2025,
        month = mar,
       volume = {695},
          eid = {A143},
        pages = {A143},
          doi = {10.1051/0004-6361/202452114},
       adsurl = {https://ui.adsabs.harvard.edu/abs/2025A&A...695A.143W}
}

@ARTICLE{Ross1992,
       author = {{Ross}, R.~R. and {Fabian}, A.~C. and {Mineshige}, S.},
        title = "{The spectra of accretion discs in active galactic nuclei.}",
      journal = {\mnras},
         year = 1992,
        month = sep,
       volume = {258},
        pages = {189-197},
          doi = {10.1093/mnras/258.1.189},
       adsurl = {https://ui.adsabs.harvard.edu/abs/1992MNRAS.258..189R}
}

@ARTICLE{Yoshioka2024,
       author = {{Yoshioka}, Shogo and {Mineshige}, Shin and {Ohsuga}, Ken and {Kawashima}, Tomohisa and {Kitaki}, Takaaki},
        title = "{Radiation and outflow properties of super-Eddington accretion flows around various mass classes of black holes: Dependence on the accretion rates}",
      journal = {\pasj},
         year = 2024,
        month = oct,
       volume = {76},
       number = {5},
        pages = {1015-1025},
          doi = {10.1093/pasj/psae067},
archivePrefix = {arXiv},
       eprint = {2407.15927},
 primaryClass = {astro-ph.HE},
       adsurl = {https://ui.adsabs.harvard.edu/abs/2024PASJ...76.1015Y}
}

@ARTICLE{Mineshige2000,
       author = {{Mineshige}, Shin and {Kawaguchi}, Toshihiro and {Takeuchi}, Mitsuru and {Hayashida}, Kiyoshi},
        title = "{Slim-Disk Model for Soft X-Ray Excess and Variability of Narrow-Line Seyfert 1 Galaxies}",
      journal = {\pasj},
         year = 2000,
        month = jun,
       volume = {52},
        pages = {499-508},
          doi = {10.1093/pasj/52.3.499},
archivePrefix = {arXiv},
       eprint = {astro-ph/0003017},
 primaryClass = {astro-ph},
       adsurl = {https://ui.adsabs.harvard.edu/abs/2000PASJ...52..499M}
}

@ARTICLE{Abramowicz1988,
       author = {{Abramowicz}, M.~A. and {Czerny}, B. and {Lasota}, J.~P. and {Szuszkiewicz}, E.},
        title = "{Slim Accretion Disks}",
      journal = {\apj},
         year = 1988,
        month = sep,
       volume = {332},
        pages = {646},
          doi = {10.1086/166683},
       adsurl = {https://ui.adsabs.harvard.edu/abs/1988ApJ...332..646A}
}

@ARTICLE{Wang1999b,
       author = {{Wang}, Jian-Min and {Zhou}, You-Yuan},
        title = "{Self-similar Solution of Optically Thick Advection-dominated Flows}",
      journal = {\apj},
         year = 1999,
        month = may,
       volume = {516},
       number = {1},
        pages = {420-424},
          doi = {10.1086/307080},
       adsurl = {https://ui.adsabs.harvard.edu/abs/1999ApJ...516..420W}
}

@ARTICLE{Sirko2003,
       author = {{Sirko}, Edwin and {Goodman}, Jeremy},
        title = "{Spectral energy distributions of marginally self-gravitating quasi-stellar object discs}",
      journal = {\mnras},
         year = 2003,
        month = may,
       volume = {341},
       number = {2},
        pages = {501-508},
          doi = {10.1046/j.1365-8711.2003.06431.x},
archivePrefix = {arXiv},
       eprint = {astro-ph/0209469},
 primaryClass = {astro-ph},
       adsurl = {https://ui.adsabs.harvard.edu/abs/2003MNRAS.341..501S}
}

@ARTICLE{Zhou2024,
       author = {{Zhou}, Shuying and {Sun}, Mouyuan and {Liu}, Tong and {Wang}, Jian-Min and {Wang}, Jun-Xian and {Xue}, Yongquan},
        title = "{Stellar Black Holes Can ``Stretch'' Supermassive Black Hole Accretion Disks}",
      journal = {\apjl},
         year = 2024,
        month = may,
       volume = {966},
       number = {1},
          eid = {L9},
        pages = {L9},
          doi = {10.3847/2041-8213/ad3c3f},
archivePrefix = {arXiv},
       eprint = {2404.07407},
 primaryClass = {astro-ph.HE},
       adsurl = {https://ui.adsabs.harvard.edu/abs/2024ApJ...966L...9Z}
}

@ARTICLE{Chen2025,
       author = {{Chen}, Yi-Xian and {Jiang}, Yan-Fei and {Goodman}, Jeremy},
        title = "{Accretion of AGN Stars under Influence of Disk Geometry}",
      journal = {arXiv e-prints},
         year = 2025,
        month = may,
          eid = {arXiv:2505.13951},
        pages = {arXiv:2505.13951},
          doi = {10.48550/arXiv.2505.13951},
archivePrefix = {arXiv},
       eprint = {2505.13951},
 primaryClass = {astro-ph.HE},
       adsurl = {https://ui.adsabs.harvard.edu/abs/2025arXiv250513951C}
}

@ARTICLE{Chen2024,
       author = {{Chen}, Yi-Xian and {Jiang}, Yan-Fei and {Goodman}, Jeremy and {Lin}, Douglas N.~C.},
        title = "{Radiation Hydrodynamic Simulations of Massive Stars in Gas-rich Environments: Accretion of AGN Stars Suppressed by Thermal Feedback}",
      journal = {\apj},
         year = 2024,
        month = oct,
       volume = {974},
       number = {1},
          eid = {106},
        pages = {106},
          doi = {10.3847/1538-4357/ad6dd4},
archivePrefix = {arXiv},
       eprint = {2408.12017},
 primaryClass = {astro-ph.HE},
       adsurl = {https://ui.adsabs.harvard.edu/abs/2024ApJ...974..106C}
}

@ARTICLE{Luo2025,
       author = {{Luo}, Yang and {Wang}, Jian-Min},
        title = "{Numerical modeling the mass feeding rates onto accretion-modifieds stars embedded within AGN disks}",
      journal = {\mnras},
         year = 2025,
        month = may,
          doi = {10.1093/mnras/staf841},
archivePrefix = {arXiv},
       eprint = {2505.15048},
 primaryClass = {astro-ph.GA},
       adsurl = {https://ui.adsabs.harvard.edu/abs/2025MNRAS.tmp..796L}
}

@ARTICLE{Wang2025,
       author = {{Wang}, Jian-Min and {Hu}, Chen and {Chen}, Yong-Jie and {Songsheng}, Yu-Yang and {Wang}, Yi-Lin and {Zhang}, Hao and {Du}, Pu and {Li}, Yan-Rong and {Luo}, Bin and {Brotherton}, Michael S. and {Bai}, Jin-Ming and {Guo}, Wei-Jian and {Yang}, Seng and {Yao}, Zhu-Heng and {Aceituno}, Jesus},
        title = "{Detection of unexpected leading reverberations of broad H$\beta$ line in the quasar PHL\,1092}",
      journal = {arXiv e-prints},
         year = 2025,
        month = nov,
          eid = {arXiv:2511.07716},
        pages = {arXiv:2511.07716},
archivePrefix = {arXiv},
       eprint = {2511.07716},
 primaryClass = {astro-ph.GA},
       adsurl = {https://ui.adsabs.harvard.edu/abs/2025arXiv2511.07716}
}

@ARTICLE{Wang2014,
       author = {{Wang}, Jian-Min and {Du}, Pu and {Hu}, Chen and {Netzer}, Hagai and {Bai}, Jin-Ming and {Lu}, Kai-Xing and {Kaspi}, Shai and {Qiu}, Jie and {Li}, Yan-Rong and {Wang}, Fang and {SEAMBH Collaboration}},
        title = "{Supermassive Black Holes with High Accretion Rates in Active Galactic Nuclei. II. The Most Luminous Standard Candles in the Universe}",
      journal = {\apj},
         year = 2014,
        month = oct,
       volume = {793},
       number = {2},
          eid = {108},
        pages = {108},
          doi = {10.1088/0004-637X/793/2/108},
archivePrefix = {arXiv},
       eprint = {1408.2337},
 primaryClass = {astro-ph.HE},
       adsurl = {https://ui.adsabs.harvard.edu/abs/2014ApJ...793..108W}
}

@ARTICLE{Du2019,
       author = {{Du}, Pu and {Wang}, Jian-Min},
        title = "{The Radius-Luminosity Relationship Depends on Optical Spectra in Active Galactic Nuclei}",
      journal = {\apj},
         year = 2019,
        month = nov,
       volume = {886},
       number = {1},
          eid = {42},
        pages = {42},
          doi = {10.3847/1538-4357/ab4908},
archivePrefix = {arXiv},
       eprint = {1909.06735},
 primaryClass = {astro-ph.GA},
       adsurl = {https://ui.adsabs.harvard.edu/abs/2019ApJ...886...42D}
}

@ARTICLE{Zhuang2025,
       author = {{Zhuang}, Ming-Yang and {Li}, Junyao and {Shen}, Yue and {Lin}, Xiaojing and {Shapley}, Alice E. and {Wang}, Feige and {Wu}, Qiaoya and {Yang}, Qian},
        title = "{NEXUS: A Spectroscopic Census of Broad-line AGNs and Little Red Dots at $3\lesssim z\lesssim 6$}",
      journal = {arXiv e-prints},
         year = 2025,
        month = may,
          eid = {arXiv:2505.20393},
        pages = {arXiv:2505.20393},
          doi = {10.48550/arXiv.2505.20393},
archivePrefix = {arXiv},
       eprint = {2505.20393},
 primaryClass = {astro-ph.GA},
       adsurl = {https://ui.adsabs.harvard.edu/abs/2025arXiv250520393Z}
}

\newpage



\clearpage

\clearpage

\end{document}